\documentclass{report}

\title{Domain Specific Language for Modular Knitting Pattern Definitions: \\ Purl}
\author{Chelsea Battell \\ \small \textit{McMaster University}}
\date{April 2014}

\usepackage{listings}
\usepackage{syntax}
\usepackage{fullpage}
\usepackage[toc,page]{appendix}
\usepackage{setspace}
\usepackage{cite}
\usepackage{url}

\lstdefinelanguage{Purl}{
  keywords={pattern, section, sample, repeat, CO, BO, circular, provisional, row, rnd, MC, CC, with, to, last, end},
  keywordstyle=\bfseries,
  sensitive=false,
  comment=[l]{//},
  morecomment=[s]{/*}{*/},
  morestring=[b]',
  morestring=[b]"
}

\lstnewenvironment{purlex}[2][]%
{%
\noindent

\lstset{language=Purl,basicstyle=\onehalfspacing\slshape\normalsize,xleftmargin=0.2\textwidth,xrightmargin=0.2\textwidth,breaklines=true,numbers=none,captionpos=b,frame=single,extendedchars=true,caption=#1,label=#2}
\vspace{10px}
}
{
}

\lstnewenvironment{patternex}[2][]%
{%
\noindent

\lstset{language={},basicstyle=\normalsize,breaklines=true,numbers=none,captionpos=b,frame=single,extendedchars=true,showspaces=false,showstringspaces=false,tabsize=2,showtabs=false,caption=#1,label=#2}
\vspace{10px}
}
{
}

\lstdefinelanguage{JavaScript}{
  keywords={typeof, new, true, false, catch, function, return, null, catch, switch, var, if, in, while, do, else, case, break},
  keywordstyle=\bfseries,
  identifierstyle=\rmfamily,
  sensitive=false,
  comment=[l]{//},
  morecomment=[s]{/*}{*/},
  commentstyle=\ttfamily,
  stringstyle=\ttfamily,
  morestring=[b]',
  morestring=[b]"
}

\usepackage[T1]{fontenc}
\usepackage[utf8]{inputenc}
\usepackage{mathpazo}
\usepackage[linktocpage=true,pdftex=true,colorlinks=false]{hyperref}
\begin{document}

\maketitle
\tableofcontents

\begin{abstract}

\flushleft

Purl is a language to be used for modular definition and verification of knitting patterns. The syntax is similar to the standard
knitting pattern notation provided by the Craft Yarn Council (see~\cite{cycstandards}). Purl provides constructs not
available in the standard notation to allow reuse of segments of patterns.

\bigskip

This report was written using the literate programming (see~\cite{literate}) tool Literate Programming for Eclipse
(LEP, see~\cite{lep}). The compiler source code, and HTML and CSS for a web page with rudimentary IDE for Purl are ``tangled'' by
the LEP plugin in the Eclipse IDE~\cite{eclipse}.

\end{abstract}

\chapter{Introduction}

Knitting is a process by which a strand of yarn can be turned into flexible fabric. Patterns are written to record knitting
designs and techniques. Simple project patterns using only a few different stitches may be shared orally, but to knit more
complicated objects it becomes necessary to document these instructions. The earliest known example of an object being knit
using stitches other than the knit stitch dates from the 16th century (purl stitch and yarn over,
see~\cite{historyhandknit}). A popular and once standard instruction reference for knitting~\cite{encyclopedianeedlework}
written in 1886 lists 12 stitches. The stitch reference~\cite{stitchcollection} published in the last five years lists 30
stitches basic (not all included in this project). Improvements in pattern documentation techniques and tools is both necessary
for, and conducive to, increased complexity and innovation in knitting patterns.

Patterns are typically written according to the standard~\cite{cycstandards}, and may include extra information, such as the
number of active stitches that should be on the needles at the end of each row. This extra information guides the knitter to
help avoid mistakes, but this assumes that the pattern has no errors. In the most common case (not including industrial
settings), for every pattern written, a pattern designer has to experiment and perform many tedious computations to make
sure their pattern is correct. Also, even though segments of a pattern may be reused in many others, they are rewritten in
each new pattern.

This project implements a compiler for a language called Purl. The syntax is similar to the standard knitting pattern
notation~\cite{cycstandards}. Purl has features to increase reusability of segments of knitting patterns. The compiler
performs automatic verification of the number of stitches in each row and displays this guiding information in the output
pattern.

\section{Knitting Pattern Example}

Below is a knitting pattern for a bag~\cite{marketbagpattern} in the standard notation. This pattern is broken into two
sections: body and handle. Each section begins with how to add the initial stitches to the needle (cast-on for the body,
pick-up stitches for the handles). The rows of the pattern are defined next, and each section ends with either a bind-off or
a join. A single asterisk means to repeat stitches as directed. Two asterisks means to repeat the rows betwen the double
asterisk and the row repeat instructions.

\begin{patternex}[Market Bag]{marketbagex}
Market Bag

Body

Cast-on 8 sts circular.

Rnd 1 (RS): *K, yo, K; rep from * to end.
Rnd 2 (RS): *K; rep from * to end.
Rnd 3 (RS): (K, yo, K, yo, K) 4 times.
Rnd 4 (RS): *K; rep from * to end.
Rnd 5 (RS): (K, yo, K3, yo, K) 4 times.
Rnd 6 (RS): *K; rep from * to end.
Rnd 7 (RS): (K, yo, K5, yo, K) 4 times.
Rnd 8 (RS): *K; rep from * to end.
Rnd 9 (RS): (K, yo, K7, yo, K) 4 times.
Rnd 10 (RS): *K; rep from * to end.
Rnd 11 (RS): (K, yo, K9, yo, K) 4 times.
Rnd 12 (RS): *K; rep from * to end.
Rnd 13 (RS): (K, yo, K11, yo, K) 4 times.
Rnd 14 (RS): *K; rep from * to end.
Rnd 15 (RS): (K, yo, K13, yo, K) 4 times.
Rnd 16 (RS): *K; rep from * to end.
Rnd 17 (RS): (K, yo, K15, yo, K) 4 times.
Rnd 18 (RS): *K; rep from * to end.
Rnd 19 (RS): (K, yo, K17, yo, K) 4 times.
Rnd 20 (RS): *K; rep from * to end.
Rnd 21 (RS): (K, yo, K19, yo, K) 4 times.
Rnd 22 (RS): *K; rep from * to end.
Rnd 23 (RS): (K, yo, K21, yo, K) 4 times.
Rnd 24 (RS): *K; rep from * to end

**
Rnd 25 (RS): *k2tog, yo; rep from * to end.
Rnd 26 (RS): *K; rep from * to end.
rep from ** 30 times

**
Rnd 27 (CC) (RS): *K; rep from * to end.
Rnd 28 (CC) (RS): *P; rep from * to end.
rep from ** 4 times

Bind-off 100 sts.


Handle

Pick-up 10 sts from body top.

**
Row 1 (CC) (RS): *K; rep from * to end.
Row 2 (CC) (WS): *P; rep from * to end.
rep from ** 2 times

Row 3 (RS): K, k2tog, K4, k2tog, K.

**
Row 4 (CC) (WS): *K; rep from * to end.
Row 5 (CC) (RS): *P; rep from * to end.
rep from ** 100 times

Row 6 (WS): K, M1, K6, M1, K.

**
Row 7 (CC) (RS): *K; rep from * to end.
Row 8 (CC) (WS): *P; rep from * to end.
rep from ** 2 times

Join 10 sts to opposite side of body top.
\end{patternex}

The program in Purl to generate this pattern is below. All of the blocks beginning with ``sample'' are a new construct
introduced to increase reusability of parts of knitting patterns. These samples could be used in many other patterns and the
definition of the market bag pattern (below the samples) becomes much simpler.

\begin{purlex}[Market bag pattern]{marketbagpurlex}
sample circle with n, max
| n < max:
    rnd : [K, YO, K n, YO, K] 4.
    rnd : *K; to end.
    circle with n + 2, max.

sample diagonalLace with n:
    **
    rnd : *K2T, YO; to end.
    rnd : *K; to end.
    repeat n

sample garterStitchCC with n, type
| type = 0:
    **
        row CC : *K; to end.
        row CC : *P; to end.
    repeat n
| type = 1:
    **
        rnd CC : *K; to end.
        rnd CC : *P; to end.
    repeat n


pattern "Market Bag":

section "Body":
CO 8 circular.
rnd : *K, YO, K; to end.
rnd : *K; to end.
circle with 1, 23.
diagonalLace with 30.
garterStitchCC with 4, 1.
BO 100.

section "Handle":
PU 10 from "Body top".
garterStitchCC with 2, 0.
row : K, K2T, K 4, K2T, K.
garterStitchCC with 100, 0.
row : K, M1, K 6, M1, K.
garterStitchCC with 2, 0.
Join 10 to "Body top".
\end{purlex}

The constructs used in this example are discussed in more detail throughout this report.

\subsection{Verification Example}

In the section for the \emph{body} of the bag, the number of stitches is increasing for all but the last couple of rows. It
can be difficult for a knitter to keep track of the number of stitches they should have on the needle. Rows in this pattern
in the standard notation could be written with the stitch count at the end of each row as:

\begin{patternex}[Stitch count]{stcountex}
Row 1 : *K, yo, K; rep from * to end. (12 sts)
\end{patternex}

This is helpful, but as discussed in the intro, it is tedious (but necessary) for the pattern designer to figure out this value
for every row. The Purl compiler computes this value, ensures that each row uses all of the stitches of the row before it,
and displays it at the end of every row for reference for the knitter.

\section{Purl Compiler Overview}

The Purl compiler consists of a three-pass top-down parser and a back-end that generates a knitting pattern in the standard
notation in HTML. The first pass constructs an abstract syntax tree according to the provided grammar and is implemented
using recursive descent parsing techniques as described in~\cite{dragonbook}. The lexer is a module used by this pass.
Errors and warnings reported by the first pass are only lexical and syntactic errors. Pass two traverses the syntax tree
depth first, replacing all variables and constructs that are purely elements of Purl and are not expressible in the standard
notation. These constructs will be discussed when exploring the individual pattern elements. In the third pass, the syntax
tree is again traversed depth first. All verification occurs in this pass and errors indicate problems in the structure of
the knitting pattern. A global \texttt{State} object is used throughout parsing to track information necessary for error
reporting, such as section name, position in code, and row number in the generated pattern. It is also used in the
verification pass to track the pattern orientation, width, and row index, and to update nodes with these values as necessary.

The reason for breaking up parsing into three passes is because a syntax tree representation of the pattern is much easier
to manipulate and verify. A main feature of Purl is the ability to define modular and parametrized segments of patterns,
through the \emph{pattern sample} construct introduced by this language (see~\ref{sec:sample}), so a second pass is used for
trimming nodes representing sample calls. Also, there are some challenges in verifying a pattern. It is necessary that every
row works all of the stitches of the previous row, but there are some pattern constructs which work a number of stitches
that depends on the width of the current row. Since we allow modular pattern definitions and parameterized segments of
patterns, this verification cannot be done in a single pass over the source language.

\section{Error Handling}
 
Below are explanations of many of the strategies used by the compiler for handling errors, including panic-mode recovery
approximately as discussed in~\cite{dragonbook}. They are covered here to avoid redundant explanations later.

\begin{enumerate}

\item Whenever a character symbol is expected, there are some characters that are considered likely errors. These likely
errors will generate a \emph{warning}, and compilation will continue as usual.

\begin{center}
\begin{tabular}{|c|c|}
\hline Expected & Likely typo\\ \hline
: & ;\\ \hline
; & :\\ \hline
. & ,\\ \hline
, & .\\ \hline
** & *\\ \hline
\end{tabular}
\end{center}

\item If a keyword is expected and an ident is found, then a typo is assumed. In this case a \emph{warning} is created, and compilation
continues.

\item Any other unexpected symbols generate an \emph{error} and the lexer will scan to the end of the production (usually the
period or comma symbols) and return the node to resume compilation at the next sibling. For certain stitches (as will be
noted later), there is no reliable symbol delimiting siblings. In this case, the lexer will scan to the end of the
parent production.

\begin{itemize}
	\item If an unexpected keyword is found, then the error message reports an invalid use of keyword.
	\item If an unexpected ident symbol is found, then the error message reports an invalid use of ident.
	\item If an unexpected character symbol is found, then the error message reports an invalid use of character.
\end{itemize}

\begin{lstlisting}[title={<Unexpected Symbol Error 1>}, label=Listing1]
 if (Sym.type == SymType.Ident) {
	AddMsg(MsgType.Error, node, "Invalid use of ident " + Sym.value + ".");
} else if (hasOwnValue(KeywordSym, Sym.type)) {
	AddMsg(MsgType.Error, node, "Invalid use of keyword " + Sym.value + ".");
} else if (hasOwnValue(CharSym, Sym.type)) {
	AddMsg(MsgType.Error, node, "Invalid use of \'" + Sym.value + "\' character.");
}
\end{lstlisting}\begin{footnotesize} \textsc{Used in}: \hyperref[Listing125]{A\-s\-t\- \-C\-o\-n\-s\-t\-r\-u\-c\-t\-i\-o\-n\- \-P\-a\-s\-s on page} \pageref{Listing125}, \hyperref[Listing86]{P\-a\-t\-t\-e\-r\-n\- \-P\-a\-r\-s\-e\- \-P\-a\-t\-t\-e\-r\-n on page} \pageref{Listing86}, \hyperref[Listing87]{P\-a\-t\-t\-e\-r\-n\- \-P\-a\-r\-s\-e\- \-T\-i\-t\-l\-e on page} \pageref{Listing87}, \hyperref[Listing18]{C\-o\-l\-o\-n\- \-S\-e\-p\-a\-r\-a\-t\-o\-r on page} \pageref{Listing18}, \hyperref[Listing3]{C\-a\-s\-t\--\-O\-n\- \-P\-a\-r\-s\-e\- \-K\-e\-y\-w\-o\-r\-d on page} \pageref{Listing3}, \hyperref[Listing4]{C\-a\-s\-t\--\-O\-n\- \-P\-a\-r\-s\-e\- \-V\-a\-l\-u\-e on page} \pageref{Listing4}, \hyperref[Listing6]{P\-e\-r\-i\-o\-d\- \-T\-e\-r\-m\-i\-n\-a\-t\-o\-r on page} \pageref{Listing6}, \hyperref[Listing10]{P\-i\-c\-k\--\-U\-p\- \-P\-a\-r\-s\-e\- \-K\-e\-y\-w\-o\-r\-d on page} \pageref{Listing10}, \hyperref[Listing11]{P\-i\-c\-k\--\-U\-p\- \-P\-a\-r\-s\-e\- \-V\-a\-l\-u\-e on page} \pageref{Listing11}, \hyperref[Listing12]{P\-i\-c\-k\--\-U\-p\- \-P\-a\-r\-s\-e\- \-O\-r\-i\-g\-i\-n on page} \pageref{Listing12}, \hyperref[Listing12]{P\-i\-c\-k\--\-U\-p\- \-P\-a\-r\-s\-e\- \-O\-r\-i\-g\-i\-n on page} \pageref{Listing12}, \hyperref[Listing16]{R\-o\-w\- \-D\-e\-f\-i\-n\-i\-t\-i\-o\-n\- \-P\-a\-r\-s\-e\- \-R\-o\-w\- \-T\-y\-p\-e on page} \pageref{Listing16}, \hyperref[Listing27]{R\-o\-w\- \-E\-l\-e\-m\-e\-n\-t\- \-P\-a\-r\-s\-e on page} \pageref{Listing27}, \hyperref[Listing28]{S\-t\-i\-t\-c\-h\- \-O\-p\- \-P\-a\-r\-s\-e on page} \pageref{Listing28}, \hyperref[Listing67]{U\-n\-d\-e\-t\-e\-r\-m\-i\-n\-e\-d\- \-S\-t\-i\-t\-c\-h\- \-R\-e\-p\-e\-a\-t\- \-P\-a\-r\-s\-e\- \-O\-p\-e\-n on page} \pageref{Listing67}, \hyperref[Listing69]{U\-n\-d\-e\-t\-e\-r\-m\-i\-n\-e\-d\- \-S\-t\-i\-t\-c\-h\- \-R\-e\-p\-e\-a\-t\- \-P\-a\-r\-s\-e\- \-C\-l\-o\-s\-e on page} \pageref{Listing69}, \hyperref[Listing70]{U\-n\-d\-e\-t\-e\-r\-m\-i\-n\-e\-d\- \-S\-t\-i\-t\-c\-h\- \-R\-e\-p\-e\-a\-t\- \-P\-a\-r\-s\-e\- \-T\-o on page} \pageref{Listing70}, \hyperref[Listing71]{U\-n\-d\-e\-t\-e\-r\-m\-i\-n\-e\-d\- \-S\-t\-i\-t\-c\-h\- \-R\-e\-p\-e\-a\-t\- \-P\-a\-r\-s\-e\- \-R\-e\-p\-e\-a\-t\- \-I\-n\-s\-t\-r\-u\-c\-t\-i\-o\-n on page} \pageref{Listing71}, \hyperref[Listing61]{F\-i\-x\-e\-d\- \-S\-t\-i\-t\-c\-h\- \-R\-e\-p\-e\-a\-t\- \-P\-a\-r\-s\-e\- \-O\-p\-e\-n on page} \pageref{Listing61}, \hyperref[Listing62]{F\-i\-x\-e\-d\- \-S\-t\-i\-t\-c\-h\- \-R\-e\-p\-e\-a\-t\- \-P\-a\-r\-s\-e\- \-C\-l\-o\-s\-e on page} \pageref{Listing62}, \hyperref[Listing55]{C\-o\-m\-p\-o\-u\-n\-d\- \-S\-t\-i\-t\-c\-h\- \-P\-a\-r\-s\-e\- \-O\-p\-e\-n on page} \pageref{Listing55}, \hyperref[Listing56]{C\-o\-m\-p\-o\-u\-n\-d\- \-S\-t\-i\-t\-c\-h\- \-P\-a\-r\-s\-e\- \-C\-l\-o\-s\-e on page} \pageref{Listing56}, \hyperref[Listing75]{B\-i\-n\-d\--\-O\-f\-f\- \-P\-a\-r\-s\-e\- \-B\-O on page} \pageref{Listing75}, \hyperref[Listing76]{B\-i\-n\-d\--\-O\-f\-f\- \-P\-a\-r\-s\-e\- \-C\-o\-u\-n\-t on page} \pageref{Listing76}, \hyperref[Listing80]{J\-o\-i\-n\- \-P\-a\-r\-s\-e\- \-K\-e\-y\-w\-o\-r\-d on page} \pageref{Listing80}, \hyperref[Listing81]{J\-o\-i\-n\- \-P\-a\-r\-s\-e\- \-C\-o\-u\-n\-t on page} \pageref{Listing81}, \hyperref[Listing82]{J\-o\-i\-n\- \-P\-a\-r\-s\-e\- \-D\-e\-s\-t\-i\-n\-a\-t\-i\-o\-n on page} \pageref{Listing82}, \hyperref[Listing82]{J\-o\-i\-n\- \-P\-a\-r\-s\-e\- \-D\-e\-s\-t\-i\-n\-a\-t\-i\-o\-n on page} \pageref{Listing82}, \hyperref[Listing95]{R\-o\-w\-R\-e\-p\-e\-a\-t\-P\-a\-r\-s\-e\_Open on page} \pageref{Listing95}, \hyperref[Listing96]{R\-o\-w\-R\-e\-p\-e\-a\-t\-P\-a\-r\-s\-e\_Close on page} \pageref{Listing96}, \hyperref[Listing100]{S\-e\-c\-t\-i\-o\-n\- \-P\-a\-r\-s\-e\- \-S\-e\-c\-t\-i\-o\-n on page} \pageref{Listing100}, \hyperref[Listing101]{S\-e\-c\-t\-i\-o\-n\- \-P\-a\-r\-s\-e\- \-T\-i\-t\-l\-e on page} \pageref{Listing101}, \hyperref[Listing102]{S\-e\-c\-t\-i\-o\-n\- \-C\-o\-n\-t\-e\-n\-t\- \-P\-a\-r\-s\-e on page} \pageref{Listing102}, \hyperref[Listing102]{S\-e\-c\-t\-i\-o\-n\- \-C\-o\-n\-t\-e\-n\-t\- \-P\-a\-r\-s\-e on page} \pageref{Listing102}, \hyperref[Listing106]{S\-a\-m\-p\-l\-e\- \-D\-e\-f\- \-P\-a\-r\-s\-e\- \-S\-a\-m\-p\-l\-e on page} \pageref{Listing106}, \hyperref[Listing108]{S\-a\-m\-p\-l\-e\- \-D\-e\-f\- \-P\-a\-r\-s\-e\- \-P\-a\-r\-a\-m\-s on page} \pageref{Listing108}, \hyperref[Listing108]{S\-a\-m\-p\-l\-e\- \-D\-e\-f\- \-P\-a\-r\-s\-e\- \-P\-a\-r\-a\-m\-s on page} \pageref{Listing108}, \hyperref[Listing111]{S\-a\-m\-p\-l\-e\- \-C\-a\-l\-l\- \-P\-a\-r\-s\-e\- \-I\-d\-e\-n\-t on page} \pageref{Listing111}, \hyperref[Listing112]{S\-a\-m\-p\-l\-e\- \-C\-a\-l\-l\- \-P\-a\-r\-s\-e\- \-P\-a\-r\-a\-m\-s on page} \pageref{Listing112}  \end{footnotesize}\vskip 5mm\noindent

\end{enumerate}

\chapter{Elements of Knitting}
\label{chap:knitelements}

This section explores each element of a knitting pattern and its corresponding representation throughout compilation. This
organization allows for additions of new features in a single location, and provides isolated explanations of each knitting concept.

\section{Cast-On}
\label{sec:caston}

To begin a knitting project, it is necessary to add stitches on to the needles. This is called a
``cast-on'', and is essentially a number of loops made on one of the needles such that each loop is
connected to its adjacent loops.

\subsection{AST Node}

\begin{grammar}
<co> ::= `CO' <Nat> [`circular' | `provisional'] `.'
\end{grammar}

The syntax for a cast-on is similar to the standard notation, but replacing cast-on with CO (also commonly used in patterns)
and not requiring ``sts'' to be explicit.

\begin{purlex}[Market Bag Body Cast-On]{coex1}
CO 8 circular.
\end{purlex}

\begin{lstlisting}[title={<Cast-On Parse 2>}, label=Listing2]
var CoParse = function(){
	var node = { type : NodeType.CastOn, value : 0, line : State.line };
	<<Cast-On Parse Keyword 3>>
	<<Cast-On Parse Value 4>>	
	<<Cast-On Parse CoType 5>>
	<<Period Terminator 6>>
	return node;
};
\end{lstlisting}\begin{footnotesize} \textsc{Used in}: \hyperref[Listing125]{A\-s\-t\- \-C\-o\-n\-s\-t\-r\-u\-c\-t\-i\-o\-n\- \-P\-a\-s\-s on page} \pageref{Listing125}  \textsc{Included Blocks}: \hyperref[Listing3]{3 on page} \pageref{Listing3}, \hyperref[Listing4]{4 on page} \pageref{Listing4}, \hyperref[Listing5]{5 on page} \pageref{Listing5}, \hyperref[Listing6]{6 on page} \pageref{Listing6}\end{footnotesize}\vskip 5mm\noindent

The ``CO'' keyword is used to declare a cast-on for a pattern.

\begin{lstlisting}[title={<Cast-On Parse Keyword 3>}, label=Listing3]
if (Sym.type == KeywordSym.CastOn) {
	nextSym();
} else if (Sym.type == SymType.Ident) {
	var msg = "A cast-on dedlaration must start with \'" + KeywordSym.CastOn + "\'.";
	AddMsg(MsgType.Warning, node, msg);
	nextSym();
} else {
	<<Unexpected Symbol Error 1>>
	var msg = "Missing \'" + KeywordSym.CastOn + "\' at start of cast-on declaration.";
	AddMsg(MsgType.Error, node, msg);
	scanToSym(CharSym.Period);
	nextSym();
	return node;
}
\end{lstlisting}\begin{footnotesize} \textsc{Used in}: \hyperref[Listing2]{C\-a\-s\-t\--\-O\-n\- \-P\-a\-r\-s\-e on page} \pageref{Listing2}  \textsc{Included Blocks}: \hyperref[Listing1]{1 on page} \pageref{Listing1}\end{footnotesize}\vskip 5mm\noindent

Following the cast-on keyword, a natural number is given as the number of stitches that will be added to the needle.

\begin{lstlisting}[title={<Cast-On Parse Value 4>}, label=Listing4]
if (Sym.type == SymType.Nat) {
	node.value = Sym.value;
	nextSym();
} else {
	<<Unexpected Symbol Error 1>>
	AddMsg(MsgType.Error, node, "Missing cast-on count.");
	scanToSym(CharSym.Period);
	nextSym();
	return node;
}
\end{lstlisting}\begin{footnotesize} \textsc{Used in}: \hyperref[Listing2]{C\-a\-s\-t\--\-O\-n\- \-P\-a\-r\-s\-e on page} \pageref{Listing2}  \textsc{Included Blocks}: \hyperref[Listing1]{1 on page} \pageref{Listing1}\end{footnotesize}\vskip 5mm\noindent

The default cast-on type is flat, but the grammar also allows a circular or provisional cast-on. A circular cast-on is used
when knitting a circular-shaped object and beginning knitting from the center of the circle. A provisional cast-on is used
if the cast-on will be removed later and the project will be worked in the opposite direction.

\begin{lstlisting}[title={<Cast-On Parse CoType 5>}, label=Listing5]
if (Sym.type == KeywordSym.CastOnCirc) {
	node.coType = CoType.Circular;
	nextSym();
} else if (Sym.type == KeywordSym.CastOnProv) {
	node.coType = CoType.Provisional;
	nextSym();
} else {
	node.coType = CoType.Flat;
}
\end{lstlisting}\begin{footnotesize} \textsc{Used in}: \hyperref[Listing2]{C\-a\-s\-t\--\-O\-n\- \-P\-a\-r\-s\-e on page} \pageref{Listing2}  \end{footnotesize}\vskip 5mm\noindent

The end of a cast-on is marked by a period.

\begin{lstlisting}[title={<Period Terminator 6>}, label=Listing6]
if (Sym.type == CharSym.Period) {
	nextSym();
} else if (Sym.type == CharSym.Comma) {
	AddMsg(MsgType.Warning, node, "Use \'.\' symbol at end of " + node.type + ".");
	nextSym();
} else {
	<<Unexpected Symbol Error 1>>
	AddMsg(MsgType.Error, node, "Missing \'.\' symbol at end of " + node.type + ".");
	scanToSym(CharSym.Period);
	nextSym();
}
\end{lstlisting}\begin{footnotesize} \textsc{Used in}: \hyperref[Listing2]{C\-a\-s\-t\--\-O\-n\- \-P\-a\-r\-s\-e on page} \pageref{Listing2}, \hyperref[Listing9]{P\-i\-c\-k\--\-U\-p\- \-P\-a\-r\-s\-e on page} \pageref{Listing9}, \hyperref[Listing15]{R\-o\-w\- \-D\-e\-f\-i\-n\-i\-t\-i\-o\-n\- \-P\-a\-r\-s\-e on page} \pageref{Listing15}, \hyperref[Listing77]{B\-i\-n\-d\--\-O\-f\-f\- \-P\-a\-r\-s\-e on page} \pageref{Listing77}, \hyperref[Listing83]{J\-o\-i\-n\- \-P\-a\-r\-s\-e on page} \pageref{Listing83}, \hyperref[Listing113]{S\-a\-m\-p\-l\-e\- \-C\-a\-l\-l\- \-P\-a\-r\-s\-e on page} \pageref{Listing113}  \textsc{Included Blocks}: \hyperref[Listing1]{1 on page} \pageref{Listing1}\end{footnotesize}\vskip 5mm\noindent

\subsection{Verification}

When a cast-on is verified, the initial side, row width, and row index are set in the \texttt{State} object.

\begin{lstlisting}[title={<Verify Cast-On 7>}, label=Listing7]
var VerifyCastOn = function(node) {
	State.side = SideType.RS;
	State.width = node.value;
	State.rowIndex = 1;
};
\end{lstlisting}\begin{footnotesize} \textsc{Used in}: \hyperref[Listing131]{V\-e\-r\-i\-f\-i\-c\-a\-t\-i\-o\-n\- \-P\-a\-s\-s on page} \pageref{Listing131}  \end{footnotesize}\vskip 5mm\noindent

\subsection{HTML Generation}

\begin{lstlisting}[title={<Write HTML Cast-On 8>}, label=Listing8]
var WriteCo = function(node) {
	var coType = node.coType != null && node.coType.length > 0 ? " " + node.coType : "";
	return AddElement(TagType.Div, ClassType.CastOn, "Cast-on " + node.value + " sts" + coType + ".");
};
\end{lstlisting}\begin{footnotesize} \textsc{Used in}: \hyperref[Listing132]{c\-o\-d\-e\-/\-c\-o\-d\-e\-g\-e\-n\-.\-j\-s on page} \pageref{Listing132}  \end{footnotesize}\vskip 5mm\noindent

\section{Pick-Up}
\label{sec:pickup}

It is possible to begin knitting off of a completed knitted object by \emph{picking up stitches}. This is done by using one
needle to pull loops of yarn through spaces along the edge that stitches are to be picked up from, so that new active
stitches are on the needle.

\subsection{AST Node}

\begin{grammar}
<pu> ::= `PU' <Nat> `from' <String> `.'
\end{grammar}

\begin{purlex}[Market Bag Handle Pick-Up]{puex1}
PU 10 from ``Body top''.
\end{purlex}

\begin{lstlisting}[title={<Pick-Up Parse 9>}, label=Listing9]
var PuParse = function(){
	var node = { type : NodeType.PickUp, value : 0, line : State.line };
	<<Pick-Up Parse Keyword 10>>
	<<Pick-Up Parse Value 11>>	
	<<Pick-Up Parse Origin 12>>
	<<Period Terminator 6>>
	return node;
};
\end{lstlisting}\begin{footnotesize} \textsc{Used in}: \hyperref[Listing125]{A\-s\-t\- \-C\-o\-n\-s\-t\-r\-u\-c\-t\-i\-o\-n\- \-P\-a\-s\-s on page} \pageref{Listing125}  \textsc{Included Blocks}: \hyperref[Listing10]{1\-0 on page} \pageref{Listing10}, \hyperref[Listing11]{1\-1 on page} \pageref{Listing11}, \hyperref[Listing12]{1\-2 on page} \pageref{Listing12}, \hyperref[Listing6]{6 on page} \pageref{Listing6}\end{footnotesize}\vskip 5mm\noindent

The ``PU'' keyword is used at the beginning of a declaration to pick-up stitches.

\begin{lstlisting}[title={<Pick-Up Parse Keyword 10>}, label=Listing10]
if (Sym.type == KeywordSym.PickUp) {
	nextSym();
} else if (Sym.type == SymType.Ident) {
	var msg = "A pick-up dedlaration must start with \'" + KeywordSym.PickUp + "\'.";
	AddMsg(MsgType.Warning, node, msg);
	nextSym();
} else {
	<<Unexpected Symbol Error 1>>
	var msg = "Missing \'" + KeywordSym.PickUp + "\' at start of pick-up declaration.";
	AddMsg(MsgType.Error, node, msg);
	scanToSym(CharSym.Period);
	nextSym();
	return node;
}
\end{lstlisting}\begin{footnotesize} \textsc{Used in}: \hyperref[Listing9]{P\-i\-c\-k\--\-U\-p\- \-P\-a\-r\-s\-e on page} \pageref{Listing9}  \textsc{Included Blocks}: \hyperref[Listing1]{1 on page} \pageref{Listing1}\end{footnotesize}\vskip 5mm\noindent

Following the pick-up keyword, a natural number is given as the number of stitches that will be added to the needle.

\begin{lstlisting}[title={<Pick-Up Parse Value 11>}, label=Listing11]
if (Sym.type == SymType.Nat) {
	node.value = Sym.value;
	nextSym();
} else {
	<<Unexpected Symbol Error 1>>
	AddMsg(MsgType.Error, node, "Missing pick-up count.");
	scanToSym(CharSym.Period);
	nextSym();
	return node;
}
\end{lstlisting}\begin{footnotesize} \textsc{Used in}: \hyperref[Listing9]{P\-i\-c\-k\--\-U\-p\- \-P\-a\-r\-s\-e on page} \pageref{Listing9}  \textsc{Included Blocks}: \hyperref[Listing1]{1 on page} \pageref{Listing1}\end{footnotesize}\vskip 5mm\noindent

Next we expect the keyword ``from'', followed by a string directing where the stitches should be picked up from.

\begin{lstlisting}[title={<Pick-Up Parse Origin 12>}, label=Listing12]
if (Sym.type == KeywordSym.From) {
	nextSym();
} else {
	<<Unexpected Symbol Error 1>>
	scanToSym(CharSym.Period);
	nextSym();
	return node;
}

if (Sym.type == SymType.String) {
	node.origin = Sym.value;
	nextSym();
} else {
	<<Unexpected Symbol Error 1>>
	AddMsg(MsgType.Error, node, "Missing pick-up origin.");
	scanToSym(CharSym.Period);
	nextSym();
	return node;
}
\end{lstlisting}\begin{footnotesize} \textsc{Used in}: \hyperref[Listing9]{P\-i\-c\-k\--\-U\-p\- \-P\-a\-r\-s\-e on page} \pageref{Listing9}  \textsc{Included Blocks}: \hyperref[Listing1]{1 on page} \pageref{Listing1}, \hyperref[Listing1]{1 on page} \pageref{Listing1}\end{footnotesize}\vskip 5mm\noindent

The end of a pick-up is marked by a period.

\subsection{Verification}

When a pick-up is verified, the initial side, row width, and row index are set in the \texttt{State} object.

\begin{lstlisting}[title={<Verify Pick-Up 13>}, label=Listing13]
var VerifyPickUp = function(node) {
	State.side = SideType.RS;
	State.width = node.value;
	State.rowIndex = 1;
};
\end{lstlisting}\begin{footnotesize} \textsc{Used in}: \hyperref[Listing131]{V\-e\-r\-i\-f\-i\-c\-a\-t\-i\-o\-n\- \-P\-a\-s\-s on page} \pageref{Listing131}  \end{footnotesize}\vskip 5mm\noindent

\subsection{HTML Generation}

\begin{lstlisting}[title={<Write HTML Pick-Up 14>}, label=Listing14]
var WritePu = function(node) {
	return AddElement(TagType.Div, ClassType.CastOn, "Pick-up " + node.value + " sts from " + node.origin + ".");
};
\end{lstlisting}\begin{footnotesize} \textsc{Used in}: \hyperref[Listing132]{c\-o\-d\-e\-/\-c\-o\-d\-e\-g\-e\-n\-.\-j\-s on page} \pageref{Listing132}  \end{footnotesize}\vskip 5mm\noindent

\section{Row}
\label{sec:row}

Once stitches have been added to the needles, it is possible to begin the body of the pattern. Typically knitting is worked
from the left needle to the right needle. Knitting by hand may be done flat or in the round. Both of these methods are
worked ``horizontally'', meaning an object is created as a sequence of knit rows. Flat knitting is done with two straight
needles which can be thought of as two stacks since the first stitch worked in each row is the last created in the previous
row. Circular knitting with a circular needle (two needles connected with a cable) can be thought of as two queues since the
first stitch worked in each row is the first stitch that was created in the previous row.

The ``active'' stitches are the loops on the needles. These are on the left needle at the start of a row. They are
considered active because if removed from the needle, they could be pulled out from whatever stitches they support below. A
stitch is worked by pulling a loop of yarn through an active stitch and then dropping that active stitch from the left
needle. This previously active stitch will be anchored by the loop pulled through it, and that loop is now an active stitch
on the right needle. A row will list the stitches required to work all of the active stitches from the left needle to the
right needle.

\subsection{AST Node}

\begin{grammar}
<rowDef> ::= (`row' | `rnd') [`MC' | `CC'] `:' <rowElem> (`,' <rowElem>)* `.'
\end{grammar}

\begin{lstlisting}[title={<Row Definition Parse 15>}, label=Listing15]
var RowDefParse = function(){
	
	var node = { type : NodeType.Row, children : [], line : State.line };

	<<Row Definition Parse Row Type 16>>
	<<Row Definition Parse Color 17>>
	<<Colon Separator 18>>	
	<<Row Definition Parse Row Elements 19>>
	<<Period Terminator 6>>
	
	return node;
};
\end{lstlisting}\begin{footnotesize} \textsc{Used in}: \hyperref[Listing125]{A\-s\-t\- \-C\-o\-n\-s\-t\-r\-u\-c\-t\-i\-o\-n\- \-P\-a\-s\-s on page} \pageref{Listing125}  \textsc{Included Blocks}: \hyperref[Listing16]{1\-6 on page} \pageref{Listing16}, \hyperref[Listing17]{1\-7 on page} \pageref{Listing17}, \hyperref[Listing18]{1\-8 on page} \pageref{Listing18}, \hyperref[Listing19]{1\-9 on page} \pageref{Listing19}, \hyperref[Listing6]{6 on page} \pageref{Listing6}\end{footnotesize}\vskip 5mm\noindent

A row begins with the row type of either ``row'' or ``rnd''. In flat knitting the row type ``row'' is used, meaning once all
the stitches on the left needle have been worked (or popped), the whole object is turned over, the left and right needles
are swapped, and the process can begin again (the last stitch pushed on the right needle is now the first popped off the left). In
circular knitting the row type ``rnd'' is used, meaning once the stitches on the left needle have been worked and new
stitches are on the right needle, all the stitches are moved along the cable so that they are worked in a FIFO fashion. A
default row type of ``row" is assumed if an unexpected symbol is found.

The body section in the market bag pattern example is knit in the round, and the handle is knit flat.

\begin{purlex}[Market Bag Body Row]{rowex1}
rnd : *K, YO, K; to end.
\end{purlex}

\begin{purlex}[Market Bag Handle Row]{rowex2}
row : K, K2T, K 4, K2T, K.
\end{purlex}

\begin{lstlisting}[title={<Row Definition Parse Row Type 16>}, label=Listing16]
if (Sym.type == KeywordSym.Row) {
	node.rowType = RowType.Row;
	nextSym();
} else if (Sym.type == KeywordSym.Rnd) {
	node.rowType = RowType.Rnd;
	nextSym();
} else {
	<<Unexpected Symbol Error 1>>
	AddMsg(MsgType.Error, node, "Invalid row type specified.");
	node.rowType = RowType.Rnd;
	nextSym();
}
\end{lstlisting}\begin{footnotesize} \textsc{Used in}: \hyperref[Listing15]{R\-o\-w\- \-D\-e\-f\-i\-n\-i\-t\-i\-o\-n\- \-P\-a\-r\-s\-e on page} \pageref{Listing15}  \textsc{Included Blocks}: \hyperref[Listing1]{1 on page} \pageref{Listing1}\end{footnotesize}\vskip 5mm\noindent

The pattern writer can optionally specify which yarn color is to be used for a row. Currently
only a main color, `MC', and a contrasting color, `CC' are available in the language. The main
color is assumed by default.

\begin{lstlisting}[title={<Row Definition Parse Color 17>}, label=Listing17]
if (Sym.type == KeywordSym.ColorMain) {
	node.color = ColorType.Main;
	nextSym();
} else if (Sym.type == KeywordSym.ColorContrast) {
	node.color = ColorType.Contrast;
	nextSym();
}
\end{lstlisting}\begin{footnotesize} \textsc{Used in}: \hyperref[Listing15]{R\-o\-w\- \-D\-e\-f\-i\-n\-i\-t\-i\-o\-n\- \-P\-a\-r\-s\-e on page} \pageref{Listing15}  \end{footnotesize}\vskip 5mm\noindent

A colon symbol separates the row declaration from its content.

\begin{lstlisting}[title={<Colon Separator 18>}, label=Listing18]
if (Sym.type == CharSym.Colon){
	nextSym();
} else if (Sym.type == CharSym.Semicolon) {
	var msg = "Use \':\' symbol before listing " + node.type + " elements.";
	AddMsg(MsgType.Warning, node, msg);
	nextSym();
} else {
	<<Unexpected Symbol Error 1>>
	var msg = "Missing \':\' symbol before listing " + node.type + " elements.";
	AddMsg(MsgType.Error, node, msg);
	scanToSym(CharSym.Period);
}
\end{lstlisting}\begin{footnotesize} \textsc{Used in}: \hyperref[Listing90]{P\-a\-t\-t\-e\-r\-n\- \-P\-a\-r\-s\-e on page} \pageref{Listing90}, \hyperref[Listing15]{R\-o\-w\- \-D\-e\-f\-i\-n\-i\-t\-i\-o\-n\- \-P\-a\-r\-s\-e on page} \pageref{Listing15}, \hyperref[Listing103]{S\-e\-c\-t\-i\-o\-n\- \-P\-a\-r\-s\-e on page} \pageref{Listing103}, \hyperref[Listing110]{S\-a\-m\-p\-l\-e\- \-B\-r\-a\-n\-c\-h\-e\-s\- \-P\-a\-r\-s\-e on page} \pageref{Listing110}, \hyperref[Listing109]{S\-a\-m\-p\-l\-e\- \-D\-e\-f\- \-P\-a\-r\-s\-e on page} \pageref{Listing109}  \textsc{Included Blocks}: \hyperref[Listing1]{1 on page} \pageref{Listing1}\end{footnotesize}\vskip 5mm\noindent

The elements of a row of a pattern are separated by commas. The syntax tree node for each row
element is added to an array representing the children of the row.

\begin{lstlisting}[title={<Row Definition Parse Row Elements 19>}, label=Listing19]
node.children.push(RowElemParse());
	
while (Sym.type == CharSym.Comma) {
	nextSym();
	node.children.push(RowElemParse());
}
\end{lstlisting}\begin{footnotesize} \textsc{Used in}: \hyperref[Listing15]{R\-o\-w\- \-D\-e\-f\-i\-n\-i\-t\-i\-o\-n\- \-P\-a\-r\-s\-e on page} \pageref{Listing15}  \end{footnotesize}\vskip 5mm\noindent

The end of a row definition is marked by a period.

\subsection{Verification}

To verify a row, first the node needs to have row index and side properties set, as these are used in pattern output.

\begin{lstlisting}[title={<Verify Row Setup 20>}, label=Listing20]
node.index = State.rowIndex;
node.side = State.side;
\end{lstlisting}\begin{footnotesize} \textsc{Used in}: \hyperref[Listing25]{V\-e\-r\-i\-f\-y\- \-R\-o\-w on page} \pageref{Listing25}  \end{footnotesize}\vskip 5mm\noindent

The row verification function uses a \texttt{rowState} object that is updated upon verification of each child of the row
(stitches, compound stitches, and stitch repeats).

\begin{lstlisting}[title={<Verify Row Children 21>}, label=Listing21]
var rowState = { initialWidth : State.width, workedSt : 0, stChange : 0 };
VerifyRowElemChildren(node, rowState);
\end{lstlisting}\begin{footnotesize} \textsc{Used in}: \hyperref[Listing25]{V\-e\-r\-i\-f\-y\- \-R\-o\-w on page} \pageref{Listing25}  \end{footnotesize}\vskip 5mm\noindent

Details on verification of specific row elements is discussed in the row elements section (see~\ref{sec:rowelem}).

\begin{lstlisting}[title={<Verify Row Elem Children of Node 22>}, label=Listing22]
var VerifyRowElemChildren = function(node, rowState_0) {
	var rowState_1 = {initialWidth : rowState_0.initialWidth, workedSt : 0, stChange : 0};
	
	if (node.children != null) {
		for (var i = 0; i < node.children.length; i++) {
			VerifyRowElem(node.children[i], rowState_1);
		}
	}
	
	rowState_0.workedSt += rowState_1.workedSt;
	rowState_0.stChange += rowState_1.stChange;
};
\end{lstlisting}\begin{footnotesize} \textsc{Used in}: \hyperref[Listing131]{V\-e\-r\-i\-f\-i\-c\-a\-t\-i\-o\-n\- \-P\-a\-s\-s on page} \pageref{Listing131}  \end{footnotesize}\vskip 5mm\noindent

The verification error that will be caught at this level is an \emph{incorrect number of worked stitches}. This means that the
stitches specified for this row either use less or require more stitches than exist from the previous row.

\begin{lstlisting}[title={<Verify Row Worked Stitches Error 23>}, label=Listing23]
if (rowState.workedSt != State.width) {
	var msg = rowState.workedSt + " sts worked over " + State.width + " sts.";
	AddMsg(MsgType.Verification, node, msg);
}
\end{lstlisting}\begin{footnotesize} \textsc{Used in}: \hyperref[Listing25]{V\-e\-r\-i\-f\-y\- \-R\-o\-w on page} \pageref{Listing25}  \end{footnotesize}\vskip 5mm\noindent

Once the children of the row have been verified, the row width of the node and the \texttt{State} object are updated
according to the stitch change caused by this row. If the row type of this row is ``row'', then the project is to be flipped
at the end of the row, so the side is changed. The \texttt{State.rowIndex} is also incremented by one before we move to the
next sibling.

\begin{lstlisting}[title={<Verify Row State Update 24>}, label=Listing24]
node.width = rowState.workedSt + rowState.stChange;
State.width = node.width;

if (node.rowType == RowType.Row) {
	if (State.side == SideType.RS){
		State.side = SideType.WS;
	} else if (State.side == SideType.WS) {
		State.side = SideType.RS;
	}
}

State.rowIndex += 1;
\end{lstlisting}\begin{footnotesize} \textsc{Used in}: \hyperref[Listing25]{V\-e\-r\-i\-f\-y\- \-R\-o\-w on page} \pageref{Listing25}  \end{footnotesize}\vskip 5mm\noindent

\begin{lstlisting}[title={<Verify Row 25>}, label=Listing25]
var VerifyRow = function(node) {
	<<Verify Row Setup 20>>
	<<Verify Row Children 21>>
	<<Verify Row Worked Stitches Error 23>>
	<<Verify Row State Update 24>>
};
\end{lstlisting}\begin{footnotesize} \textsc{Used in}: \hyperref[Listing131]{V\-e\-r\-i\-f\-i\-c\-a\-t\-i\-o\-n\- \-P\-a\-s\-s on page} \pageref{Listing131}  \textsc{Included Blocks}: \hyperref[Listing20]{2\-0 on page} \pageref{Listing20}, \hyperref[Listing21]{2\-1 on page} \pageref{Listing21}, \hyperref[Listing23]{2\-3 on page} \pageref{Listing23}, \hyperref[Listing24]{2\-4 on page} \pageref{Listing24}\end{footnotesize}\vskip 5mm\noindent

\subsection{HTML Generation}

The written row includes the row type, side, color if explicitly given, row index, children, and the number of active
stitches at the end of the row.

\begin{lstlisting}[title={<Write HTML Row 26>}, label=Listing26]
var WriteRow = function(node) {
	
	var result = [];
	
	result.push(OpenElement(TagType.Div, ClassType.Row));
	
	if (node.rowType == RowType.Row) {
		result.push("Row");
	} else if (node.rowType == RowType.Rnd) {
		result.push("Rnd");
	}
	
	result.push(node.index);
	
	if (node.color == ColorType.Main) {
		result.push("(MC)");
	} else if (node.color == ColorType.Contrast) {
		result.push("(CC)");
	}
	
	result.push("(" + node.side + ")" + ":");
	
	var content = [];
	
	if (node.children != null) {
		for (var i = 0; i < node.children.length; i++) {
			content.push(WriteNode(node.children[i]));
		}
	}
	
	result.push(content.join(", ") + ".");
	
	var count = "(" + node.width + " sts)";
	result.push(AddElement(TagType.Span, ClassType.StitchCount, count));
	
	result.push(CloseElement(TagType.Div));
	
	return result.join(" ");
};
\end{lstlisting}\begin{footnotesize} \textsc{Used in}: \hyperref[Listing132]{c\-o\-d\-e\-/\-c\-o\-d\-e\-g\-e\-n\-.\-j\-s on page} \pageref{Listing132}  \end{footnotesize}\vskip 5mm\noindent

\section{Row Elements}
\label{sec:rowelem}

Children of a row are grouped as \emph{row elements}. The <rowElem> production is used to provide structure to the language:
it is necessary to not allow nesting of undetermined stitch repeats (see~\ref{sec:ustrep}), but it is desirable to allow
fixed repeats (see~\ref{sec:fstrep}) to have fixed stitch repeats as children. The bottom-level row elements are
\hyperref[sec:basicst]{basic stitch}, \hyperref[sec:compst]{compound stitch}, \hyperref[sec:fstrep]{fixed stitch repeat},
and \hyperref[sec:ustrep]{undetermined stitch repeat}.

\subsection{AST Node}

\begin{grammar}
<rowElem> ::= <stitchOp> | <uStRep>
\end{grammar}

\begin{grammar}
<stitchOp> ::= <fixedStRep> | <compSt> | <basicSt>
\end{grammar}

The parse function corresponding to the <rowElem> production determine what <rowElem> the current symbol is a first symbol
of, and passes up the result of the appropriate parse function. If the current symbol is a basic stitch, an open
angle bracket, or open parentheses, then we need to then parse according to the <stitchOp> production. An asterisk is the
only possible first symbol of an undetermined stitch repeat, so for an asterisk we parse according to the <uStRep>
production.

\begin{lstlisting}[title={<Row Element Parse 27>}, label=Listing27]
var RowElemParse = function(){
	
	var node = {};
	
	if (hasOwnValue(StitchSym, Sym.type)
		|| Sym.type == CharSym.OpenAngle
		|| Sym.type == CharSym.OpenBrack) {
		node = StitchOpParse();
	} else if (Sym.type == CharSym.Asterisk) {
		node = UStRepParse();
	} else {
		<<Unexpected Symbol Error 1>>
		AddMsg(MsgType.Error, node, "Invalid row element.");
		scanToSym(CharSym.Period);
	}
	
	return node;
};
\end{lstlisting}\begin{footnotesize} \textsc{Used in}: \hyperref[Listing125]{A\-s\-t\- \-C\-o\-n\-s\-t\-r\-u\-c\-t\-i\-o\-n\- \-P\-a\-s\-s on page} \pageref{Listing125}  \textsc{Included Blocks}: \hyperref[Listing1]{1 on page} \pageref{Listing1}\end{footnotesize}\vskip 5mm\noindent

\begin{lstlisting}[title={<Stitch Op Parse 28>}, label=Listing28]
var StitchOpParse = function() {
	
	var node = {};
	
	if (Sym.type == CharSym.OpenBrack) {
		node = FixedStRepParse();
	} else if (Sym.type == CharSym.OpenAngle) {
		node = CompStParse();
	} else if (hasOwnValue(StitchSym, Sym.type)) {
		node = BasicStParse();
	} else {
		<<Unexpected Symbol Error 1>>
		var msg = sym.value + " is not a known stitch, start of stitch repeat or compound stitch.";
		AddMsg(MsgType.Error, node, msg);
		scanToSym(CharSym.Period);
	}
	
	return node;
};
\end{lstlisting}\begin{footnotesize} \textsc{Used in}: \hyperref[Listing125]{A\-s\-t\- \-C\-o\-n\-s\-t\-r\-u\-c\-t\-i\-o\-n\- \-P\-a\-s\-s on page} \pageref{Listing125}  \textsc{Included Blocks}: \hyperref[Listing1]{1 on page} \pageref{Listing1}\end{footnotesize}\vskip 5mm\noindent

\subsection{Verification}

Row elements are verified according to the row element node type. Details are covered in their respective sections.

\begin{lstlisting}[title={<Verify Row Elem 29>}, label=Listing29]
var VerifyRowElem = function(node, rowState_0) {
	
	var rowState_1 = {initialWidth : rowState_0.initialWidth, workedSt : 0, stChange : 0};
	
	var rep = 1;
	if (node.repCount != null) {
		VerifyExpression(node.repCount);
		if (node.repCount.value > 1) {
			rep = node.repCount.value;
		}
	}
	
	var num = 0;
	if (node.num != null) {
		VerifyExpression(node.num);
		if (node.num.value > 0) {
			num = node.num.value;
		}
	}
	
	switch (node.type) {
		
	case NodeType.FixedStRep:
			<<Verify Fixed Stitch Repeat 65>>
			break;
		
		case NodeType.UStRep:
			<<Verify Undetermined Stitch Repeat 73>>
			break;
		
		case NodeType.CompSt:
			<<Verify Compound Stitch 59>>
			break;
		
		case NodeType.Knit:
		case NodeType.Purl:
		case NodeType.KnitTBL:
		case NodeType.PurlTBL:
		case NodeType.KnitBelow:
		case NodeType.PurlBelow:
		case NodeType.Slip:
		case NodeType.SlipKW:
		case NodeType.SlipPW:
		case NodeType.YarnOver:
		case NodeType.KnitFB:
		case NodeType.PurlFB:
		case NodeType.Make:
		case NodeType.MakeL:
		case NodeType.MakeR:
		case NodeType.KnitTog:
		case NodeType.PurlTog:
		case NodeType.SSK:
		case NodeType.SSP:
		case NodeType.PSSO:
			<<Verify Basic Stitch 53>>
			break;
		
		default:
			break;
	}
};
\end{lstlisting}\begin{footnotesize} \textsc{Used in}: \hyperref[Listing131]{V\-e\-r\-i\-f\-i\-c\-a\-t\-i\-o\-n\- \-P\-a\-s\-s on page} \pageref{Listing131}  \textsc{Included Blocks}: \hyperref[Listing65]{6\-5 on page} \pageref{Listing65}, \hyperref[Listing73]{7\-3 on page} \pageref{Listing73}, \hyperref[Listing59]{5\-9 on page} \pageref{Listing59}, \hyperref[Listing53]{5\-3 on page} \pageref{Listing53}\end{footnotesize}\vskip 5mm\noindent

\section{Basic Stitches}
\label{sec:basicst}

As discussed in~\ref{sec:row}, a stitch is typically formed by pulling the trailing yarn through an active stitch on the
left needle, creating a new stitch to the right needle. Any active stitches on the left needle that the right needle passes
through are dropped off the left needle once the stitch is complete. Variations in the direction the right needle passes
through the last active stitch and whether the yarn is in front or behind the needle allow for different stitches to be
created. More complex stitches are created by pulling the trailing yarn through some other location in the fabric, or
through multiple active stitches, rather than a single active stitch.

A stitch that adds more new stitches to the right needle than are removed from the left is called an increase. Similarly, a
stitch that adds fewer stitches to the right needle than it drops off the left is called a decrease. We will say that the
number of stitches dropped off of the left needle is the number of \emph{worked} stitches, since it is the number of
stitches from the previous row that have been processed. We say that the difference a stitch makes to the width is the
\emph{stitch change}.

Due to a desire to adhere to the standard knitting pattern and stitch notation~\cite{cycstandards} and allow flexibility in
the naming of identifiers in the language, lexing and parsing of stitches has been treated differently than the rest of the
language. Some stitches have numeric parameters in the middle of the stitch notation. Consider the stitch K2T. This
represents a \emph{knit 2 together} stitch, but any natural number is valid. A string of this form is allowed in the
structure of identifiers (e.g. K1abc is an acceptable identifier). To reserve certain strings with parameterized segments as
stitches, an attempt is made to match an identifier string to regular expressions for each stitch before assuming it is an
ident symbol. So the stitch type is determined by the lexer, but the specific information contained in the stitch string is
then extracted in the stitch parse function.

The example below is a row from the market bag pattern. After the colon is the list of stitches to be created for this row.

\begin{purlex}[Market Bag Handle Row]{rowex3}
row : K, K2T, K 4, K2T, K.
\end{purlex}

\subsection{AST Node}

\begin{lstlisting}[title={<Stitch Information 30>}, label=Listing30]
var stParts = Sym.value.split(/([1-9][0-9]*)/);
\end{lstlisting}\begin{footnotesize} \textsc{Used in}: \hyperref[Listing32]{B\-a\-s\-i\-c\- \-S\-t\-i\-t\-c\-h\- \-P\-a\-r\-s\-e on page} \pageref{Listing32}  \end{footnotesize}\vskip 5mm\noindent

A stitch may be optionally followed by a natural number expression to indicate the number of times
a stitch should be repeated.

\begin{lstlisting}[title={<Node Rep Count Optional 31>}, label=Listing31]
if (Sym.type == SymType.Ident || Sym.type == SymType.Nat) {
	node.repCount = ExpressionParse();
}
\end{lstlisting}\begin{footnotesize} \textsc{Used in}: \hyperref[Listing58]{C\-o\-m\-p\-o\-u\-n\-d\- \-S\-t\-i\-t\-c\-h\- \-P\-a\-r\-s\-e on page} \pageref{Listing58}, \hyperref[Listing32]{B\-a\-s\-i\-c\- \-S\-t\-i\-t\-c\-h\- \-P\-a\-r\-s\-e on page} \pageref{Listing32}  \end{footnotesize}\vskip 5mm\noindent

\begin{lstlisting}[title={<Basic Stitch Parse 32>}, label=Listing32]
var BasicStParse = function() {
	
	var node = { line : State.line };
	
	<<Stitch Information 30>>
	
	switch (Sym.type) {
		case StitchSym.Knit:
			<<Knit 33>>
			break;
		case StitchSym.Purl:
			<<Purl 36>>
			break;
		case StitchSym.KnitTBL:
			<<KnitTBL 34>>
			break;
		case StitchSym.PurlTBL:
			<<PurlTBL 37>>
			break;
		case StitchSym.KnitBelow:
			<<KnitBelow 35>>
			break;
		case StitchSym.PurlBelow:
			<<PurlBelow 38>>
			break;
		case StitchSym.Slip:
			<<Slip 39>>
			break;
		case StitchSym.SlipKW:
			<<SlipKW 40>>
			break;
		case StitchSym.SlipPW:
			<<SlipPW 41>>
			break;
		case StitchSym.YarnOver:
			<<YarnOver 42>>
			break;
		case StitchSym.KnitFB:
			<<KnitFB 43>>
			break;
		case StitchSym.PurlFB:
			<<PurlFB 44>>
			break;
		case StitchSym.Make:
			<<Make 45>>
			break;
		case StitchSym.MakeL:
			<<MakeL 46>>
			break;
		case StitchSym.MakeR:
			<<MakeR 47>>
			break;
		case StitchSym.KnitTog:
			<<KnitTog 48>>
			break;
		case StitchSym.PurlTog:
			<<PurlTog 49>>
			break;
		case StitchSym.SSK:
			<<SSK 50>>
			break;
		case StitchSym.SSP:
			<<SSP 51>>
			break;
		case StitchSym.PSSO:
			<<PSSO 52>>
			break;
		default:
			node.workedSt = 0;
			node.stChange = 0;
			break;
	}
	
	nextSym();
	
	<<Node Rep Count Optional 31>>
	
	return node;
};
\end{lstlisting}\begin{footnotesize} \textsc{Used in}: \hyperref[Listing125]{A\-s\-t\- \-C\-o\-n\-s\-t\-r\-u\-c\-t\-i\-o\-n\- \-P\-a\-s\-s on page} \pageref{Listing125}  \textsc{Included Blocks}: \hyperref[Listing30]{3\-0 on page} \pageref{Listing30}, \hyperref[Listing33]{3\-3 on page} \pageref{Listing33}, \hyperref[Listing36]{3\-6 on page} \pageref{Listing36}, \hyperref[Listing34]{3\-4 on page} \pageref{Listing34}, \hyperref[Listing37]{3\-7 on page} \pageref{Listing37}, \hyperref[Listing35]{3\-5 on page} \pageref{Listing35}, \hyperref[Listing38]{3\-8 on page} \pageref{Listing38}, \hyperref[Listing39]{3\-9 on page} \pageref{Listing39}, \hyperref[Listing40]{4\-0 on page} \pageref{Listing40}, \hyperref[Listing41]{4\-1 on page} \pageref{Listing41}, \hyperref[Listing42]{4\-2 on page} \pageref{Listing42}, \hyperref[Listing43]{4\-3 on page} \pageref{Listing43}, \hyperref[Listing44]{4\-4 on page} \pageref{Listing44}, \hyperref[Listing45]{4\-5 on page} \pageref{Listing45}, \hyperref[Listing46]{4\-6 on page} \pageref{Listing46}, \hyperref[Listing47]{4\-7 on page} \pageref{Listing47}, \hyperref[Listing48]{4\-8 on page} \pageref{Listing48}, \hyperref[Listing49]{4\-9 on page} \pageref{Listing49}, \hyperref[Listing50]{5\-0 on page} \pageref{Listing50}, \hyperref[Listing51]{5\-1 on page} \pageref{Listing51}, \hyperref[Listing52]{5\-2 on page} \pageref{Listing52}, \hyperref[Listing31]{3\-1 on page} \pageref{Listing31}\end{footnotesize}\vskip 5mm\noindent

\newpage

\begin{description}
\item[Knit] The most ubiquitous stitch. \hfill

\begin{description}
\item[Needle Entry] \hfill \\ Top active stitch, from left
\item[Yarn Position] \hfill \\ Back
\item[Effect] \hfill
\begin{lstlisting}[title={<Knit 33>}, label=Listing33]
node.type = NodeType.Knit;
node.workedSt = 1;
node.stChange = 0;
\end{lstlisting}\begin{footnotesize} \textsc{Used in}: \hyperref[Listing32]{B\-a\-s\-i\-c\- \-S\-t\-i\-t\-c\-h\- \-P\-a\-r\-s\-e on page} \pageref{Listing32}  \end{footnotesize}\vskip 5mm\noindent
\end{description}

\item[Knit Through Back Loop] \hfill

\begin{description}
\item[Needle Entry] \hfill \\ Top active stitch, from right
\item[Yarn Position] \hfill \\ Back
\item[Effect] \hfill
\begin{lstlisting}[title={<KnitTBL 34>}, label=Listing34]
node.type = NodeType.KnitTBL;
node.workedSt = 1;
node.stChange = 0;
\end{lstlisting}\begin{footnotesize} \textsc{Used in}: \hyperref[Listing32]{B\-a\-s\-i\-c\- \-S\-t\-i\-t\-c\-h\- \-P\-a\-r\-s\-e on page} \pageref{Listing32}  \end{footnotesize}\vskip 5mm\noindent
\end{description}

\item[Knit Below] Stitch with parameter \textit{num} \hfill

\begin{description}
\item[Needle Entry] \hfill \\ \textit{num} stitches below top active stitch, from right
\item[Yarn Position] \hfill \\ Back
\item[Effect] \hfill
\begin{lstlisting}[title={<KnitBelow 35>}, label=Listing35]
node.type = NodeType.KnitBelow;
node.num = stParts[1];
node.workedSt = 1;
node.stChange = 0;
\end{lstlisting}\begin{footnotesize} \textsc{Used in}: \hyperref[Listing32]{B\-a\-s\-i\-c\- \-S\-t\-i\-t\-c\-h\- \-P\-a\-r\-s\-e on page} \pageref{Listing32}  \end{footnotesize}\vskip 5mm\noindent
\end{description}

\item[Purl] \hfill

\begin{description}
\item[Needle Entry] \hfill \\ Top active stitch, from right
\item[Yarn Position] \hfill \\ Front
\item[Effect] \hfill
\begin{lstlisting}[title={<Purl 36>}, label=Listing36]
node.type = NodeType.Purl;
node.workedSt = 1;
node.stChange = 0;
\end{lstlisting}\begin{footnotesize} \textsc{Used in}: \hyperref[Listing32]{B\-a\-s\-i\-c\- \-S\-t\-i\-t\-c\-h\- \-P\-a\-r\-s\-e on page} \pageref{Listing32}  \end{footnotesize}\vskip 5mm\noindent
\end{description}

\item[Purl Through Back Loop] \hfill

\begin{description}
\item[Needle Entry] \hfill \\ Top active stitch, from left
\item[Yarn Position] \hfill \\ Front
\item[Effect] \hfill
\begin{lstlisting}[title={<PurlTBL 37>}, label=Listing37]
node.type = NodeType.PurlTBL;
node.workedSt = 1;
node.stChange = 0;
\end{lstlisting}\begin{footnotesize} \textsc{Used in}: \hyperref[Listing32]{B\-a\-s\-i\-c\- \-S\-t\-i\-t\-c\-h\- \-P\-a\-r\-s\-e on page} \pageref{Listing32}  \end{footnotesize}\vskip 5mm\noindent
\end{description}

\item[Purl Below] Stitch with parameter \textit{num} \hfill

\begin{description}
\item[Needle Entry] \hfill \\ \textit{num} below top active stitch, from right
\item[Yarn Position] \hfill \\ Front
\item[Effect] \hfill
\begin{lstlisting}[title={<PurlBelow 38>}, label=Listing38]
node.type = NodeType.PurlBelow;
node.num = stParts[1];
node.workedSt = 1;
node.stChange = 0;
\end{lstlisting}\begin{footnotesize} \textsc{Used in}: \hyperref[Listing32]{B\-a\-s\-i\-c\- \-S\-t\-i\-t\-c\-h\- \-P\-a\-r\-s\-e on page} \pageref{Listing32}  \end{footnotesize}\vskip 5mm\noindent
\end{description}

\item[Slip] \hfill

\begin{description}
\item[Needle Entry] \hfill \\ Top active stitch, from left
\item[Yarn Position] \hfill \\ Back
\item[Notes] \hfill \\ No loop pulled through
\item[Effect] \hfill
\begin{lstlisting}[title={<Slip 39>}, label=Listing39]
node.type = NodeType.Slip;
node.workedSt = 1;
node.stChange = 0;
\end{lstlisting}\begin{footnotesize} \textsc{Used in}: \hyperref[Listing32]{B\-a\-s\-i\-c\- \-S\-t\-i\-t\-c\-h\- \-P\-a\-r\-s\-e on page} \pageref{Listing32}  \end{footnotesize}\vskip 5mm\noindent
\end{description}

\item[Slip Knitwise] The default slip stitch \hfill

\begin{description}
\item[Needle Entry] \hfill \\ Top active stitch, from left
\item[Yarn Position] \hfill \\ Back
\item[Notes] \hfill \\ No loop pulled through
\item[Effect] \hfill
\begin{lstlisting}[title={<SlipKW 40>}, label=Listing40]
node.type = NodeType.SlipKW;
node.workedSt = 1;
node.stChange = 0;
\end{lstlisting}\begin{footnotesize} \textsc{Used in}: \hyperref[Listing32]{B\-a\-s\-i\-c\- \-S\-t\-i\-t\-c\-h\- \-P\-a\-r\-s\-e on page} \pageref{Listing32}  \end{footnotesize}\vskip 5mm\noindent
\end{description}

\item[Slip Purlwise] \hfill

\begin{description}
\item[Needle Entry] \hfill \\ Top active stitch, from right
\item[Yarn Position] \hfill \\ Front
\item[Notes] \hfill \\ No loop pulled through
\item[Effect] \hfill
\begin{lstlisting}[title={<SlipPW 41>}, label=Listing41]
node.type = NodeType.SlipPW;
node.workedSt = 1;
node.stChange = 0;
\end{lstlisting}\begin{footnotesize} \textsc{Used in}: \hyperref[Listing32]{B\-a\-s\-i\-c\- \-S\-t\-i\-t\-c\-h\- \-P\-a\-r\-s\-e on page} \pageref{Listing32}  \end{footnotesize}\vskip 5mm\noindent
\end{description}

\item[Yarn Over] An increase \hfill

\begin{description}
\item[Yarn Position] \hfill \\ Back
\item[Instruction] \hfill \\ Wrap yarn counterclockwise arount the right needle
\item[Effect] \hfill
\begin{lstlisting}[title={<YarnOver 42>}, label=Listing42]
node.type = NodeType.YarnOver;
node.workedSt = 0;
node.stChange = 1;
\end{lstlisting}\begin{footnotesize} \textsc{Used in}: \hyperref[Listing32]{B\-a\-s\-i\-c\- \-S\-t\-i\-t\-c\-h\- \-P\-a\-r\-s\-e on page} \pageref{Listing32}  \end{footnotesize}\vskip 5mm\noindent
\end{description}

\item[Knit Front and Back] An increase \hfill

\begin{description}
\item[Instruction] \hfill \\ In the same active stitch, knit then knit through back loop
\item[Effect] \hfill
\begin{lstlisting}[title={<KnitFB 43>}, label=Listing43]
node.type = NodeType.KnitFB;
node.workedSt = 1;
node.stChange = 1;
\end{lstlisting}\begin{footnotesize} \textsc{Used in}: \hyperref[Listing32]{B\-a\-s\-i\-c\- \-S\-t\-i\-t\-c\-h\- \-P\-a\-r\-s\-e on page} \pageref{Listing32}  \end{footnotesize}\vskip 5mm\noindent
\end{description}

\item[Purl Front and Back] An increase \hfill

\begin{description}
\item[Instruction] \hfill \\ In the same active stitch, purl then purl through back loop 
\item[Effect] \hfill
\begin{lstlisting}[title={<PurlFB 44>}, label=Listing44]
node.type = NodeType.PurlFB;
node.workedSt = 1;
node.stChange = 1;
\end{lstlisting}\begin{footnotesize} \textsc{Used in}: \hyperref[Listing32]{B\-a\-s\-i\-c\- \-S\-t\-i\-t\-c\-h\- \-P\-a\-r\-s\-e on page} \pageref{Listing32}  \end{footnotesize}\vskip 5mm\noindent
\end{description}

\item[Make] An increase with parameter \textit{num} \hfill

\begin{description}
\item[Setup] \hfill \\ From back, left needle picks up vertical bar between top left and top right active stitches
\item[Instruction] \hfill \\ Knit the stitch picked up by the left needle, repeat \textit{num} times
\item[Effect] \hfill
\begin{lstlisting}[title={<Make 45>}, label=Listing45]
node.type = NodeType.Make;
node.num = stParts[1];
node.workedSt = 0;
node.stChange = 1;
\end{lstlisting}\begin{footnotesize} \textsc{Used in}: \hyperref[Listing32]{B\-a\-s\-i\-c\- \-S\-t\-i\-t\-c\-h\- \-P\-a\-r\-s\-e on page} \pageref{Listing32}  \end{footnotesize}\vskip 5mm\noindent
\end{description}

\item[Make Left] The default ``make'' stitch, an increase with parameter \textit{num} \hfill

\begin{description}
\item[Setup] \hfill \\ From back, left needle picks up vertical bar between top left and top right active stitches
\item[Instruction] \hfill \\ Knit the stitch picked up by the left needle, repeat \textit{num} times
\item[Effect] \hfill
\begin{lstlisting}[title={<MakeL 46>}, label=Listing46]
node.type = NodeType.MakeL;
node.num = stParts[1];
node.workedSt = 0;
node.stChange = 1;
\end{lstlisting}\begin{footnotesize} \textsc{Used in}: \hyperref[Listing32]{B\-a\-s\-i\-c\- \-S\-t\-i\-t\-c\-h\- \-P\-a\-r\-s\-e on page} \pageref{Listing32}  \end{footnotesize}\vskip 5mm\noindent
\end{description}

\item[Make Right] An increase with parameter \textit{num} \hfill

\begin{description}
\item[Setup] \hfill \\ From front, left needle picks up vertical bar between top left and top right active stitches
\item[Instruction] \hfill \\ Knit through the back loop the stitch picked up by the left needle, repeat \textit{num} times 
\item[Effect] \hfill
\begin{lstlisting}[title={<MakeR 47>}, label=Listing47]
node.type = NodeType.MakeR;
node.num = stParts[1];
node.workedSt = 0;
node.stChange = 1;
\end{lstlisting}\begin{footnotesize} \textsc{Used in}: \hyperref[Listing32]{B\-a\-s\-i\-c\- \-S\-t\-i\-t\-c\-h\- \-P\-a\-r\-s\-e on page} \pageref{Listing32}  \end{footnotesize}\vskip 5mm\noindent
\end{description}

\item[Knit Together] A decrease with parameter \textit{num} \hfill

\begin{description}
\item[Needle Entry] \hfill \\ Top \textit{num} active stitches, from left
\item[Yarn Position] \hfill \\ Back
\item[Effect] \hfill
\begin{lstlisting}[title={<KnitTog 48>}, label=Listing48]
node.type = NodeType.KnitTog;
node.num = stParts[1];
node.workedSt = node.num;
node.stChange = (-1) * (node.num - 1);
\end{lstlisting}\begin{footnotesize} \textsc{Used in}: \hyperref[Listing32]{B\-a\-s\-i\-c\- \-S\-t\-i\-t\-c\-h\- \-P\-a\-r\-s\-e on page} \pageref{Listing32}  \end{footnotesize}\vskip 5mm\noindent
\end{description}

\item[Purl Together] A decrease with parameter \textit{num} \hfill

\begin{description}
\item[Needle Entry] \hfill \\ Top \textit{num} active stitches, from right
\item[Yarn Position] \hfill \\ Front
\item[Effect] \hfill
\begin{lstlisting}[title={<PurlTog 49>}, label=Listing49]
node.type = NodeType.PurlTog;
node.num = stParts[1];
node.workedSt = node.num;
node.stChange = (-1) * (node.num - 1);
\end{lstlisting}\begin{footnotesize} \textsc{Used in}: \hyperref[Listing32]{B\-a\-s\-i\-c\- \-S\-t\-i\-t\-c\-h\- \-P\-a\-r\-s\-e on page} \pageref{Listing32}  \end{footnotesize}\vskip 5mm\noindent
\end{description}

\item[Slip Slip Knit] A decrease \hfill

\begin{description}
\item[Instruction] \hfill \\ Slip knitwise then slip purlwise and knit the two slipped stitches together
\item[Effect] \hfill
\begin{lstlisting}[title={<SSK 50>}, label=Listing50]
node.type = NodeType.SSK;
node.workedSt = 2;
node.stChange = -1;
\end{lstlisting}\begin{footnotesize} \textsc{Used in}: \hyperref[Listing32]{B\-a\-s\-i\-c\- \-S\-t\-i\-t\-c\-h\- \-P\-a\-r\-s\-e on page} \pageref{Listing32}  \end{footnotesize}\vskip 5mm\noindent
\end{description}

\item[Slip Slip Purl] A decrease \hfill

\begin{description}
\item[Instruction] \hfill \\ Slip knitwise then slip purlwise and purl the two slipped stitches together
\item[Effect] \hfill
\begin{lstlisting}[title={<SSP 51>}, label=Listing51]
node.type = NodeType.SSP;
node.workedSt = 2;
node.stChange = -1;
\end{lstlisting}\begin{footnotesize} \textsc{Used in}: \hyperref[Listing32]{B\-a\-s\-i\-c\- \-S\-t\-i\-t\-c\-h\- \-P\-a\-r\-s\-e on page} \pageref{Listing32}  \end{footnotesize}\vskip 5mm\noindent
\end{description}

\item[Pass Slip Stitch Over] A decrease \hfill

\begin{description}
\item[Needle Entry] \hfill \\ On right needle, second active stitch from top, from left
\item[Yarn Position] \hfill \\ Back
\item[Instruction] \hfill \\ Pass stitch over top active stitch on right needle, and off needle
\item[Notes] \hfill \\ Most commonly used after a slip then knit, hence passing the slipped stitch over
\item[Effect] \hfill
\begin{lstlisting}[title={<PSSO 52>}, label=Listing52]
node.type = NodeType.PSSO;
node.workedSt = 0;
node.stChange = -1;
\end{lstlisting}\begin{footnotesize} \textsc{Used in}: \hyperref[Listing32]{B\-a\-s\-i\-c\- \-S\-t\-i\-t\-c\-h\- \-P\-a\-r\-s\-e on page} \pageref{Listing32}  \end{footnotesize}\vskip 5mm\noindent
\end{description}

\end{description}

\subsection{Verification}

The role of a basic stitch during the verification pass is to update the row state with the stitch change and number of
worked stitches by the given stitch.

\begin{lstlisting}[title={<Verify Basic Stitch 53>}, label=Listing53]
rowState_0.workedSt += node.workedSt * rep;
rowState_0.stChange += node.stChange * rep;
\end{lstlisting}\begin{footnotesize} \textsc{Used in}: \hyperref[Listing29]{V\-e\-r\-i\-f\-y\- \-R\-o\-w\- \-E\-l\-e\-m on page} \pageref{Listing29}  \end{footnotesize}\vskip 5mm\noindent

\subsection{HTML Generation}

The writing of stitches follows the standard notation~\cite{cycstandards}.

\begin{lstlisting}[title={<Write HTML Basic Stitch 54>}, label=Listing54]
var WriteBasicStitch = function(node) {
	
	var result = [];
	
	var rep = "";
	var num = "";
	
	if (node.repCount != null && node.repCount.value > 1) {
		rep = node.repCount.value;
	}
	
	if (node.num != null) {
		if (node.num.value > 0) {
			num = node.num.value;
		} else if (node.num > 0) {
			num = node.num;
		} 
	}
	
	switch (node.type) {
		
		case NodeType.Knit:
			result.push(AddElement(TagType.Span, ClassType.Stitch, "K" + rep));
			break;
		
		case NodeType.Purl:
			result.push(AddElement(TagType.Span, ClassType.Stitch, "P" + rep));
			break;
		
		case NodeType.KnitTBL:
			result.push(AddElement(TagType.Span, ClassType.Stitch, "K" + rep + " tbl"));
			break;
		
		case NodeType.PurlTBL:
			result.push(AddElement(TagType.Span, ClassType.Stitch, "P" + rep + " tbl"));
			break;
		
		case NodeType.KnitBelow:
			result.push(AddElement(TagType.Span, ClassType.Stitch, "K" + num + "B" + rep));
			break;
		
		case NodeType.PurlBelow:
			result.push(AddElement(TagType.Span, ClassType.Stitch, "P" + num + "B" + rep));
			break;
		
		case NodeType.Slip:
			result.push(AddElement(TagType.Span, ClassType.Stitch, "sl" + rep));
			break;
		
		case NodeType.SlipKW:
			result.push(AddElement(TagType.Span, ClassType.Stitch, "sl" + rep + "k"));
			break;
		
		case NodeType.SlipPW:
			result.push(AddElement(TagType.Span, ClassType.Stitch, "sl" + rep + "p"));
			break;
		
		case NodeType.YarnOver:
			result.push(AddElement(TagType.Span, ClassType.Stitch, "yo" + rep));
			break;
		
		case NodeType.KnitFB:
			result.push(AddElement(TagType.Span, ClassType.Stitch, "KFB"));
			if (rep > 0) { result.push(rep); }
			break;
		
		case NodeType.PurlFB:
			result.push(AddElement(TagType.Span, ClassType.Stitch, "PFB"));
			if (rep > 0) { result.push(rep); }
			break;
		
		case NodeType.Make:
			result.push(AddElement(TagType.Span, ClassType.Stitch, "M" + num));
			if (rep > 0) { result.push(rep); }
			break;
		
		case NodeType.MakeL:
			result.push(AddElement(TagType.Span, ClassType.Stitch, "M" + num + "L"));
			if (rep > 0) { result.push(rep); }
			break;
		
		case NodeType.MakeR:
			result.push(AddElement(TagType.Span, ClassType.Stitch, "M" + num + "R"));
			if (rep > 0) { result.push(rep); }
			break;
		
		case NodeType.KnitTog:
			result.push(AddElement(TagType.Span, ClassType.Stitch, "k" + num + "tog"));
			if (rep > 0) { result.push(rep); }
			break;
		
		case NodeType.PurlTog:
			result.push(AddElement(TagType.Span, ClassType.Stitch, "k" + num + "tog"));
			if (rep > 0) { result.push(rep); }
			break;
		
		case NodeType.SSK:
			result.push(AddElement(TagType.Span, ClassType.Stitch, "ssk"));
			if (rep > 0) { result.push(rep); }
			break;
		
		case NodeType.SSP:
			result.push(AddElement(TagType.Span, ClassType.Stitch, "ssp"));
			if (rep > 0) { result.push(rep); }
			break;
		
		case NodeType.PSSO:
			result.push(AddElement(TagType.Span, ClassType.Stitch, "psso"));
			if (rep > 0) { result.push(rep); }
			break;
		
		default:
			break;
	}
	
	return result.join(" ");
};
\end{lstlisting}\begin{footnotesize} \textsc{Used in}: \hyperref[Listing132]{c\-o\-d\-e\-/\-c\-o\-d\-e\-g\-e\-n\-.\-j\-s on page} \pageref{Listing132}  \end{footnotesize}\vskip 5mm\noindent

\section{Compound Stitch}
\label{sec:compst}

It is possible to work a number of stitches in a single stitch. This means that after the loop has been pulled through an
active stitch, that stitch is not dropped off the left needle, but remains until the sequence of stitches have been
performed.

\subsection{AST Node}

\begin{grammar}
<compSt> ::= `<' <basicSt> ( `,' <basicSt> )* `>'  [<expr>]
\end{grammar}

The below example of a compound stitch means to perform the stitches between the angle brackets in one stitch.

\begin{purlex}[Compound stitch]{compex}
<K, P, K>
\end{purlex}

The sequence of stitches of a compound stitch is contained between angle brackets. The use of any other bracket symbols
will generate a warning and allow compilation to continue. A compound is separated from its siblings by a comma, but its
children are also comma-separated. This is a case where on an error it is necessary for the lexer to scan to the terminator of the
parent node production before continuing parsing.

\begin{lstlisting}[title={<Compound Stitch Parse Open 55>}, label=Listing55]
if (Sym.type == CharSym.OpenAngle) {
	nextSym();
} else if (Sym.type == CharSym.OpenParen
			|| Sym.type == CharSym.OpenBrace
			|| Sym.type == CharSym.OpenBrack) {
	AddMsg(MsgType.Warning, node, "Use \'<\' symbol at start of compound stitch.");
	nextSym();
} else {
	 <<Unexpected Symbol Error 1>>
	AddMsg(MsgType.Error, node, "Missing \'<\' symbol at start of compound stitch.");
	scanToSym(CharSym.Period);
	return node;
}
\end{lstlisting}\begin{footnotesize} \textsc{Used in}: \hyperref[Listing58]{C\-o\-m\-p\-o\-u\-n\-d\- \-S\-t\-i\-t\-c\-h\- \-P\-a\-r\-s\-e on page} \pageref{Listing58}  \textsc{Included Blocks}: \hyperref[Listing1]{1 on page} \pageref{Listing1}\end{footnotesize}\vskip 5mm\noindent

\begin{lstlisting}[title={<Compound Stitch Parse Close 56>}, label=Listing56]
if (Sym.type == CharSym.CloseAngle) {
	nextSym();
} else if (Sym.type == CharSym.CloseParen
			|| Sym.type == CharSym.CloseBrace
			|| Sym.type == CharSym.CloseBrack) {
	AddMsg(MsgType.Warning, node, "Use \'>\' symbol at end of compound stitch.");
	nextSym();
} else {
	<<Unexpected Symbol Error 1>>
	AddMsg(MsgType.Error, node, "Missing \'>\' symbol at end of compound stitch.");
	scanToSym(CharSym.Period);
	return node;
}
\end{lstlisting}\begin{footnotesize} \textsc{Used in}: \hyperref[Listing58]{C\-o\-m\-p\-o\-u\-n\-d\- \-S\-t\-i\-t\-c\-h\- \-P\-a\-r\-s\-e on page} \pageref{Listing58}  \textsc{Included Blocks}: \hyperref[Listing1]{1 on page} \pageref{Listing1}\end{footnotesize}\vskip 5mm\noindent

Between the compound stitch brackets is a comma separated list of basic stitches.

\begin{lstlisting}[title={<Compound Stitch Parse Children 57>}, label=Listing57]
node.children.push(BasicStParse());

while (Sym.type == CharSym.Comma) {
	nextSym();
	node.children.push(BasicStParse());
}
\end{lstlisting}\begin{footnotesize} \textsc{Used in}: \hyperref[Listing58]{C\-o\-m\-p\-o\-u\-n\-d\- \-S\-t\-i\-t\-c\-h\- \-P\-a\-r\-s\-e on page} \pageref{Listing58}  \end{footnotesize}\vskip 5mm\noindent

A compound stitch optionally ends with a natural number or variable indicating how many times the
compound stitch should be repeated. Note that a compound stitch works one stitch for every repeat,
since all stitches in the sequence are worked into a single active stitch on the left needle.

\begin{lstlisting}[title={<Compound Stitch Parse 58>}, label=Listing58]
var CompStParse = function() {
	
	var node = { type : NodeType.CompSt, children : [], line : State.line };
	
	<<Compound Stitch Parse Open 55>>
	<<Compound Stitch Parse Children 57>>
	<<Compound Stitch Parse Close 56>>
	<<Node Rep Count Optional 31>>
	
	return node;
};
\end{lstlisting}\begin{footnotesize} \textsc{Used in}: \hyperref[Listing125]{A\-s\-t\- \-C\-o\-n\-s\-t\-r\-u\-c\-t\-i\-o\-n\- \-P\-a\-s\-s on page} \pageref{Listing125}  \textsc{Included Blocks}: \hyperref[Listing55]{5\-5 on page} \pageref{Listing55}, \hyperref[Listing57]{5\-7 on page} \pageref{Listing57}, \hyperref[Listing56]{5\-6 on page} \pageref{Listing56}, \hyperref[Listing31]{3\-1 on page} \pageref{Listing31}\end{footnotesize}\vskip 5mm\noindent

\subsection{Verification}

\begin{lstlisting}[title={<Verify Compound Stitch 59>}, label=Listing59]
VerifyRowElemChildren(node, rowState_1);
rowState_0.workedSt += rep;
rowState_0.stChange += rowState_1.stChange * rep;
\end{lstlisting}\begin{footnotesize} \textsc{Used in}: \hyperref[Listing29]{V\-e\-r\-i\-f\-y\- \-R\-o\-w\- \-E\-l\-e\-m on page} \pageref{Listing29}  \end{footnotesize}\vskip 5mm\noindent

\subsection{HTML Generation}

\begin{lstlisting}[title={<Write HTML Compound Stitch 60>}, label=Listing60]
var WriteCompSt = function(node) {
	
	var result = [];
	var stitches = [];
	
	if (node.children != null) {
		for (var i = 0; i < node.children.length; i++) {
			stitches.push(WriteNode(node.children[i]));
		}
	}
	
	result.push("(" + stitches.join(", ") + ")");
	
	if (node.repCount != null && node.repCount.value > 1) {
		result.push(node.repCount.value + " times ");
	}
	
	result.push("in next st");
	
	return result.join(" ");
};
\end{lstlisting}\begin{footnotesize} \textsc{Used in}: \hyperref[Listing132]{c\-o\-d\-e\-/\-c\-o\-d\-e\-g\-e\-n\-.\-j\-s on page} \pageref{Listing132}  \end{footnotesize}\vskip 5mm\noindent

\section{Fixed Stitch Repeat}
\label{sec:fstrep}

A sequence of stitches may be repeated a fixed number of times.

\subsection{AST Node}

\begin{grammar}
<fixedStRep> ::= `[' <stitchOp> ( ',' <stitchOp> )* `]' <expr>
\end{grammar}

The below example means to perform the sequence between brackets (knit then purl) three times.

\begin{purlex}[Fixed repeat]{fixedex}
(K, P) 3
\end{purlex}

The children of a fixed stitch repeat is a sequence of fixed stitch repeats, \hyperref[sec:compst]{compound stitches}, and
\hyperref[sec:basicst]{basic stitches} contained within parentheses. As we saw for compound stitches, we cannot confidently
assume the location of the next sibling in an error situation. We again scan to the terminator of the parent node
production.

\begin{lstlisting}[title={<Fixed Stitch Repeat Parse Open 61>}, label=Listing61]
if (Sym.type == CharSym.OpenBrack) {
	nextSym();
} else if (Sym.type == CharSym.OpenAngle
			|| Sym.type == CharSym.OpenBrace
			|| Sym.type == CharSym.OpenParen) {
	AddMsg(MsgType.Warning, node, "Use \'[\' symbol to start fixed stitch repeat.");
	nextSym();
} else {
	<<Unexpected Symbol Error 1>>
	AddMsg(MsgType.Error, node, "Missing \'[\' symbol to start fixed stitch repeat.");
	scanToSym(CharSym.Period);
	return node;
}
\end{lstlisting}\begin{footnotesize} \textsc{Used in}: \hyperref[Listing64]{F\-i\-x\-e\-d\- \-S\-t\-i\-t\-c\-h\- \-R\-e\-p\-e\-a\-t\- \-P\-a\-r\-s\-e on page} \pageref{Listing64}  \textsc{Included Blocks}: \hyperref[Listing1]{1 on page} \pageref{Listing1}\end{footnotesize}\vskip 5mm\noindent

\begin{lstlisting}[title={<Fixed Stitch Repeat Parse Close 62>}, label=Listing62]
if (Sym.type == CharSym.CloseBrack) {
	nextSym();
} else if (Sym.type == CharSym.CloseAngle
			|| Sym.type == CharSym.CloseBrace
			|| Sym.type == CharSym.CloseParen) {
	var msg = "Use \']\' symbol to end fixed stitch repeat stitches.";
	AddMsg(MsgType.Warning, node, msg);
	nextSym();
} else {
	<<Unexpected Symbol Error 1>>
	var msg = "Missing \']\' symbol to end fixed stitch repeat stitches.";
	AddMsg(MsgType.Error, node, msg);
	scanToSym(CharSym.Period);
	return node;
}
\end{lstlisting}\begin{footnotesize} \textsc{Used in}: \hyperref[Listing64]{F\-i\-x\-e\-d\- \-S\-t\-i\-t\-c\-h\- \-R\-e\-p\-e\-a\-t\- \-P\-a\-r\-s\-e on page} \pageref{Listing64}  \textsc{Included Blocks}: \hyperref[Listing1]{1 on page} \pageref{Listing1}\end{footnotesize}\vskip 5mm\noindent

The children of a fixed stitch repeat node are commma separated.

\begin{lstlisting}[title={<Fixed Stitch Repeat Parse Children 63>}, label=Listing63]
node.children.push(StitchOpParse());
	
while (Sym.type == CharSym.Comma) {	
	nextSym();
	node.children.push(StitchOpParse());
}
\end{lstlisting}\begin{footnotesize} \textsc{Used in}: \hyperref[Listing64]{F\-i\-x\-e\-d\- \-S\-t\-i\-t\-c\-h\- \-R\-e\-p\-e\-a\-t\- \-P\-a\-r\-s\-e on page} \pageref{Listing64}  \end{footnotesize}\vskip 5mm\noindent

A fixed stitch repeat must end with a natural number expression.

\begin{lstlisting}[title={<Fixed Stitch Repeat Parse 64>}, label=Listing64]
var FixedStRepParse = function() {
	
	var node = { type : NodeType.FixedStRep, children : [], line : State.line };
	
	<<Fixed Stitch Repeat Parse Open 61>>
	<<Fixed Stitch Repeat Parse Children 63>>
	<<Fixed Stitch Repeat Parse Close 62>>
	node.repCount = ExpressionParse(CharSym.Period);
	
	return node;
};
\end{lstlisting}\begin{footnotesize} \textsc{Used in}: \hyperref[Listing125]{A\-s\-t\- \-C\-o\-n\-s\-t\-r\-u\-c\-t\-i\-o\-n\- \-P\-a\-s\-s on page} \pageref{Listing125}  \textsc{Included Blocks}: \hyperref[Listing61]{6\-1 on page} \pageref{Listing61}, \hyperref[Listing63]{6\-3 on page} \pageref{Listing63}, \hyperref[Listing62]{6\-2 on page} \pageref{Listing62}\end{footnotesize}\vskip 5mm\noindent

\subsection{Verification}

A fixed stitch repeat adds the number of worked stitches and stitch change multiplied by the specified repeat amount to
these values for its parent node.

\begin{lstlisting}[title={<Verify Fixed Stitch Repeat 65>}, label=Listing65]
VerifyRowElemChildren(node, rowState_1);
rowState_0.workedSt += rowState_1.workedSt * rep;
rowState_0.stChange += rowState_1.stChange * rep;
\end{lstlisting}\begin{footnotesize} \textsc{Used in}: \hyperref[Listing29]{V\-e\-r\-i\-f\-y\- \-R\-o\-w\- \-E\-l\-e\-m on page} \pageref{Listing29}  \end{footnotesize}\vskip 5mm\noindent

\subsection{HTML Generation}

\begin{lstlisting}[title={<Write HTML Fixed Stitch Repeat 66>}, label=Listing66]
var WriteFixedStRep = function(node) {
	
	var stitches = [];
	
	if (node.children != null) {
		for (var i = 0; i < node.children.length; i++) {
			stitches.push(WriteNode(node.children[i]));
		}
	}
	
	return "[" + stitches.join(", ") + "] " + node.repCount.value + " times";
};
\end{lstlisting}\begin{footnotesize} \textsc{Used in}: \hyperref[Listing132]{c\-o\-d\-e\-/\-c\-o\-d\-e\-g\-e\-n\-.\-j\-s on page} \pageref{Listing132}  \end{footnotesize}\vskip 5mm\noindent

\section{Undetermined Stitch Repeat}
\label{sec:ustrep} 

A sequence of stitches may be repeated a number of times that depends on the current length of the
row. We will call this an undetermined stitch repeat.

\subsection{AST Node}

\begin{grammar}
<uStRep> : `*' <stitchOp> ( ',' <stitchOp> )* `;' `to' (`last' <expr> `| `end')
\end{grammar}

The first example below means perform the knit stitch as many times as required to get to the end of the row. The second
example means perform the knit stitch to the last two stitches, then purl the last two.

\begin{purlex}[Undetermined Stitch Repeat 1]{ustrepex1}
*K; to end
\end{purlex}

\begin{purlex}[Undetermined Stitch Repeat 2]{ustrepex2}
*K; to last 2, P 2.
\end{purlex}

An undetermined stitch repeat must begin with an asterisk. As in \hyperref[sec:compst]{compound stitches} and
\hyperref[sec:fstrep]{fixed repeats}, if an invalid symbol is used, the lexer scans to the terminator of the parent
production to continue parsing.

\begin{lstlisting}[title={<Undetermined Stitch Repeat Parse Open 67>}, label=Listing67]
if (Sym.type == CharSym.Asterisk) {
	nextSym();
} else {
	<<Unexpected Symbol Error 1>>
	var msg = "Missing \'*\' symbol at start of undetermined stitch repeat.";
	AddMsg(MsgType.Error, node, msg);
	scanToSym(CharSym.Period);
	return node;
}
\end{lstlisting}\begin{footnotesize} \textsc{Used in}: \hyperref[Listing72]{U\-n\-d\-e\-t\-e\-r\-m\-i\-n\-e\-d\- \-S\-t\-i\-t\-c\-h\- \-R\-e\-p\-e\-a\-t\- \-P\-a\-r\-s\-e on page} \pageref{Listing72}  \textsc{Included Blocks}: \hyperref[Listing1]{1 on page} \pageref{Listing1}\end{footnotesize}\vskip 5mm\noindent

Following the asterisk is a sequence of comma-separated \hyperref[sec:basicst]{basic stitches},
\hyperref[sec:compst]{compound stitches}, and \hyperref[sec:fstrep]{fixed stitch repeats}.

\begin{lstlisting}[title={<Undetermined Stitch Repeat Parse Children 68>}, label=Listing68]
node.children.push(StitchOpParse());
	
while (Sym.type == CharSym.Comma) {	
	nextSym();
	node.children.push(StitchOpParse());
}
\end{lstlisting}\begin{footnotesize} \textsc{Used in}: \hyperref[Listing72]{U\-n\-d\-e\-t\-e\-r\-m\-i\-n\-e\-d\- \-S\-t\-i\-t\-c\-h\- \-R\-e\-p\-e\-a\-t\- \-P\-a\-r\-s\-e on page} \pageref{Listing72}  \end{footnotesize}\vskip 5mm\noindent

The sequence of children is terminated with a semicolon. 

\begin{lstlisting}[title={<Undetermined Stitch Repeat Parse Close 69>}, label=Listing69]
if (Sym.type == CharSym.Semicolon) {
	nextSym();
} else if (Sym.type == CharSym.Colon) {
	var msg = "Use \';\' symbol at the end of undetermined stitch repeat stitches.";
	AddMsg(MsgType.Warning, node, msg);
	nextSym();
} else {
	<<Unexpected Symbol Error 1>>
	var msg = "Missing \';\' symbol at the end of undetermined stitch repeat stitches.";
	AddMsg(MsgType.Error, node, msg);
	scanToSym(CharSym.Period);
	return node;
}
\end{lstlisting}\begin{footnotesize} \textsc{Used in}: \hyperref[Listing72]{U\-n\-d\-e\-t\-e\-r\-m\-i\-n\-e\-d\- \-S\-t\-i\-t\-c\-h\- \-R\-e\-p\-e\-a\-t\- \-P\-a\-r\-s\-e on page} \pageref{Listing72}  \textsc{Included Blocks}: \hyperref[Listing1]{1 on page} \pageref{Listing1}\end{footnotesize}\vskip 5mm\noindent

If there is an error in the remainder of the undeterminded stitch repeat, then since we are past the sequence of children we
can assume that the next comma delimits the next sibling, and the next period is the terminator of the parent production.

Next, the pattern writer must specify how far along the row this repeat should be performed,
beginning with the keyword ``to".

\begin{lstlisting}[title={<Undetermined Stitch Repeat Parse To 70>}, label=Listing70]
if (Sym.type == KeywordSym.To) {
	nextSym();
} else if (Sym.type == SymType.Ident) {
	var msg = "Use \'" + KeywordSym.To + "\' after \';\' for undetermined stitch repeat.";
	AddMsg(MsgType.Warning, node, msg);
	nextSym();
} else {
	<<Unexpected Symbol Error 1>>
	var msg = "Missing \'" + KeywordSym.To + "\' after \';\' for undetermined stitch repeat.";
	AddMsg(MsgType.Error, node, msg);
	scanToSym(CharSym.Comma || CharSym.Period);
}
\end{lstlisting}\begin{footnotesize} \textsc{Used in}: \hyperref[Listing72]{U\-n\-d\-e\-t\-e\-r\-m\-i\-n\-e\-d\- \-S\-t\-i\-t\-c\-h\- \-R\-e\-p\-e\-a\-t\- \-P\-a\-r\-s\-e on page} \pageref{Listing72}  \textsc{Included Blocks}: \hyperref[Listing1]{1 on page} \pageref{Listing1}\end{footnotesize}\vskip 5mm\noindent

The keyword ``end'' is used if the sequence should be repeated to the end of the row, and the keyword
``last'' is used if the sequence should be repeated to within a given number of stitches from the
end of the row. For this reason, nesting of undetermined stitch repeats is not possible. An
important note is that the number of stitches remaining on the left needle at the end of the last
repeat must coincide exactly with the number specified here (0 for ``end'').

\begin{lstlisting}[title={<Undetermined Stitch Repeat Parse Repeat Instruction 71>}, label=Listing71]
if (Sym.type == KeywordSym.Last) {
	nextSym();
	node.num = ExpressionParse(CharSym.Comma || CharSym.Period);
} else if (Sym.type == KeywordSym.End) {
	node.num = { type : NodeType.NatLiteral, value : 0 };
	nextSym();
} else {
	<<Unexpected Symbol Error 1>>
	var msg = "Missing repeat instructions. Expecting \'" + KeywordSym.Last + "\' or \'" + KeywordSym.End + "\'.";
	AddMsg(MsgType.Error, node, msg);
	scanToSym(CharSym.Comma || CharSym.Period);
}
\end{lstlisting}\begin{footnotesize} \textsc{Used in}: \hyperref[Listing72]{U\-n\-d\-e\-t\-e\-r\-m\-i\-n\-e\-d\- \-S\-t\-i\-t\-c\-h\- \-R\-e\-p\-e\-a\-t\- \-P\-a\-r\-s\-e on page} \pageref{Listing72}  \textsc{Included Blocks}: \hyperref[Listing1]{1 on page} \pageref{Listing1}\end{footnotesize}\vskip 5mm\noindent

\begin{lstlisting}[title={<Undetermined Stitch Repeat Parse 72>}, label=Listing72]
var UStRepParse = function() {
	
	var node = { type : NodeType.UStRep, children : [], line : State.line };
	
	<<Undetermined Stitch Repeat Parse Open 67>>
	<<Undetermined Stitch Repeat Parse Children 68>>
	<<Undetermined Stitch Repeat Parse Close 69>>
	<<Undetermined Stitch Repeat Parse To 70>>
	<<Undetermined Stitch Repeat Parse Repeat Instruction 71>>
	
	return node;
};
\end{lstlisting}\begin{footnotesize} \textsc{Used in}: \hyperref[Listing125]{A\-s\-t\- \-C\-o\-n\-s\-t\-r\-u\-c\-t\-i\-o\-n\- \-P\-a\-s\-s on page} \pageref{Listing125}  \textsc{Included Blocks}: \hyperref[Listing67]{6\-7 on page} \pageref{Listing67}, \hyperref[Listing68]{6\-8 on page} \pageref{Listing68}, \hyperref[Listing69]{6\-9 on page} \pageref{Listing69}, \hyperref[Listing70]{7\-0 on page} \pageref{Listing70}, \hyperref[Listing71]{7\-1 on page} \pageref{Listing71}\end{footnotesize}\vskip 5mm\noindent

\subsection{Verification}

To verify an undetermined stitch repeat, we first determine the number of stitches the children will be repeated over. If
the number of stitches worked over a single repeat does not divide this value, then there will be a number of unworked
stitches equal to the division remainder at the end of the row. In this case, an error is generated. Otherwise, the row state
of the parent node is updated with the number of worked stitches and stitch change caused by the undetermined stitch repeat.

\begin{lstlisting}[title={<Verify Undetermined Stitch Repeat 73>}, label=Listing73]
VerifyRowElemChildren(node, rowState_1);
				
var stToWork = rowState_1.initialWidth - rowState_0.workedSt - node.num.value;

var remainSt = stToWork % rowState_1.workedSt;
if (remainSt != 0) {
	var msg = remainSt + " st will remain after the last possible repeat.";
	AddMsg(MsgType.Verification, node, msg);
} else {
	rep = (stToWork - (stToWork % rowState_1.workedSt)) / rowState_1.workedSt;
}

rowState_0.workedSt += rowState_1.workedSt * rep;
rowState_0.stChange += rowState_1.stChange * rep;
\end{lstlisting}\begin{footnotesize} \textsc{Used in}: \hyperref[Listing29]{V\-e\-r\-i\-f\-y\- \-R\-o\-w\- \-E\-l\-e\-m on page} \pageref{Listing29}  \end{footnotesize}\vskip 5mm\noindent

\subsection{HTML Generation}

\begin{lstlisting}[title={<Write HTML Undetermined Stitch Repeat 74>}, label=Listing74]
var WriteUStRep = function(node) {
	
	var result = [];
	var stitches = [];
	
	if (node.children != null) {
		for (var i = 0; i < node.children.length; i++) {
			stitches.push(WriteNode(node.children[i]));
		}
	}
	
	result.push("*" + stitches.join(", ") + "; rep from * to");
	
	var rem = node.num.value;
	if (rem == 0) {
		result.push("end");
	} else if (rem == 1) {
		result.push("last " + rem + " st");
	} else if (rem > 1) {
		result.push("last " + rem + " sts");
	} else {
		result.push("invalid value");
	}
	
	return result.join(" ");
};
\end{lstlisting}\begin{footnotesize} \textsc{Used in}: \hyperref[Listing132]{c\-o\-d\-e\-/\-c\-o\-d\-e\-g\-e\-n\-.\-j\-s on page} \pageref{Listing132}  \end{footnotesize}\vskip 5mm\noindent

\section{Bind-Off}
\label{sec:bindoff}

We have considered cast-ons, rows and stitches. These elements, followed by a bind-off, are sufficient to construct a simple
pattern.
 
\subsection{AST Node}

\begin{grammar}
<bo> ::= `BO' <Nat> `.'
\end{grammar}

\begin{purlex}[Market Bag Body Bind-off]{boex}
BO 100.
\end{purlex}

The ``BO'' keyword is used to declare a bind-off for the pattern.

\begin{lstlisting}[title={<Bind-Off Parse BO 75>}, label=Listing75]
if (Sym.type == KeywordSym.BindOff) {
	nextSym();
} else if (Sym.type == SymType.Ident) {
	var msg = "A bind-off declaration must start with \'" + KeywordSym.BindOff + "\'.";
	AddMsg(MsgType.Warning, node, msg);
} else {
	<<Unexpected Symbol Error 1>>
	var msg = "Missing \'"+ KeywordSym.BindOff + "\' at start of bind-off declaration.";
	AddMsg(MsgType.Error, node, msg);
	scanToSym(CharSym.Period);
	nextSym();
	return node;
}
\end{lstlisting}\begin{footnotesize} \textsc{Used in}: \hyperref[Listing77]{B\-i\-n\-d\--\-O\-f\-f\- \-P\-a\-r\-s\-e on page} \pageref{Listing77}  \textsc{Included Blocks}: \hyperref[Listing1]{1 on page} \pageref{Listing1}\end{footnotesize}\vskip 5mm\noindent

Next, a natural number is given as the number of stitches to bind-off.

\begin{lstlisting}[title={<Bind-Off Parse Count 76>}, label=Listing76]
if (Sym.type == SymType.Nat) {
	node.value = Sym.value;
	nextSym();
} else {
	<<Unexpected Symbol Error 1>>
	AddMsg(MsgType.Error, node, "Bind-off count not specified.");
	scanToSym(CharSym.Period);
	nextSym();
	return node;
}
\end{lstlisting}\begin{footnotesize} \textsc{Used in}: \hyperref[Listing77]{B\-i\-n\-d\--\-O\-f\-f\- \-P\-a\-r\-s\-e on page} \pageref{Listing77}  \textsc{Included Blocks}: \hyperref[Listing1]{1 on page} \pageref{Listing1}\end{footnotesize}\vskip 5mm\noindent

A bind-off ends with a period as terminator.

\begin{lstlisting}[title={<Bind-Off Parse 77>}, label=Listing77]
var BoParse = function() {

	var node = { type : NodeType.BindOff, value : 0, line : State.line };
	
	<<Bind-Off Parse BO 75>>
	<<Bind-Off Parse Count 76>>
	<<Period Terminator 6>>
	
	return node;
};
\end{lstlisting}\begin{footnotesize} \textsc{Used in}: \hyperref[Listing125]{A\-s\-t\- \-C\-o\-n\-s\-t\-r\-u\-c\-t\-i\-o\-n\- \-P\-a\-s\-s on page} \pageref{Listing125}  \textsc{Included Blocks}: \hyperref[Listing75]{7\-5 on page} \pageref{Listing75}, \hyperref[Listing76]{7\-6 on page} \pageref{Listing76}, \hyperref[Listing6]{6 on page} \pageref{Listing6}\end{footnotesize}\vskip 5mm\noindent

\subsection{Verification}

The value of a bind-off node must be equivalent to the final width of the row before it, otherwise either some stitches will
still be active after completing a pattern, or else there will be too few stitches to bind-off (the former error being much
more severe).

\begin{lstlisting}[title={<Verify Bind-Off 78>}, label=Listing78]
var VerifyBindOff = function(node) {
	
	if (node.value != State.width) {
		var msg = "Binding off " + node.value + " sts over " + State.width + " sts.";
		AddMsg(MsgType.Verification, node, msg);
	}
};
\end{lstlisting}\begin{footnotesize} \textsc{Used in}: \hyperref[Listing131]{V\-e\-r\-i\-f\-i\-c\-a\-t\-i\-o\-n\- \-P\-a\-s\-s on page} \pageref{Listing131}  \end{footnotesize}\vskip 5mm\noindent

\subsection{HTML Generation}

\begin{lstlisting}[title={<Write HTML Bind-Off 79>}, label=Listing79]
var WriteBo = function(node) {
	var msg = "Bind-off  " + node.value + " sts.";
	return AddElement(TagType.Div, ClassType.BindOff, msg);
}; 
\end{lstlisting}\begin{footnotesize} \textsc{Used in}: \hyperref[Listing132]{c\-o\-d\-e\-/\-c\-o\-d\-e\-g\-e\-n\-.\-j\-s on page} \pageref{Listing132}  \end{footnotesize}\vskip 5mm\noindent

\section{Join}
\label{sec:join}

An alternative to using a bind-off to finish a pattern is to \emph{join} the remaining active stitches to some other
location on the same section, or another knitted object. If active stitches are to be joined to other active stitches, this
is called \emph{grafting}.

\subsection{AST Node}

\begin{grammar}
<join> ::= `Join' <Nat> `to' <String> `.'
\end{grammar}

\begin{purlex}[Market Bag Handle Join]{joinex}
Join 10 to ``Body top''.
\end{purlex}

The ``Join'' keyword is used to declare a join for the pattern.

\begin{lstlisting}[title={<Join Parse Keyword 80>}, label=Listing80]
if (Sym.type == KeywordSym.Join) {
	nextSym();
} else if (Sym.type == SymType.Ident) {
	var msg = "A join declaration must start with \'" + KeywordSym.Join + "\'.";
	AddMsg(MsgType.Warning, node, msg);
} else {
	<<Unexpected Symbol Error 1>>
	var msg = "Missing \'"+ KeywordSym.Join + "\' at start of join declaration.";
	AddMsg(MsgType.Error, node, msg);
	scanToSym(CharSym.Period);
	nextSym();
	return node;
}
\end{lstlisting}\begin{footnotesize} \textsc{Used in}: \hyperref[Listing83]{J\-o\-i\-n\- \-P\-a\-r\-s\-e on page} \pageref{Listing83}  \textsc{Included Blocks}: \hyperref[Listing1]{1 on page} \pageref{Listing1}\end{footnotesize}\vskip 5mm\noindent

Next, a natural number is given as the number of stitches to join.

\begin{lstlisting}[title={<Join Parse Count 81>}, label=Listing81]
if (Sym.type == SymType.Nat) {
	node.value = Sym.value;
	nextSym();
} else {
	<<Unexpected Symbol Error 1>>
	AddMsg(MsgType.Error, node, "Join count not specified.");
	scanToSym(CharSym.Period);
	nextSym();
	return node;
}
\end{lstlisting}\begin{footnotesize} \textsc{Used in}: \hyperref[Listing83]{J\-o\-i\-n\- \-P\-a\-r\-s\-e on page} \pageref{Listing83}  \textsc{Included Blocks}: \hyperref[Listing1]{1 on page} \pageref{Listing1}\end{footnotesize}\vskip 5mm\noindent

It is necessary to state where the active stitches should be joined. This begins with the keyword ``to'' followed by a
string with directions on the join.

\begin{lstlisting}[title={<Join Parse Destination 82>}, label=Listing82]
if (Sym.type == KeywordSym.To) {
	nextSym();
} else {
	<<Unexpected Symbol Error 1>>
	scanToSym(CharSym.Period);
	nextSym();
	return node;
}

if (Sym.type == SymType.String) {
	node.destination = Sym.value;
	nextSym();
} else {
	<<Unexpected Symbol Error 1>>
	AddMsg(MsgType.Error, node, "Missing join destination.");
	scanToSym(CharSym.Period);
	nextSym();
	return node;
}
\end{lstlisting}\begin{footnotesize} \textsc{Used in}: \hyperref[Listing83]{J\-o\-i\-n\- \-P\-a\-r\-s\-e on page} \pageref{Listing83}  \textsc{Included Blocks}: \hyperref[Listing1]{1 on page} \pageref{Listing1}, \hyperref[Listing1]{1 on page} \pageref{Listing1}\end{footnotesize}\vskip 5mm\noindent

A join ends with a period as terminator.

\begin{lstlisting}[title={<Join Parse 83>}, label=Listing83]
var JoinParse = function() {

	var node = { type : NodeType.Join, value : 0, line : State.line };
	
	<<Join Parse Keyword 80>>
	<<Join Parse Count 81>>
	<<Join Parse Destination 82>>
	<<Period Terminator 6>>
	
	return node;
};
\end{lstlisting}\begin{footnotesize} \textsc{Used in}: \hyperref[Listing125]{A\-s\-t\- \-C\-o\-n\-s\-t\-r\-u\-c\-t\-i\-o\-n\- \-P\-a\-s\-s on page} \pageref{Listing125}  \textsc{Included Blocks}: \hyperref[Listing80]{8\-0 on page} \pageref{Listing80}, \hyperref[Listing81]{8\-1 on page} \pageref{Listing81}, \hyperref[Listing82]{8\-2 on page} \pageref{Listing82}, \hyperref[Listing6]{6 on page} \pageref{Listing6}\end{footnotesize}\vskip 5mm\noindent

\subsection{Verification}

The value of a join node must be equivalent to the final width of the row before it, otherwise either some stitches will
still be active after completing a pattern, or else there will be too few stitches to bind-off (the former error being much
more severe).

\begin{lstlisting}[title={<Verify Join 84>}, label=Listing84]
var VerifyJoin = function(node) {
	
	if (node.value != State.width) {
		var msg = "Joining " + node.value + " sts of " + State.width + " sts.";
		AddMsg(MsgType.Verification, node, msg);
	}
};
\end{lstlisting}\begin{footnotesize} \textsc{Used in}: \hyperref[Listing131]{V\-e\-r\-i\-f\-i\-c\-a\-t\-i\-o\-n\- \-P\-a\-s\-s on page} \pageref{Listing131}  \end{footnotesize}\vskip 5mm\noindent

\subsection{HTML Generation}

\begin{lstlisting}[title={<Write HTML Join 85>}, label=Listing85]
var WriteJoin = function(node) {
	var msg = "Join  " + node.value + " sts to " + node.destination + ".";
	return AddElement(TagType.Div, ClassType.Join, msg);
}; 
\end{lstlisting}\begin{footnotesize} \textsc{Used in}: \hyperref[Listing132]{c\-o\-d\-e\-/\-c\-o\-d\-e\-g\-e\-n\-.\-j\-s on page} \pageref{Listing132}  \end{footnotesize}\vskip 5mm\noindent

\section{Pattern}
\label{sec:pattern}

Since we have now seen enough to construct a simple patern, we will look at the syntax of a knitting pattern written in Purl.

\subsection{AST Node}

\begin{grammar}
<pattern> ::= `pattern' <String> `:' (<co> <body> <bo> | <section>+)
\end{grammar}

\begin{purlex}[Simple single row pattern]{pattex1}
pattern ``One Row'':
CO 1.
row : K.
BO 1.
\end{purlex}

A pattern definition begins with the keyword ``pattern''.

\begin{lstlisting}[title={<Pattern Parse Pattern 86>}, label=Listing86]
if (Sym.type == KeywordSym.Pattern) {
	nextSym();
} else if (Sym.type == SymType.Ident) {
	var msg = "A pattern declaration must start with \'" + KeywordSym.Pattern + "\'.";
	AddMsg(MsgType.Warning, node, msg);
	nextSym();
} else {
	<<Unexpected Symbol Error 1>>
	var msg = "Expecting \'" + KeywordSym.Pattern + "\' to start pattern declaration.";
	AddMsg(MsgType.Error, node, msg);
	scanToSym(CharSym.Colon);
}
\end{lstlisting}\begin{footnotesize} \textsc{Used in}: \hyperref[Listing90]{P\-a\-t\-t\-e\-r\-n\- \-P\-a\-r\-s\-e on page} \pageref{Listing90}  \textsc{Included Blocks}: \hyperref[Listing1]{1 on page} \pageref{Listing1}\end{footnotesize}\vskip 5mm\noindent

The next requirement is a string in double quotes. This is the title of the patern. The reason for using a string rather
than a single identifier is to allow greater flexibility in the naming of patterns. In this way, a pattern can have a
multi-word name that can also include any reserved keywords. If an identifier is provided rather than a string, a warning is
created and compilation continues. Any other symbol causes an error and the lexer scans to the next colon symbol, which is
required following the pattern title.

\begin{lstlisting}[title={<Pattern Parse Title 87>}, label=Listing87]
if (Sym.type == SymType.String) {
	node.name = Sym.value;
	nextSym();
} else if (Sym.type == SymType.Ident) {
	node.name = Sym.value;
	var msg = "Remember to use double quotes around the name of your pattern.";
	AddMsg(MsgType.Warning, node, msg);
	nextSym(); 
} else {
	<<Unexpected Symbol Error 1>>
	AddMsg(MsgType.Error, node, "The pattern name is not specified.");
	scanToSym(CharSym.Colon);
}
\end{lstlisting}\begin{footnotesize} \textsc{Used in}: \hyperref[Listing90]{P\-a\-t\-t\-e\-r\-n\- \-P\-a\-r\-s\-e on page} \pageref{Listing90}  \textsc{Included Blocks}: \hyperref[Listing1]{1 on page} \pageref{Listing1}\end{footnotesize}\vskip 5mm\noindent

The main content of the pattern may by a cast-on, pattern body, then bind-off for a single-section pattern, or a number of
defined sections. If the symbol after the colon is the cast-on keyword ``CO'', then we are parsing a simple pattern. A
simple pattern provides instructions for a single discrete object. In contrast, a composite pattern provides instructions to
knit multiple objects to be joined to form a larger structure.

\begin{lstlisting}[title={<Pattern Parse Main 88>}, label=Listing88]
if (Sym.type == KeywordSym.CastOn) {
	<<Pattern Content Parse 89>>
} else {
	<<Pattern Parse Composite 99>>
}
\end{lstlisting}\begin{footnotesize} \textsc{Used in}: \hyperref[Listing90]{P\-a\-t\-t\-e\-r\-n\- \-P\-a\-r\-s\-e on page} \pageref{Listing90}  \textsc{Included Blocks}: \hyperref[Listing89]{8\-9 on page} \pageref{Listing89}, \hyperref[Listing99]{9\-9 on page} \pageref{Listing99}\end{footnotesize}\vskip 5mm\noindent

A simple pattern begins with a \hyperref[sec:caston]{cast-on} or \hyperref[sec:pickup]{pick-up}, followed by a \hyperref[sec:patternbody]{pattern body}, and
finally a \hyperref[sec:bindoff]{bind-off} or a \hyperref[sec:join]{join}.

\begin{lstlisting}[title={<Pattern Content Parse 89>}, label=Listing89]
node.start = CoParse();
node.children = BodyParse();
node.finish = BoParse();
\end{lstlisting}\begin{footnotesize} \textsc{Used in}: \hyperref[Listing88]{P\-a\-t\-t\-e\-r\-n\- \-P\-a\-r\-s\-e\- \-M\-a\-i\-n on page} \pageref{Listing88}  \end{footnotesize}\vskip 5mm\noindent

\begin{lstlisting}[title={<Pattern Parse 90>}, label=Listing90]
var PatternParse = function(){
	
	var node = { type : NodeType.Pattern, children : [], line : State.line };
	
	<<Pattern Parse Pattern 86>>
	<<Pattern Parse Title 87>>
	<<Colon Separator 18>>

	<<Pattern Parse Main 88>>
	
	return node;
};
\end{lstlisting}\begin{footnotesize} \textsc{Used in}: \hyperref[Listing125]{A\-s\-t\- \-C\-o\-n\-s\-t\-r\-u\-c\-t\-i\-o\-n\- \-P\-a\-s\-s on page} \pageref{Listing125}  \textsc{Included Blocks}: \hyperref[Listing86]{8\-6 on page} \pageref{Listing86}, \hyperref[Listing87]{8\-7 on page} \pageref{Listing87}, \hyperref[Listing18]{1\-8 on page} \pageref{Listing18}, \hyperref[Listing88]{8\-8 on page} \pageref{Listing88}\end{footnotesize}\vskip 5mm\noindent

\subsection{Verification}

When the \texttt{co} property is not null, we are verifying a simple pattern, so we verify the cast-on, the children, and
then the bind-off. Otherwise, we have a composite pattern and we verify the children (which are pattern sections in this
case).

\begin{lstlisting}[title={<Verify Pattern 91>}, label=Listing91]
var VerifyPattern = function(node) {
	
	if (node.start != null) {
		VerifyNode(node.start);
		VerifyChildren(node);
		VerifyNode(node.finish);
	} else {
		VerifyChildren(node);
	}
};
\end{lstlisting}\begin{footnotesize} \textsc{Used in}: \hyperref[Listing131]{V\-e\-r\-i\-f\-i\-c\-a\-t\-i\-o\-n\- \-P\-a\-s\-s on page} \pageref{Listing131}  \end{footnotesize}\vskip 5mm\noindent

\subsection{HTML Generation}

\begin{lstlisting}[title={<Write HTML Pattern 92>}, label=Listing92]
var WritePattern = function(node) {
	
	var result = [];
	
	result.push(OpenElement(TagType.Div, ClassType.Pattern));
	result.push(AddElement(TagType.Div, ClassType.PatternName, node.name));
	
	if (node.start != null) {
		result.push(WriteNode(node.start));
		result.push(WriteBody(node));
		result.push(WriteNode(node.finish));
	} else if (node.children != null) {
		for (var i = 0; i < node.children.length; i++) {
			result.push(WriteNode(node.children[i]));
		}
	}
	
	result.push(CloseElement(TagType.Div));
	
	return result.join("")
};
\end{lstlisting}\begin{footnotesize} \textsc{Used in}: \hyperref[Listing132]{c\-o\-d\-e\-/\-c\-o\-d\-e\-g\-e\-n\-.\-j\-s on page} \pageref{Listing132}  \end{footnotesize}\vskip 5mm\noindent

\section{Pattern Body}
\label{sec:patternbody}

The body of a pattern is the main content of the pattern (everything between cast-on and bind-off). This consists of zero or
more \hyperref[sec:row]{rows}, row repeats (see~\ref{sec:rowrep}), and sample calls (see~\ref{subsec:samplecall}). Note that
<pattern> is not the only production that uses the <body> production to create its child nodes. \hyperref[sec:rowrep]{Row
repeats} and sample definitions (see~\ref{subsec:sampledef}) are other constructs that use the <body> production.

\subsection{AST Node}

\begin{grammar}
<body> ::= (<rowDef> | <rowRep> | <sampleCall>)*
\end{grammar}

The body parse method builds up and returns an array of nodes which make up a pattern body.

\begin{lstlisting}[title={<Body Parse 93>}, label=Listing93]
var BodyParse = function(){
	
	var bodyElems = [];
	
	while (Sym.type != SymType.EOF) {
		switch (Sym.type) {
			case KeywordSym.Row:
			case KeywordSym.Rnd:
				bodyElems.push(RowDefParse());
				break;
			
			case SymType.RowRep:
				bodyElems.push(RowRepeatParse());
				break;
			
			case SymType.Ident:
				bodyElems.push(SampleCallParse());
				break;
			
			default:
				return bodyElems;
		}
	}
	
	return bodyElems;
};
\end{lstlisting}\begin{footnotesize} \textsc{Used in}: \hyperref[Listing125]{A\-s\-t\- \-C\-o\-n\-s\-t\-r\-u\-c\-t\-i\-o\-n\- \-P\-a\-s\-s on page} \pageref{Listing125}  \end{footnotesize}\vskip 5mm\noindent

\subsection{HTML Generation}

\begin{lstlisting}[title={<Write HTML Body 94>}, label=Listing94]
var WriteBody = function(node) {
	
	var result = [];
	result.push(OpenElement(TagType.Div, ClassType.Body));
	
	if (node.children != null) {
		for (var i = 0; i < node.children.length; i++) {
			result.push(WriteNode(node.children[i]));
		}
	}
	
	result.push(CloseElement(TagType.Div));
	
	return result.join("");
};
\end{lstlisting}\begin{footnotesize} \textsc{Used in}: \hyperref[Listing132]{c\-o\-d\-e\-/\-c\-o\-d\-e\-g\-e\-n\-.\-j\-s on page} \pageref{Listing132}  \end{footnotesize}\vskip 5mm\noindent

\section{Row Repeats}
\label{sec:rowrep}

If elements of a pattern body are to be repeated a number of times, rather than rewriting those elements, a row repeat may
be used for more concise notation.

\subsection{AST Node}

\begin{grammar}
<rowRepeat> ::= `**' <body> `repeat' <expr>
\end{grammar}

The body of the \texttt{diagonalLace} sample used by the market bag pattern uses a row repeat. This means the two rows
between the double asterisk and \texttt{repeat n} are to be repeated n times.

\begin{purlex}[Ribbing repeat]{rowrepex}
**
rnd : *K2T, YO; to end.
rnd : *K; to end.
repeat n
\end{purlex}

A row repeat begins with two asterisks.

\begin{lstlisting}[title={<RowRepeatParse\_Open 95>}, label=Listing95]
if (Sym.type == SymType.RowRep) {
	nextSym();
} else if (Sym.type == CharSym.Asterisk) {
	var msg = "Row repeat must begin with \'" + SymType.RowRep + "\'.";
	AddMsg(MsgType.Warning, node, msg);
	nextSym();
} else {
	<<Unexpected Symbol Error 1>>
	var msg = "Missing \'" + SymType.RowRep + "\' at start of row repeat.";
	AddMsg(MsgType.Error, node, msg);
	scanToSym(CharSym.Period);
	nextSym();
}
\end{lstlisting}\begin{footnotesize} \textsc{Used in}: \hyperref[Listing97]{R\-o\-w\- \-R\-e\-p\-e\-a\-t\- \-P\-a\-r\-s\-e on page} \pageref{Listing97}  \textsc{Included Blocks}: \hyperref[Listing1]{1 on page} \pageref{Listing1}\end{footnotesize}\vskip 5mm\noindent

The end of a row repeat body is marked by the ``repeat'' keyword. This must then be followed by a natural number or variable.

\begin{lstlisting}[title={<RowRepeatParse\_Close 96>}, label=Listing96]
if (Sym.type == KeywordSym.Repeat) {
	nextSym();
} else if (Sym.type == SymType.Ident) {
	var msg = "Row repeat body must be followed by \'" + KeywordSym.Repeat + "\'.";
	AddMsg(MsgType.Warning, node, msg);
	nextSym();
} else {
	<<Unexpected Symbol Error 1>>
	var msg = "Missing \'" + KeywordSym.Repeat + "\' after row repeat body.";
	AddMsg(MsgType.Error, node, msg);
	scanToSym(CharSym.Period);
	nextSym();
}
\end{lstlisting}\begin{footnotesize} \textsc{Used in}: \hyperref[Listing97]{R\-o\-w\- \-R\-e\-p\-e\-a\-t\- \-P\-a\-r\-s\-e on page} \pageref{Listing97}  \textsc{Included Blocks}: \hyperref[Listing1]{1 on page} \pageref{Listing1}\end{footnotesize}\vskip 5mm\noindent

The children of a row repeat are represented as an array of nodes. We use the same body parse method as simple
\hyperref[sec:pattern]{patterns} and pattern \hyperref[sec:section]{sections}.

\begin{lstlisting}[title={<Row Repeat Parse 97>}, label=Listing97]
var RowRepeatParse = function() {
	
	var node = { type : NodeType.RowRep, children : [], line : State.line };
	
	<<RowRepeatParse_Open 95>>
	node.children = BodyParse();
	<<RowRepeatParse_Close 96>>
	node.repCount = ExpressionParse(CharSym.Period);
	
	return node;
};
\end{lstlisting}\begin{footnotesize} \textsc{Used in}: \hyperref[Listing125]{A\-s\-t\- \-C\-o\-n\-s\-t\-r\-u\-c\-t\-i\-o\-n\- \-P\-a\-s\-s on page} \pageref{Listing125}  \textsc{Included Blocks}: \hyperref[Listing95]{9\-5 on page} \pageref{Listing95}, \hyperref[Listing96]{9\-6 on page} \pageref{Listing96}\end{footnotesize}\vskip 5mm\noindent

\subsection{HTML Generation}

\begin{lstlisting}[title={<Write HTML Row Repeat 98>}, label=Listing98]
var WriteRowRep = function(node) {
	
	var result = [];
	
	result.push(OpenElement(TagType.Div, ClassType.RowRep));
	result.push("**");
	result.push(WriteBody(node));
	result.push("rep from ** " + node.repCount.value + " times");
	result.push(CloseElement(TagType.Div));
	
	return result.join("");
};
\end{lstlisting}\begin{footnotesize} \textsc{Used in}: \hyperref[Listing132]{c\-o\-d\-e\-/\-c\-o\-d\-e\-g\-e\-n\-.\-j\-s on page} \pageref{Listing132}  \end{footnotesize}\vskip 5mm\noindent

\section{Pattern Section}
\label{sec:section}

If writing a pattern to create multiple discrete objects, then the pattern must be made up of a
number of \emph{pattern sections}. In this case, we write a composite pattern, which is a pattern with zero or more
pattern sections as its children.

\begin{lstlisting}[title={<Pattern Parse Composite 99>}, label=Listing99]
while (Sym.type == KeywordSym.Section) {
	node.children.push(SectionParse());
}
\end{lstlisting}\begin{footnotesize} \textsc{Used in}: \hyperref[Listing88]{P\-a\-t\-t\-e\-r\-n\- \-P\-a\-r\-s\-e\- \-M\-a\-i\-n on page} \pageref{Listing88}  \end{footnotesize}\vskip 5mm\noindent

\subsection{AST Node}

\begin{grammar}
<section> ::= `section' <String> `:' (<co> | <pu>) <body> (<bo> | <join>)
\end{grammar}

The market bag body is a section of the market bag pattern.

\begin{purlex}[Market Bag Body]{sectionex}
section "Body":
CO 8 circular.
rnd : *K, YO, K; to end.
rnd : *K; to end.
circleX with 1, 23.
diagonalLace with 30.
garterStitchCC with 4, 1.
BO 100.
\end{purlex}

A section definition begins with the ``section'' keyword.

\begin{lstlisting}[title={<Section Parse Section 100>}, label=Listing100]
if (Sym.type == KeywordSym.Section) {
	nextSym();
} else if (Sym.type == SymType.Ident) {
	var msg = "A section declaration must start with \'" + KeywordSym.Section + "\'.";
	AddMsg(MsgType.Warning, node, msg);
	nextSym();
} else {
	<<Unexpected Symbol Error 1>>
	var msg = "Missing \'" + KeywordSym.Section + "\' at start of section declaration.";
	AddMsg(MsgType.error, node, msg);
	scanToSym(CharSym.Colon);
}
\end{lstlisting}\begin{footnotesize} \textsc{Used in}: \hyperref[Listing103]{S\-e\-c\-t\-i\-o\-n\- \-P\-a\-r\-s\-e on page} \pageref{Listing103}  \textsc{Included Blocks}: \hyperref[Listing1]{1 on page} \pageref{Listing1}\end{footnotesize}\vskip 5mm\noindent

The title of the section follows, which should be a string enclosed in double quotes. As in the title of a pattern, if an
identifier symbol is found instead, then a warning is generated, the name of the section is set to the identifier, and
compilation can continue. Any other symbol will cause an error and the lexer will scan to the colon separator that should
occur before the section content.

\begin{lstlisting}[title={<Section Parse Title 101>}, label=Listing101]
if (Sym.type == SymType.String) {
	node.name = Sym.value;
	State.sectionName = node.name;
	nextSym();
} else if (Sym.type == SymType.Ident) {
	node.name = Sym.value;
	var msg = "Remember to use double quotes around the name of a section.";
	AddMsg(MsgType.Warning, node, msg);
	nextSym(); 
} else {
	<<Unexpected Symbol Error 1>>
	AddMsg(MsgType.Error, node, "The section name is not specified.");
	scanToSym(CharSym.Colon);
}
\end{lstlisting}\begin{footnotesize} \textsc{Used in}: \hyperref[Listing103]{S\-e\-c\-t\-i\-o\-n\- \-P\-a\-r\-s\-e on page} \pageref{Listing103}  \textsc{Included Blocks}: \hyperref[Listing1]{1 on page} \pageref{Listing1}\end{footnotesize}\vskip 5mm\noindent

Parsing of the content of a section is similar to parsing the content of a simple \hyperref[sec:pattern]{pattern}, except
that picking up stitches and joining are alternatives for cast-on and bind-off, respectively.

\begin{lstlisting}[title={<Section Content Parse 102>}, label=Listing102]
if (Sym.type == KeywordSym.CastOn) {
	node.start = CoParse();
} else if (Sym.type == KeywordSym.PickUp) {
	node.start = PuParse();
} else {
	<<Unexpected Symbol Error 1>>
	node.start = {};
	scanToSym(CharSym.Period);
	nextSym();
}

node.children = BodyParse();

if (Sym.type == KeywordSym.BindOff) {
	node.finish = BoParse();
} else if (Sym.type == KeywordSym.Join) {
	node.finish = JoinParse();
} else {
	<<Unexpected Symbol Error 1>>
	node.finish = {};
	scanToSym(CharSym.Period);
	nextSym();
}
\end{lstlisting}\begin{footnotesize} \textsc{Used in}: \hyperref[Listing103]{S\-e\-c\-t\-i\-o\-n\- \-P\-a\-r\-s\-e on page} \pageref{Listing103}  \textsc{Included Blocks}: \hyperref[Listing1]{1 on page} \pageref{Listing1}, \hyperref[Listing1]{1 on page} \pageref{Listing1}\end{footnotesize}\vskip 5mm\noindent

\begin{lstlisting}[title={<Section Parse 103>}, label=Listing103]
var SectionParse = function() {
	
	var node = { type : NodeType.Section, line : State.line };
	
	<<Section Parse Section 100>>
	<<Section Parse Title 101>>
	<<Colon Separator 18>>
	<<Section Content Parse 102>>
	
	return node;
};
\end{lstlisting}\begin{footnotesize} \textsc{Used in}: \hyperref[Listing125]{A\-s\-t\- \-C\-o\-n\-s\-t\-r\-u\-c\-t\-i\-o\-n\- \-P\-a\-s\-s on page} \pageref{Listing125}  \textsc{Included Blocks}: \hyperref[Listing100]{1\-0\-0 on page} \pageref{Listing100}, \hyperref[Listing101]{1\-0\-1 on page} \pageref{Listing101}, \hyperref[Listing18]{1\-8 on page} \pageref{Listing18}, \hyperref[Listing102]{1\-0\-2 on page} \pageref{Listing102}\end{footnotesize}\vskip 5mm\noindent

\subsection{Verification}

Pattern section verification is very similar to verification of a simple pattern, but first the \texttt{sectionName}
property of the \texttt{State} object is set to the name of the current section. This is used in reporting of errors.

\begin{lstlisting}[title={<Verify Section 104>}, label=Listing104]
var VerifySection = function(node) {
	
	State.sectionName = node.name;
	
	VerifyNode(node.start);
	VerifyChildren(node);
	VerifyNode(node.finish);
};
\end{lstlisting}\begin{footnotesize} \textsc{Used in}: \hyperref[Listing131]{V\-e\-r\-i\-f\-i\-c\-a\-t\-i\-o\-n\- \-P\-a\-s\-s on page} \pageref{Listing131}  \end{footnotesize}\vskip 5mm\noindent

\subsection{HTML Generation}

\begin{lstlisting}[title={<Write HTML Section 105>}, label=Listing105]
var WriteSection = function(node) {
		
		var result = [];
		
		result.push(OpenElement(TagType.Div, ClassType.Section));
		result.push(AddElement(TagType.Div, ClassType.SectionName, node.name));
		
		result.push(WriteNode(node.start));
		result.push(WriteBody(node));
		result.push(WriteNode(node.finish));
		
		result.push(CloseElement(TagType.Div));
		
		return result.join("");
	};
\end{lstlisting}\begin{footnotesize} \textsc{Used in}: \hyperref[Listing132]{c\-o\-d\-e\-/\-c\-o\-d\-e\-g\-e\-n\-.\-j\-s on page} \pageref{Listing132}  \end{footnotesize}\vskip 5mm\noindent

\section{Pattern Sample}
\label{sec:sample}

A pattern sample may be thought of as any segment of a knitting pattern between the cast-on and bind-off. In Purl this
construct allows pattern samples to be defined with natural number parameters. A pattern sample can be called from any
pattern body. There is no analogous concept in the standard notation.

\subsection{Sample Definition}
\label{subsec:sampledef}

In order to use a sample, it must first be defined outside of the pattern definition.

\begin{grammar}
<sampleDef> ::= `sample' <Ident> [`with' <Ident> (`,' <Ident>)*]  (<sampleBranch> | ':' <body>)
\end{grammar}

\begin{grammar}
<sampleBranch> ::= (`|' <condition> `:' <body>)+
\end{grammar}

The ``circle'' sample used by the market bag pattern is an example of a sample with a branch. For the two parameters
\texttt{n} and \texttt{max} given, rows are added to the calling pattern only if \texttt{n < max}.

\begin{purlex}[Circle Sample for Market Bag]{sampleex}
sample circle with n, max
| n < max:
    rnd : [K, YO, K n, YO, K] 4.
    rnd : *K; to end.
    circle with n + 2, max.
\end{purlex}

A pattern sample definition begins with the ``sample'' keyword.

\begin{lstlisting}[title={<Sample Def Parse Sample 106>}, label=Listing106]
if (Sym.type == KeywordSym.Sample) {
	nextSym();
} else if (Sym.type == SymType.Ident) {
	var msg = "A sample definition must start with \'" + KeywordSym.Sample + "\'.";
	AddMsg(MsgType.Warning, node, msg);
	nextSym();
} else {
	<<Unexpected Symbol Error 1>>
	var msg = "Missing \'" + KeywordSym.Sample + "\' at start of sample definition";
	AddMsg(MsgType.Error, node, msg);
	scanToSym(SymType.Ident);
}
\end{lstlisting}\begin{footnotesize} \textsc{Used in}: \hyperref[Listing109]{S\-a\-m\-p\-l\-e\- \-D\-e\-f\- \-P\-a\-r\-s\-e on page} \pageref{Listing109}  \textsc{Included Blocks}: \hyperref[Listing1]{1 on page} \pageref{Listing1}\end{footnotesize}\vskip 5mm\noindent

Next is the sample identifier. Since the sample is not part of the target language, the identifier will not be seen after
compilation. So there is no reason to consider aesthetic motivations in sample naming (as there is in pattern and section
naming). If the current symbol is not an identifier, then an error is created and the lexer scans to the next colon symbol to
continue compilation.

\begin{lstlisting}[title={<Sample Def Parse Ident 107>}, label=Listing107]
if (Sym.type == SymType.Ident) {
	node.name = Sym.value;
	nextSym();
} else if (hasOwnValue(KeywordSym, Sym.type)) {
	var msg = Sym.value + " is a reserved keyword and not a valid sample identifier.";
	AddMsg(MsgType.Error, node, msg);
	scanToSym(KeywordSym.With || CharSym.Colon);
} else {
	AddMsg(MsgType.Error, node, "Missing or invalid sample identifier.");
	scanToSym(KeywordSym.With || CharSym.Colon);
}
\end{lstlisting}\begin{footnotesize} \textsc{Used in}: \hyperref[Listing109]{S\-a\-m\-p\-l\-e\- \-D\-e\-f\- \-P\-a\-r\-s\-e on page} \pageref{Listing109}  \end{footnotesize}\vskip 5mm\noindent

If a sample definition uses parameters, the parameter list is prefixed with the keyword ``with'', followed by one or more
identifiers separated by commas. These are added to the list of parameter names used in the sample definition.

\begin{lstlisting}[title={<Sample Def Parse Params 108>}, label=Listing108]
if (Sym.type == KeywordSym.With) {
	nextSym();
	
	if (Sym.type == SymType.Ident) {
		node.paramNames.push(Sym.value);
		nextSym();
	} else {
		<<Unexpected Symbol Error 1>>
	}
	
	while (Sym.type == CharSym.Comma) {
		nextSym();
		
		if (Sym.type == SymType.Ident) {
			node.paramNames.push(Sym.value);
			nextSym();
		} else {
			<<Unexpected Symbol Error 1>>
		}
	}
}
\end{lstlisting}\begin{footnotesize} \textsc{Used in}: \hyperref[Listing109]{S\-a\-m\-p\-l\-e\- \-D\-e\-f\- \-P\-a\-r\-s\-e on page} \pageref{Listing109}  \textsc{Included Blocks}: \hyperref[Listing1]{1 on page} \pageref{Listing1}, \hyperref[Listing1]{1 on page} \pageref{Listing1}\end{footnotesize}\vskip 5mm\noindent

The sample definition node is then added to the collection of samples in the global state object of the parser so that a
sample may be used in its own definition.

A sample definition can be written to use a specific branch for the sample body if the branch condition is satisfied. For a
sample without branching, we expect a colon separator after the parameters, followed by the sample definition body, which is
parsed in the same way as simple patterns, pattern sections, and row repeats. 

\begin{lstlisting}[title={<Sample Def Parse 109>}, label=Listing109]
var SampleDefParse = function() {
	
	var node = { type : NodeType.SampleDef, 
		paramNames : [],
		children : [],
		line : State.line
	};
	
	<<Sample Def Parse Sample 106>>
	<<Sample Def Parse Ident 107>>
	<<Sample Def Parse Params 108>>
	
	State.samples[node.name] = node;
	
	if (Sym.type == CharSym.VerticalBar) {
		<<Sample Branches Parse 110>>
	} else {
		<<Colon Separator 18>>
		node.children = BodyParse();
	}
};
\end{lstlisting}\begin{footnotesize} \textsc{Used in}: \hyperref[Listing125]{A\-s\-t\- \-C\-o\-n\-s\-t\-r\-u\-c\-t\-i\-o\-n\- \-P\-a\-s\-s on page} \pageref{Listing125}  \textsc{Included Blocks}: \hyperref[Listing106]{1\-0\-6 on page} \pageref{Listing106}, \hyperref[Listing107]{1\-0\-7 on page} \pageref{Listing107}, \hyperref[Listing108]{1\-0\-8 on page} \pageref{Listing108}, \hyperref[Listing110]{1\-1\-0 on page} \pageref{Listing110}, \hyperref[Listing18]{1\-8 on page} \pageref{Listing18}\end{footnotesize}\vskip 5mm\noindent

If the sample is defined with branches, then a branch begins with `|', followed by a condition, a colon separator, then the
sample body.

\begin{lstlisting}[title={<Sample Branches Parse 110>}, label=Listing110]
while (Sym.type == CharSym.VerticalBar) {
	nextSym();
	
	var branch = { type : NodeType.Branch };
	branch.condition = ConditionParse();
	<<Colon Separator 18>>
	branch.children = BodyParse();
	
	node.children.push(branch);
}
\end{lstlisting}\begin{footnotesize} \textsc{Used in}: \hyperref[Listing109]{S\-a\-m\-p\-l\-e\- \-D\-e\-f\- \-P\-a\-r\-s\-e on page} \pageref{Listing109}  \textsc{Included Blocks}: \hyperref[Listing18]{1\-8 on page} \pageref{Listing18}\end{footnotesize}\vskip 5mm\noindent

\subsection{Sample Call}
\label{subsec:samplecall}

Once a sample has been defined, it can be called from any pattern body (simple pattern, section, row repeat, sample
definition). The sample example in the above section also includes a recursive sample call.

\begin{grammar}
<sampleCall> ::= <Ident> [`with' <expr> (`,' <expr>)+] `.'
\end{grammar}

\begin{purlex}[Use stockinette stitch sample]{samplecallex}
stockinetteStitch with 10.
\end{purlex}

A sample call begins with an identifier for the sample to be used.

\begin{lstlisting}[title={<Sample Call Parse Ident 111>}, label=Listing111]
if (Sym.type == SymType.Ident) {
	node.name = Sym.value;
	nextSym();
} else {
	<<Unexpected Symbol Error 1>>
	AddMsg(MsgType.Error, node, "Missing sample call identifier.");
	scanToSym(CharSym.Period);
}
\end{lstlisting}\begin{footnotesize} \textsc{Used in}: \hyperref[Listing113]{S\-a\-m\-p\-l\-e\- \-C\-a\-l\-l\- \-P\-a\-r\-s\-e on page} \pageref{Listing113}  \textsc{Included Blocks}: \hyperref[Listing1]{1 on page} \pageref{Listing1}\end{footnotesize}\vskip 5mm\noindent

Given the identifier, the details of the corresponding sample definition can be acquired from the global \texttt{State}
object. As in a sample definition, the parameter list is comma-separated and prefixed with the keyword ``with''. We loop
through the array of parameter names from the sample definition and construct a parameter map object which assigns the
passed expressions to parameter names. If an incorrect number of parameters is given, an error is created
and the lexer scans to the end of the sample call.

\begin{lstlisting}[title={<Sample Call Parse Params 112>}, label=Listing112]
var sampleDef = State.samples[node.name];
	
if (Sym.type == KeywordSym.With) {
	nextSym();
	
	var i;
	var required = sampleDef.paramNames.length;
	for (i = 0; i < required; i++) {
		
		if (i > 0) {
			if (Sym.type == CharSym.Comma) {
				nextSym();
			} else {
				<<Unexpected Symbol Error 1>>
			}
		}
		
		node.paramMap[sampleDef.paramNames[i]] = ExpressionParse();
	}
	 
	if (i != sampleDef.paramNames.length) {
		var msg =  node.name + "parameters required: " + required + ", passed: " + i + ".";
		AddMsg(MsgType.Error, node, msg);
		scanToSym(CharSym.Period);
	}
}
\end{lstlisting}\begin{footnotesize} \textsc{Used in}: \hyperref[Listing113]{S\-a\-m\-p\-l\-e\- \-C\-a\-l\-l\- \-P\-a\-r\-s\-e on page} \pageref{Listing113}  \textsc{Included Blocks}: \hyperref[Listing1]{1 on page} \pageref{Listing1}\end{footnotesize}\vskip 5mm\noindent

A sample call is terminated by a period symbol.

\begin{lstlisting}[title={<Sample Call Parse 113>}, label=Listing113]
var SampleCallParse = function() {
	
	var node = { type : NodeType.SampleCall, paramMap : {}, line : State.line };
	
	<<Sample Call Parse Ident 111>>
	<<Sample Call Parse Params 112>>
	<<Period Terminator 6>>
	
	return node;
};
\end{lstlisting}\begin{footnotesize} \textsc{Used in}: \hyperref[Listing125]{A\-s\-t\- \-C\-o\-n\-s\-t\-r\-u\-c\-t\-i\-o\-n\- \-P\-a\-s\-s on page} \pageref{Listing125}  \textsc{Included Blocks}: \hyperref[Listing111]{1\-1\-1 on page} \pageref{Listing111}, \hyperref[Listing112]{1\-1\-2 on page} \pageref{Listing112}, \hyperref[Listing6]{6 on page} \pageref{Listing6}\end{footnotesize}\vskip 5mm\noindent

\section{Root}

We have now looked at all available knitting pattern constructs in Purl. All that remains is to combine the top level
elements of a knitting pattern program: sample definitions and a pattern definition. The market bag pattern given in the
introduction is an example of a complete Purl program.

\subsection{AST Node}

\begin{grammar}
<program> ::= (<sampleDef>)* <pattern>
\end{grammar}

In the present implementation, all sample definitions and one pattern definition must be combined in one file, with all sample
definitions above the pattern definition. This is not ideal since Purl is meant to allow modularity and reusability of
segments of knitting patterns, but it is sufficient for the initial implementation and experimentation.

\begin{lstlisting}[title={<Program Parse 114>}, label=Listing114]
var ProgramParse = function() {
	
	var program = { type : NodeType.Root, pattern : null, line : State.line };
	
	while (Sym.type == KeywordSym.Sample) {
		SampleDefParse();
	}
	
	program.pattern = PatternParse();
	
	return program;
};
\end{lstlisting}\begin{footnotesize} \textsc{Used in}: \hyperref[Listing125]{A\-s\-t\- \-C\-o\-n\-s\-t\-r\-u\-c\-t\-i\-o\-n\- \-P\-a\-s\-s on page} \pageref{Listing125}  \end{footnotesize}\vskip 5mm\noindent

\subsection{HTML Generation}

\begin{lstlisting}[title={<Write HTML Root 115>}, label=Listing115]
var WriteRoot = function(node) {
	return WriteNode(node.pattern);
}
\end{lstlisting}\begin{footnotesize} \textsc{Used in}: \hyperref[Listing132]{c\-o\-d\-e\-/\-c\-o\-d\-e\-g\-e\-n\-.\-j\-s on page} \pageref{Listing132}  \end{footnotesize}\vskip 5mm\noindent

\chapter{Lexical Analysis}

The function \texttt{CreateLexer} returns an object that consists of a function to get the next symbol and a function to get
the current line in the code. The input is first split into an array. \texttt{CreateLexer} has a private method \texttt{next} to update
the current character, position within the array, line, and position on the line.

\begin{lstlisting}[title={<code/lexer.js 116>}, label=Listing116]
var CreateLexer = function(src){
	
	var srcArr = [];
	if (src != null) {
		srcArr = src.split("");
	}
	
	var ch = srcArr[0];
	if (srcArr.length == 0) {
		ch = SymType.EOF;
	}
	var pos = 0;
	var lineNum = 1;
	var linePos = 0;
	
	var next = function() {
		
		if (pos < srcArr.length - 1) {
			pos += 1;
			linePos += 1;
			ch = srcArr[pos];
			
			if (/[\n\r]/.test(ch)) {
				lineNum += 1;
				linePos = 0;
			}
		} else {
			ch = SymType.EOF;
		}
	};
	
	return {
		<<Lexer GetLine Function 117>>
		,
		<<Lexer GetSym Function 118>>
	};
};
\end{lstlisting}\begin{footnotesize} \textsc{Included Blocks}: \hyperref[Listing117]{1\-1\-7 on page} \pageref{Listing117}, \hyperref[Listing118]{1\-1\-8 on page} \pageref{Listing118}\end{footnotesize}\vskip 5mm\noindent

\texttt{GetLine} simply returns an object with properties for line number, line position and character position.

\begin{lstlisting}[title={<Lexer GetLine Function 117>}, label=Listing117]
GetLine : function() {
	return { num : lineNum, pos : linePos, charPos : pos };
}
\end{lstlisting}\begin{footnotesize} \textsc{Used in}: \hyperref[Listing116]{c\-o\-d\-e\-/\-l\-e\-x\-e\-r\-.\-j\-s on page} \pageref{Listing116}  \end{footnotesize}\vskip 5mm\noindent

\texttt{GetSym} determines the current token by the value of the current character, skipping any whitespace characters.

\begin{lstlisting}[title={<Lexer GetSym Function 118>}, label=Listing118]
GetSym : function() {
	
	while ((/\s/g).test(ch)) {
		next();
	}
	
	if (ch == SymType.EOF) {
		<<Lex EOF 119>>
	} else if (ch == SymType.String) {
		<<Lex String 120>>
	} else if (/[0-9]/.test(ch)) {
		<<Lex Num 121>>
	} else if (/[a-zA-Z]/.test(ch)) {
		<<Lex Alphanum 122>>
	} else {
		<<Lex Char 123>>
	}
	return { type : SymType.Unknown, value : SymType.Unknown };
}
\end{lstlisting}\begin{footnotesize} \textsc{Used in}: \hyperref[Listing116]{c\-o\-d\-e\-/\-l\-e\-x\-e\-r\-.\-j\-s on page} \pageref{Listing116}  \textsc{Included Blocks}: \hyperref[Listing119]{1\-1\-9 on page} \pageref{Listing119}, \hyperref[Listing120]{1\-2\-0 on page} \pageref{Listing120}, \hyperref[Listing121]{1\-2\-1 on page} \pageref{Listing121}, \hyperref[Listing122]{1\-2\-2 on page} \pageref{Listing122}, \hyperref[Listing123]{1\-2\-3 on page} \pageref{Listing123}\end{footnotesize}\vskip 5mm\noindent

If the lexer reaches the end of the input, an end of file token is returned, and the character will no longer be updated.

\begin{lstlisting}[title={<Lex EOF 119>}, label=Listing119]
return { type : SymType.EOF, value : SymType.EOF };
\end{lstlisting}\begin{footnotesize} \textsc{Used in}: \hyperref[Listing118]{L\-e\-x\-e\-r\- \-G\-e\-t\-S\-y\-m\- \-F\-u\-n\-c\-t\-i\-o\-n on page} \pageref{Listing118}  \end{footnotesize}\vskip 5mm\noindent

If the current character is a double quote, then the symbol type is \emph{title}, and everything until the next double quote
is the value.

\begin{lstlisting}[title={<Lex String 120>}, label=Listing120]
next();
var str = "";

while (ch != SymType.String && ch != SymType.EOF) {
	str = str + ch;
	next();
}

if (ch == SymType.String) {
	next();
}

return { type : SymType.String, value : str };
\end{lstlisting}\begin{footnotesize} \textsc{Used in}: \hyperref[Listing118]{L\-e\-x\-e\-r\- \-G\-e\-t\-S\-y\-m\- \-F\-u\-n\-c\-t\-i\-o\-n on page} \pageref{Listing118}  \end{footnotesize}\vskip 5mm\noindent

If the current character is a number, then the symbol type is a natural number and the symbol value is the concatenation of
all characters until a non-numeric character.

\begin{lstlisting}[title={<Lex Num 121>}, label=Listing121]
var num = "";

while (/^[0-9]$/.test(ch)) {
	num = num + ch;
	next();
}

return { type : SymType.Nat, value : num };
\end{lstlisting}\begin{footnotesize} \textsc{Used in}: \hyperref[Listing118]{L\-e\-x\-e\-r\- \-G\-e\-t\-S\-y\-m\- \-F\-u\-n\-c\-t\-i\-o\-n on page} \pageref{Listing118}  \end{footnotesize}\vskip 5mm\noindent

If the current character is a letter, concatenate all characters until the next non-alphanumeric character. If this string
matches a reserved \hyperref[sec:symtype]{keyword}, then return a symbol representing that keyword. If the string matches a
regular expression in the \hyperref[sec:symtype]{stitch lookup}, then return a symbol representing that stitch. Otherwise,
return an ident symbol.

\begin{lstlisting}[title={<Lex Alphanum 122>}, label=Listing122]
var id = "";
		
while (/^[a-zA-Z0-9]$/.test(ch)) {
	id = id + ch;
	next();
}

for (var idSym in KeywordSym) {
	if (id == KeywordSym[idSym]) {
		return { type : id, value : id };
	}
}

for (var stSym in StitchSym)	{
	
	if ((StitchSym[stSym]).test(id)) {
		return { type : StitchSym[stSym], value : id };
	}
}

return { type : SymType.Ident, value : id };
\end{lstlisting}\begin{footnotesize} \textsc{Used in}: \hyperref[Listing118]{L\-e\-x\-e\-r\- \-G\-e\-t\-S\-y\-m\- \-F\-u\-n\-c\-t\-i\-o\-n on page} \pageref{Listing118}  \end{footnotesize}\vskip 5mm\noindent

If the current symbol is in the \hyperref[sec:symtype]{char symbol lookup}, then if it is an asterisk, first check if the
next character is also an asterisk. If a double asterisk, then we have a \hyperref[sec:rowrep]{row repeat} symbol, otherwise
return the initial character symbol type and value.

\begin{lstlisting}[title={<Lex Char 123>}, label=Listing123]
for (var chSym in CharSym) {

	if (ch == CharSym[chSym]) {
		
		var result = ch;
		next();
		
		if (result == CharSym.Asterisk && ch == CharSym.Asterisk) {
			result = SymType.RowRep;
			next();
		} else if (result == CharSym.OpenAngle && ch == CharSym.Equal) {
			result = SymType.LessEq;
			next();
		} else if (result == CharSym.CloseAngle && ch == CharSym.Equal) {
			result = SymType.GreaterEq;
			next();
		}
		
		return { type : result, value : result };
	}
}
\end{lstlisting}\begin{footnotesize} \textsc{Used in}: \hyperref[Listing118]{L\-e\-x\-e\-r\- \-G\-e\-t\-S\-y\-m\- \-F\-u\-n\-c\-t\-i\-o\-n on page} \pageref{Listing118}  \end{footnotesize}\vskip 5mm\noindent

\section{Parser}

\begin{lstlisting}[title={<code/parser.js 124>}, label=Listing124]
var Parser = (function() {
	
	<<Ast Construction Pass 125>>
	<<Sample Substitution Pass 126>>
	<<Verification Pass 131>>
	
	var AddMsg = function(msgType, node, msgStr) {
		var msgObj = {
			messageType : msgType,
			sectionName : State.sectionName,
			line : State.line,
			rowIndex : State.rowIndex,
			message : msgStr
		};
		State.messages.push(msgObj);
		
		switch (msgType) {
			case MsgType.Error:
				node.hasErrorMsg = true;
				break;
			case MsgType.Warning:
				node.hasWarningMsg = true;
				break;
			case MsgType.Verification:
				node.hasVerificationMsg = true;
				break;
			default:
				break;
		}
	};
	
	var State = {};
	
	return {
		Parse : function(input){
			State = { sectionName : null, samples : {}, messages : [] };
			
			var ast = AstConstructionPass(input);
			console.clear();
			console.log("PASS 1:------------------------------- \n" + JSON.stringify(ast, undefined, 2));
			console.log("STATE:------------------------------- \n" + JSON.stringify(State, undefined, 2));
			SampleSubstitutionPass(ast);
			console.log("PASS 2:------------------------------- \n" + JSON.stringify(ast, undefined, 2));
			console.log("STATE:------------------------------- \n" + JSON.stringify(State, undefined, 2));
			VerificationPass(ast);
			console.log("PASS 3:------------------------------- \n" + JSON.stringify(ast, undefined, 2));
			console.log("STATE:------------------------------- \n" + JSON.stringify(State, undefined, 2));
			ast.messages = State.messages;
			State = {};
			
			return ast;
		}
	};
}());
\end{lstlisting}\begin{footnotesize} \textsc{Included Blocks}: \hyperref[Listing125]{1\-2\-5 on page} \pageref{Listing125}, \hyperref[Listing126]{1\-2\-6 on page} \pageref{Listing126}, \hyperref[Listing131]{1\-3\-1 on page} \pageref{Listing131}\end{footnotesize}\vskip 5mm\noindent

\subsection{Pass 1: AST Construction}

The first pass constructs a syntax tree representing the source pattern. See~\ref{chap:knitelements} for details on parsing
specific pattern elements.

\begin{lstlisting}[title={<Ast Construction Pass 125>}, label=Listing125]
var AstConstructionPass = function(input) {
	
	var nextSym = function() {
		Sym = Lexer.GetSym();
		if (Sym.type == SymType.EOF) {
			State.line = null;
		} else {
			State.line = Lexer.GetLine();
		}
	};
	
	var scanToSym = function(symType) {
		while (Sym.type != symType && Sym.type != SymType.EOF) {
			nextSym();
		}
	};
	
	var hasOwnValue = function(obj, val) {
		for (var prop in obj) {
			if (obj.hasOwnProperty(prop) && obj[prop] === val) {
				return true;
			}
		}
		return false;
	};
	
	var getNatSym = function(terminatorSym) {
		var result = {};
		
		if (Sym.type == SymType.Nat) {
			result = { type : NodeType.NatLiteral, value : Sym.value };
			nextSym();
		} else if (Sym.type == SymType.Ident) {
			result = { type : NodeType.NatVariable, value : Sym.value };
			nextSym();
		} else {
			<<Unexpected Symbol Error 1>>
			scanToSym(terminatorSym);
		}
		
		return result;
	};
	
	<<Program Parse 114>>
	<<Pattern Parse 90>>
	<<Cast-On Parse 2>>
	<<Pick-Up Parse 9>>
	<<Body Parse 93>>
	<<Row Definition Parse 15>>
	<<Row Element Parse 27>>
	<<Stitch Op Parse 28>>
	<<Undetermined Stitch Repeat Parse 72>>
	<<Fixed Stitch Repeat Parse 64>>
	<<Compound Stitch Parse 58>>
	<<Basic Stitch Parse 32>>
	<<Bind-Off Parse 77>>
	<<Join Parse 83>>
	<<Row Repeat Parse 97>>
	<<Section Parse 103>>
	<<Sample Def Parse 109>>
	<<Sample Call Parse 113>>
	<<Expression Parse 137>>
	<<Condition Parse 140>>
	
	var Lexer = CreateLexer(input);
	var Sym;
	
	nextSym();
	
	if (Sym.type == SymType.EOF) {
		AddMsg(MsgType.Warning, {}, "No pattern to compile :(");
		return {};
	} else {
		return ProgramParse();
	}
};
\end{lstlisting}\begin{footnotesize} \textsc{Used in}: \hyperref[Listing124]{c\-o\-d\-e\-/\-p\-a\-r\-s\-e\-r\-.\-j\-s on page} \pageref{Listing124}  \textsc{Included Blocks}: \hyperref[Listing1]{1 on page} \pageref{Listing1}, \hyperref[Listing114]{1\-1\-4 on page} \pageref{Listing114}, \hyperref[Listing90]{9\-0 on page} \pageref{Listing90}, \hyperref[Listing2]{2 on page} \pageref{Listing2}, \hyperref[Listing9]{9 on page} \pageref{Listing9}, \hyperref[Listing93]{9\-3 on page} \pageref{Listing93}, \hyperref[Listing15]{1\-5 on page} \pageref{Listing15}, \hyperref[Listing27]{2\-7 on page} \pageref{Listing27}, \hyperref[Listing28]{2\-8 on page} \pageref{Listing28}, \hyperref[Listing72]{7\-2 on page} \pageref{Listing72}, \hyperref[Listing64]{6\-4 on page} \pageref{Listing64}, \hyperref[Listing58]{5\-8 on page} \pageref{Listing58}, \hyperref[Listing32]{3\-2 on page} \pageref{Listing32}, \hyperref[Listing77]{7\-7 on page} \pageref{Listing77}, \hyperref[Listing83]{8\-3 on page} \pageref{Listing83}, \hyperref[Listing97]{9\-7 on page} \pageref{Listing97}, \hyperref[Listing103]{1\-0\-3 on page} \pageref{Listing103}, \hyperref[Listing109]{1\-0\-9 on page} \pageref{Listing109}, \hyperref[Listing113]{1\-1\-3 on page} \pageref{Listing113}, \hyperref[Listing137]{1\-3\-7 on page} \pageref{Listing137}, \hyperref[Listing140]{1\-4\-0 on page} \pageref{Listing140}\end{footnotesize}\vskip 5mm\noindent

\subsection{Pass 2: Sample Substitution}

The second pass is responsible for replacing all sample call nodes with the children of the corresponding sample definition,
and updating parameter values according to the \texttt{paramMap} object of each sample call. The AST is traversed depth
first in this pass and the function \texttt{SampleSubstitutionPass} starts by traversing the children of the pattern node
with an empty object as the \texttt{paramMap}.

\begin{lstlisting}[title={<Sample Substitution Pass 126>}, label=Listing126]
var SampleSubstitutionPass = function(ast) {
	
	<<Pass 2 Traverse Children 127>>
	<<Pass 2 Sample Call Children 129>>
	<<Pass 2 Update Expressions 130>>
	<<Pass 2 Update Condition 141>>
	
	var paramMap = {};
	
	if (ast.type == NodeType.Root) {
		if (ast.pattern != null) {
			TraverseChildren(ast.pattern, paramMap);
		}
	}
};
\end{lstlisting}\begin{footnotesize} \textsc{Used in}: \hyperref[Listing124]{c\-o\-d\-e\-/\-p\-a\-r\-s\-e\-r\-.\-j\-s on page} \pageref{Listing124}  \textsc{Included Blocks}: \hyperref[Listing127]{1\-2\-7 on page} \pageref{Listing127}, \hyperref[Listing129]{1\-2\-9 on page} \pageref{Listing129}, \hyperref[Listing130]{1\-3\-0 on page} \pageref{Listing130}, \hyperref[Listing141]{1\-4\-1 on page} \pageref{Listing141}\end{footnotesize}\vskip 5mm\noindent

All row and stitch constructs require any expressions to have variables updated, and for nodes with children, the child
nodes are then traversed.

\begin{lstlisting}[title={<Pass 2 Traverse Children 127>}, label=Listing127]
var TraverseChildren = function(node, paramMap) {
	if (node.children != null) {
		for (var i = 0; i < node.children.length; i++) {
			
			var child = node.children[i];
			
			switch (child.type) {
				
				case NodeType.Section:
					TraverseChildren(child, paramMap);
					break;
				
				case NodeType.SampleCall:
					<<Pass 2 Replace Sample Call 128>>
					break;
				
				case NodeType.Branch:
					UpdateCondition(child.condition, paramMap);
					if (child.condition.doBranch) {
						TraverseChildren(child, paramMap);
						node.children = child.children;
						return;
					}
					break;
				
				case NodeType.Row:
				case NodeType.RowRep:
				case NodeType.FixedStRep:
				case NodeType.UStRep:
				case NodeType.CompSt:
					UpdateExpressions(child, paramMap);
					TraverseChildren(child, paramMap);
					break;
				
				case NodeType.Knit:
				case NodeType.Purl:
				case NodeType.KnitTBL:
				case NodeType.PurlTBL:
				case NodeType.KnitBelow:
				case NodeType.PurlBelow:
				case NodeType.Slip:
				case NodeType.SlipKW:
				case NodeType.SlipPW:
				case NodeType.YarnOver:
				case NodeType.KnitFB:
				case NodeType.PurlFB:
				case NodeType.Make:
				case NodeType.MakeL:
				case NodeType.MakeR:
				case NodeType.KnitTog:
				case NodeType.PurlTog:
				case NodeType.SSK:
				case NodeType.SSP:
				case NodeType.PSSO:
					UpdateExpressions(child, paramMap);
					break;
				
				case NodeType.NatVariable:
					var first = node.children.slice(0, i);
					var last = node.children.slice(i + 1);
					var exprChildren = paramMap[child.value].children;
					node.children = first.concat(exprChildren.concat(last));
					i += exprChildren.length - 1;
					break;
				
				default:
					break;
			}
		}
	}
};
\end{lstlisting}\begin{footnotesize} \textsc{Used in}: \hyperref[Listing126]{S\-a\-m\-p\-l\-e\- \-S\-u\-b\-s\-t\-i\-t\-u\-t\-i\-o\-n\- \-P\-a\-s\-s on page} \pageref{Listing126}  \textsc{Included Blocks}: \hyperref[Listing128]{1\-2\-8 on page} \pageref{Listing128}\end{footnotesize}\vskip 5mm\noindent

When traversing the children of a node, if we find a \hyperref[subsec:samplecall]{sample call} then we need to remove the sample call node from the tree.

\begin{lstlisting}[title={<Pass 2 Replace Sample Call 128>}, label=Listing128]
var first = node.children.slice(0, i);
var last = node.children.slice(i + 1);

var sampleChildren = GetSampleCallChildren(child, paramMap);

node.children = first.concat(sampleChildren.concat(last));

i += sampleChildren.length - 1
\end{lstlisting}\begin{footnotesize} \textsc{Used in}: \hyperref[Listing127]{P\-a\-s\-s\- \-2\- \-T\-r\-a\-v\-e\-r\-s\-e\- \-C\-h\-i\-l\-d\-r\-e\-n on page} \pageref{Listing127}  \end{footnotesize}\vskip 5mm\noindent

However, before the sample call node can be replaced we must acquire a deep copy of the children of the corresponding sample
definition. These children are set as the children of the sample call node, which we then traverse to update their
expressions and sample calls. To avoid naming conflicts, a \texttt{localMap} object is created as a deep copy of the parent
node's \texttt{paramMap}. The \texttt{paramMap} object of the current node must have its expressions updated according to
the local map. The \texttt{localMap} is then updated with the current node's updated \texttt{paramMap} and passed on when
traversing the children of the sample call.

\begin{lstlisting}[title={<Pass 2 Sample Call Children 129>}, label=Listing129]
var GetSampleCallChildren = function(node, paramMap) {
		
	var localMap = jQuery.extend(true, {}, paramMap);
	
	for (var domainVal in node.paramMap) {
		TraverseChildren(node.paramMap[domainVal], localMap);
		localMap[domainVal] = node.paramMap[domainVal];
	}
	
	var sampleDef = jQuery.extend(true, {}, State.samples[node.name]);
	node.children = sampleDef.children;
	
	TraverseChildren(node, localMap);
	
	return node.children;
};
\end{lstlisting}\begin{footnotesize} \textsc{Used in}: \hyperref[Listing126]{S\-a\-m\-p\-l\-e\- \-S\-u\-b\-s\-t\-i\-t\-u\-t\-i\-o\-n\- \-P\-a\-s\-s on page} \pageref{Listing126}  \end{footnotesize}\vskip 5mm\noindent

There are only two properties that currently allow natural number expressions: \texttt{repCount} and \texttt{num}. If a node
has either of these properties, then the expression is updated according to the passed \texttt{paramMap}.

\begin{lstlisting}[title={<Pass 2 Update Expressions 130>}, label=Listing130]
var UpdateExpressions = function(node, paramMap) {
	if (node.repCount != null && node.repCount.type == NodeType.Expression) {
		TraverseChildren(node.repCount, paramMap);
	}
	if (node.num != null && node.num.type == NodeType.Variable) {
		TraverseChildren(node.num, paramMap);
	}
};
\end{lstlisting}\begin{footnotesize} \textsc{Used in}: \hyperref[Listing126]{S\-a\-m\-p\-l\-e\- \-S\-u\-b\-s\-t\-i\-t\-u\-t\-i\-o\-n\- \-P\-a\-s\-s on page} \pageref{Listing126}  \end{footnotesize}\vskip 5mm\noindent

\subsection{Pass 3: Pattern Verification}

The third pass traverses the syntax tree and determines if there are structural issues with the pattern.
See~\ref{chap:knitelements} for details on verification of specific pattern elements.

\begin{lstlisting}[title={<Verification Pass 131>}, label=Listing131]
var VerificationPass = function(ast) {
	
	<<Verify Pattern 91>>
	<<Verify Section 104>>
	<<Verify Cast-On 7>>
	<<Verify Pick-Up 13>>
	<<Verify Bind-Off 78>>
	<<Verify Join 84>>
	<<Verify Row 25>>
	<<Verify Row Elem Children of Node 22>>
	<<Verify Row Elem 29>>
	<<Verify Expression 138>>
	
	var VerifyChildren = function(node) {
		
		if (node.children != null) {
			for (var i = 0; i < node.children.length; i++) {
				VerifyNode(node.children[i]);
			}
		}
	};
	
	var VerifyNode = function(node) {
		
		State.line = node.line;
		
		switch (node.type) {
				
				case NodeType.Root:
					VerifyNode(node.pattern);
					break;
				case NodeType.Pattern:
					VerifyPattern(node);
					break;
				case NodeType.Section:
					VerifySection(node);
					break;
				case NodeType.CastOn:
					VerifyCastOn(node);
					break;
				case NodeType.PickUp:
					VerifyPickUp(node);
					break;
				case NodeType.BindOff:
					VerifyBindOff(node);
					break;
				case NodeType.Join:
					VerifyJoin(node);
					break;
				case NodeType.Row:
					VerifyRow(node);
					break;
				case NodeType.RowRep:
					VerifyExpression(node.repCount);
					VerifyChildren(node);
					break;
				case NodeType.Expression:
					VerifyExpression(node);
					break;
				default:
					break;
			}
	};
	
	VerifyNode(ast);
};
\end{lstlisting}\begin{footnotesize} \textsc{Used in}: \hyperref[Listing124]{c\-o\-d\-e\-/\-p\-a\-r\-s\-e\-r\-.\-j\-s on page} \pageref{Listing124}  \textsc{Included Blocks}: \hyperref[Listing91]{9\-1 on page} \pageref{Listing91}, \hyperref[Listing104]{1\-0\-4 on page} \pageref{Listing104}, \hyperref[Listing7]{7 on page} \pageref{Listing7}, \hyperref[Listing13]{1\-3 on page} \pageref{Listing13}, \hyperref[Listing78]{7\-8 on page} \pageref{Listing78}, \hyperref[Listing84]{8\-4 on page} \pageref{Listing84}, \hyperref[Listing25]{2\-5 on page} \pageref{Listing25}, \hyperref[Listing22]{2\-2 on page} \pageref{Listing22}, \hyperref[Listing29]{2\-9 on page} \pageref{Listing29}, \hyperref[Listing138]{1\-3\-8 on page} \pageref{Listing138}\end{footnotesize}\vskip 5mm\noindent

\chapter{Code Generation}

\section{HTML}

The back end for the Purl compiler generates HTML code representing the pattern. Given an abstract syntax tree,
\texttt{ast}, representing a knitting pattern, \texttt{PatternTextWriterHTML.Generate(ast)} returns HTML to display the
knitting pattern according to the knitting pattern standard~\cite{cycstandards}. Each pattern element is usually contained
within a div with a corresponding class name. There are also tags added for pattern and section names. The order in which
elements of a pattern are written is the same as has been seen for all passes of the compiler. See~\ref{chap:knitelements}
for details on code generation of specific pattern elements.

\begin{lstlisting}[title={<code/codegen.js 132>}, label=Listing132]
var PatternTextWriterHTML = (function() {
	
	<<Code Gen Types 133>>
	<<Code Gen Tag Writing Functions 134>>
	<<Write HTML Node 136>>
	<<Write HTML Root 115>>
	<<Write HTML Pattern 92>>
	<<Write HTML Section 105>>
	<<Write HTML Cast-On 8>>
	<<Write HTML Pick-Up 14>>
	<<Write HTML Bind-Off 79>>
	<<Write HTML Join 85>>
	<<Write HTML Body 94>>
	<<Write HTML Row 26>>
	<<Write HTML Basic Stitch 54>>
	<<Write HTML Undetermined Stitch Repeat 74>>
	<<Write HTML Fixed Stitch Repeat 66>>
	<<Write HTML Compound Stitch 60>>
	<<Write HTML Row Repeat 98>>	
	
	return {
		Generate : function(ast) {
			return WriteRoot(ast);
		}
	}
})();
\end{lstlisting}\begin{footnotesize} \textsc{Included Blocks}: \hyperref[Listing133]{1\-3\-3 on page} \pageref{Listing133}, \hyperref[Listing134]{1\-3\-4 on page} \pageref{Listing134}, \hyperref[Listing136]{1\-3\-6 on page} \pageref{Listing136}, \hyperref[Listing115]{1\-1\-5 on page} \pageref{Listing115}, \hyperref[Listing92]{9\-2 on page} \pageref{Listing92}, \hyperref[Listing105]{1\-0\-5 on page} \pageref{Listing105}, \hyperref[Listing8]{8 on page} \pageref{Listing8}, \hyperref[Listing14]{1\-4 on page} \pageref{Listing14}, \hyperref[Listing79]{7\-9 on page} \pageref{Listing79}, \hyperref[Listing85]{8\-5 on page} \pageref{Listing85}, \hyperref[Listing94]{9\-4 on page} \pageref{Listing94}, \hyperref[Listing26]{2\-6 on page} \pageref{Listing26}, \hyperref[Listing54]{5\-4 on page} \pageref{Listing54}, \hyperref[Listing74]{7\-4 on page} \pageref{Listing74}, \hyperref[Listing66]{6\-6 on page} \pageref{Listing66}, \hyperref[Listing60]{6\-0 on page} \pageref{Listing60}, \hyperref[Listing98]{9\-8 on page} \pageref{Listing98}\end{footnotesize}\vskip 5mm\noindent

Only div and span tag types are used, and a class type is associated with each pattern node in the output language (used for
styling).

\begin{lstlisting}[title={<Code Gen Types 133>}, label=Listing133]
var ClassType = {
	Pattern	 : "pattern",
	PatternName : "patternname",
	PatternNote : "patternnote",
	SectionName : "sectionname",
	SectionNote : "sectionnote",
	CastOn : "caston",
	CastOnNote : "castonnote",
	PickUp : "pickup",
	Body : "body",
	Row : "row",
	RowNote : "rownote",
	RowRep : "rowrepeat",
	StitchCount : "stitchcount",
	Stitch : "stitch",
	BindOff : "bindoff",
	BindOffNote : "bindoffnote",
	Join : "join",
	Error : "error",
	Warning : "warning",
	Verification : "verification"
}

var TagType = {
	Div : "div",
	Span : "span"
}
\end{lstlisting}\begin{footnotesize} \textsc{Used in}: \hyperref[Listing132]{c\-o\-d\-e\-/\-c\-o\-d\-e\-g\-e\-n\-.\-j\-s on page} \pageref{Listing132}  \end{footnotesize}\vskip 5mm\noindent

The following functions are used by the code generation module to create the HTML tags.

\begin{lstlisting}[title={<Code Gen Tag Writing Functions 134>}, label=Listing134]
var AddElement = function(tag, classType, text) {
	return OpenElement(tag, classType) + text + CloseElement(tag);
}

var OpenElement = function(tag, classType) {
	return "<" + tag + " class=\"" + classType + "\">";
}

var CloseElement = function(tag) {
	return "</" + tag + ">";
}
\end{lstlisting}\begin{footnotesize} \textsc{Used in}: \hyperref[Listing132]{c\-o\-d\-e\-/\-c\-o\-d\-e\-g\-e\-n\-.\-j\-s on page} \pageref{Listing132}  \end{footnotesize}\vskip 5mm\noindent

The function \texttt{WriteNode} first checks if a node has any messages, and if so, an error tag is added to the result.

\begin{lstlisting}[title={<Mark Node Message 135>}, label=Listing135]
if (node.hasErrorMsg) {
	result = AddElement(TagType.Span, ClassType.Error, "!");
}

if (node.hasWarningMsg) {
	result = AddElement(TagType.Span, ClassType.Warning, "!");
}

if (node.hasVerificationMsg) {
	result = AddElement(TagType.Span, ClassType.Verification, "!");
}
\end{lstlisting}\begin{footnotesize} \textsc{Used in}: \hyperref[Listing136]{W\-r\-i\-t\-e\- \-H\-T\-M\-L\- \-N\-o\-d\-e on page} \pageref{Listing136}  \end{footnotesize}\vskip 5mm\noindent

The appropriate node writing function is then called based on the node type.

\begin{lstlisting}[title={<Write HTML Node 136>}, label=Listing136]
var WriteNode = function(node) {
	
	var result = "";
	
	if (node == null) {
		return result;
	}
	
	<<Mark Node Message 135>>
	
	switch (node.type) {
		
		case NodeType.Root: 
			result += WriteRoot(node);
			break;
		
		case NodeType.Pattern:
			result += WritePattern(node);
			break;
		
		case NodeType.Section:
			result += WriteSection(node);
			break;
		
		case NodeType.CastOn:
			result += WriteCo(node);
			break;
		
		case NodeType.PickUp:
			result += WritePu(node);
			break;
		
		case NodeType.BindOff:
			result += WriteBo(node);
			break;
		
		case NodeType.Join:
			result += WriteJoin(node);
			break;
		
		case NodeType.Row:
			result += WriteRow(node);
			break;
			
		case NodeType.RowRep:
			result += WriteRowRep(node);
			break;
			
		case NodeType.FixedStRep:
			result += WriteFixedStRep(node);
			break;
		
		case NodeType.UStRep:
			result += WriteUStRep(node);
			break;
		
		case NodeType.CompSt:
			result += WriteCompSt(node);
			break;
		
		case NodeType.Knit:
		case NodeType.Purl:
		case NodeType.KnitTBL:
		case NodeType.PurlTBL:
		case NodeType.KnitBelow:
		case NodeType.PurlBelow:
		case NodeType.Slip:
		case NodeType.SlipKW:
		case NodeType.SlipPW:
		case NodeType.YarnOver:
		case NodeType.KnitFB:
		case NodeType.PurlFB:
		case NodeType.Make:
		case NodeType.MakeL:
		case NodeType.MakeR:
		case NodeType.KnitTog:
		case NodeType.PurlTog:
		case NodeType.SSK:
		case NodeType.SSP:
		case NodeType.PSSO:
			result += WriteBasicStitch(node);
			break;
			
		default:
			break;
	}
	
	return result;
};
\end{lstlisting}\begin{footnotesize} \textsc{Used in}: \hyperref[Listing132]{c\-o\-d\-e\-/\-c\-o\-d\-e\-g\-e\-n\-.\-j\-s on page} \pageref{Listing132}  \textsc{Included Blocks}: \hyperref[Listing135]{1\-3\-5 on page} \pageref{Listing135}\end{footnotesize}\vskip 5mm\noindent

\chapter{Discussion}

The language Purl provides a format for reuse and efficient storage of knitting patterns, with the compiler generating an
assembled and formatted pattern for use by a knitter. In its present state, this project is a good foundation for writing
knitting patterns, but many desirable features are still missing. Some planned components were removed from this initial
implementation due to time constraints and a desire to give each component the time and thought deserved for a proper
implementation. Below are listed the postponed features and next steps of this project, in approximate order of priority.

\begin{description}
\item[Preconditions/Postconditions] \hfill \\ User defined requirements are a natural component for the knitting pattern
    samples construct. This concept requires more research and planning than initially anticipated. This is the top priority
    next large feature. 
\item[Test Page] The test page provided gives an idea of how the language works, but for practical purposes, the following will be necessary:
\begin{itemize}
\item Load pattern files for compilation
\item Allow pattern compilation over multiple files (pattern samples would then be reusable in multiple patterns)
\item Save feature to download target pattern in pdf format
\end{itemize}
\item[Charts] \hfill \\ Originally the compiler was planned to have code generation to patterns in both the standard text
    notation and the chart notation~\cite{cycstandards}. Charts are written and read in a style that is very different from
    text patterns, with column repeats as well as row repeats, and alternating rows requiring the visual reverse of the
    written stitch to be performed. A large amount of optimization would have been required to make a pattern that is
    ``concise'' in Purl be output in a chart that is a reasonable size.
\item[Small Updates] In a productive pattern design setting, the following would be desirable:
\begin{itemize}
\item Display of pattern attributes such as designer name, date, recommended yarn, and needles
\item Commenting in the pattern source file
\item Notes on the target pattern
\end{itemize}
\end{description}

\begin{appendices}

\chapter{Extra Productions}

\section{Expressions}

Expressions currently only allow for addition of natural number literals and variables.

\begin{grammar}
<expr> ::= (<Ident> | <Nat>) ( `+' (<Ident> | <Nat>) )*
\end{grammar}

An expression node contains children which are objects of type \texttt{NodeType.NatLiteral} or \texttt{NodeType.NatVariable}.

\begin{lstlisting}[title={<Expression Parse 137>}, label=Listing137]
var ExpressionParse = function(terminatorSym) {
	
	var node = { type : NodeType.Expression, children : [] };
	
	if (Sym.type == SymType.Nat || Sym.type == SymType.Ident) {
		
		node.children.push(getNatSym(terminatorSym));
		
		while (Sym.type == CharSym.PlusOp) {
			nextSym();
			node.children.push(getNatSym(terminatorSym));
		}
	}
	
	return node;
};
\end{lstlisting}\begin{footnotesize} \textsc{Used in}: \hyperref[Listing125]{A\-s\-t\- \-C\-o\-n\-s\-t\-r\-u\-c\-t\-i\-o\-n\- \-P\-a\-s\-s on page} \pageref{Listing125}  \end{footnotesize}\vskip 5mm\noindent

An expression is verified by adding the values of the children (all of which should be natural number literals at the
verification pass), and setting the value of the expression node.

\begin{lstlisting}[title={<Verify Expression 138>}, label=Listing138]
var VerifyExpression = function(node) {
	
	node.value = 0;
	
	if (node.children != null) {
		for (var i = 0; i < node.children.length; i++) {
			node.value += parseInt(node.children[i].value, 10);
		}
	}
};
\end{lstlisting}\begin{footnotesize} \textsc{Used in}: \hyperref[Listing131]{V\-e\-r\-i\-f\-i\-c\-a\-t\-i\-o\-n\- \-P\-a\-s\-s on page} \pageref{Listing131}  \end{footnotesize}\vskip 5mm\noindent

\begin{lstlisting}[title={<Write Expression 139>}, label=Listing139]
var WriteExpression = function(natObj) {
	
	var result = [];
	var resultVal = 0;
	
	if (natObj.children != null) {
		for (var i = 0; i < natObj.children.length; i++) {
			var child = natObj.children[i];
			if (child.type == NodeType.NatVar) {
				resultVal += child.value;
			} else if (child.type == NodeType.NatLit) {
				result.push(child.value);
			}
		}
	}
	
	result.push(resultVal);
	
	return result.join("+");
}
\end{lstlisting}

\section{Condition}

\begin{grammar}
<condition> ::= <expr> ( `=' | `<' | `<=' | `>' | `>=' ) <expr>
\end{grammar}

A condition node includes a property for the comparison operator and a node the the left and a node for the right of the
operator.

\begin{lstlisting}[title={<Condition Parse 140>}, label=Listing140]
var ConditionParse = function() {
	
	var node = { type : NodeType.Condition };
	
	if (Sym.type == SymType.Nat || Sym.type == SymType.Ident) {
		node.nodeL = ExpressionParse();
	}
	
	switch (Sym.type) {
		case CharSym.Equal:
			node.comparison = CompareType.Equal;
			nextSym();
			break;
		case CharSym.OpenAngle:
			node.comparison = CompareType.Less;
			nextSym();
			break;
		case CharSym.CloseAngle:
			node.comparison = CompareType.Greater;
			nextSym();
			break;
		case SymType.LessEq:
			node.comparison = CompareType.LessEq;
			nextSym();
			break;
		case SymType.GreaterEq:
			node.comparison = CompareType.GreaterEq;
			nextSym();
			break;
		default:
			break;
	}
	
	if (Sym.type == SymType.Nat || Sym.type == SymType.Ident) {
		node.nodeR = ExpressionParse();
	}
	
	return node;
};
\end{lstlisting}\begin{footnotesize} \textsc{Used in}: \hyperref[Listing125]{A\-s\-t\- \-C\-o\-n\-s\-t\-r\-u\-c\-t\-i\-o\-n\- \-P\-a\-s\-s on page} \pageref{Listing125}  \end{footnotesize}\vskip 5mm\noindent

\begin{lstlisting}[title={<Pass 2 Update Condition 141>}, label=Listing141]
var UpdateCondition = function(node, paramMap) {
	
	TraverseChildren(node.nodeL, paramMap);
	TraverseChildren(node.nodeR, paramMap);
	
	node.nodeL.value = 0;
	if (node.nodeL.children != null) {
		for (var i = 0; i < node.nodeL.children.length; i++) {
			node.nodeL.value += parseInt(node.nodeL.children[i].value);
		}
	}
	
	node.nodeR.value = 0;
	if (node.nodeR.children != null) {
		for (var i = 0; i < node.nodeR.children.length; i++) {
			node.nodeR.value += parseInt(node.nodeR.children[i].value);
		}
	}
	
	switch (node.comparison) {
		case CompareType.Equal:
			node.doBranch = (node.nodeL.value == node.nodeR.value);
			break;
		case CompareType.Less:
			node.doBranch = (node.nodeL.value < node.nodeR.value);
			break;
		case CompareType.LessEq:
			node.doBranch = (node.nodeL.value <= node.nodeR.value);
			break;
		case CompareType.Greater:
			node.doBranch = (node.nodeL.value > node.nodeR.value);
			break;
		case CompareType.GreaterEq:
			node.doBranch = (node.nodeL.value >= node.nodeR.value);
			break;
		default:
			node.doBranch = false;
			break;
	}
};
\end{lstlisting}\begin{footnotesize} \textsc{Used in}: \hyperref[Listing126]{S\-a\-m\-p\-l\-e\- \-S\-u\-b\-s\-t\-i\-t\-u\-t\-i\-o\-n\- \-P\-a\-s\-s on page} \pageref{Listing126}  \end{footnotesize}\vskip 5mm\noindent

\chapter{Compiler Types}

Below are types used throughout the compiler.

\section{Parser Types}
\label{sec:parsertype}

\begin{lstlisting}[title={<code/util.js 142>}, label=Listing142]
var SideType = {
    RS  : "RS",
    WS : "WS"
};

var CoType = {
	Flat			: "flat",
	Circular	: "circular",
	Prov			: "provisional"
};

var RowType = {
	Row : "row",
	Rnd : "rnd"
};

var ColorType = {
	Main		: "MC",
	Contrast	: "CC"
};

var YarnPosType = {
	Front		: "wyif",
	Back			: "wyib"
};

var NodeType = {
	Root				: "Root",
	Pattern			: "Pattern",
	Section			: "Section",
	CastOn			: "CO",
	PickUp			: "PU",
	BindOff		: "BO",
	Join				: "Join",
	Row 				: "Row",
	RowRep 		: "RowRepeat",
	SampleDef 	: "SampleDef",
	SampleCall 	: "SampleCall",
	FixedStRep	: "FixedStRep",
	UStRep 		: "UndeterminedStRep",
	CompSt		: "CompSt",
	Knit				: "K",
	Purl				: "P",
	KnitTBL		: "KB",
	PurlTBL		: "PB",
	KnitBelow	: "KBelow",
	PurlBelow	: "PBelow",
	Slip				: "S",
	SlipKW			: "SK",
	SlipPW			: "SP",
	YarnOver		: "YO",
	KnitFB			: "KFB",
	PurlFB			: "PFB",
	Make			: "M",
	MakeL			: "ML",
	MakeR			: "MR",
	KnitTog		: "KT",
	PurlTog		: "PT",
	SSK				: "SSK",
	SSP				: "SSP",
	PSSO			: "PSSO",
	Expression	: "expr",
	NatLiteral	: "NatLit",
	NatVariable : "NatVar",
	Branch			: "Branch"
};

var MsgType = {
	Error				:	"error-message",
	Warning			:	"warning-message",
	Verification		:	"verification-message"
};

var CompareType = {
	Equal : "eq",
	Less : "lt",
	LessEq : "leq",
	Greater : "gt",
	GreaterEq : "geq"
};
\end{lstlisting}\begin{footnotesize} \textsc{Parts of this Block}: \hyperref[Listing142]{1\-4\-2 on page} \pageref{Listing142}, \hyperref[Listing143]{1\-4\-3 on page} \pageref{Listing143}  \end{footnotesize}\vskip 5mm\noindent

\section{Symbol Lookup Types}
\label{sec:symtype}

\begin{lstlisting}[title={<code/util.js 143>}, label=Listing143]
var CharSym = {
	Comma			:	',',
	Period				:	'.',
	Colon				:	':',
	Asterisk			:	'*',
	PlusOp				:	'+',
	MinusOp			:	'-',
	Semicolon		:	';',
	OpenParen		:	'(',
	CloseParen		:	')',
	OpenBrack		:	'[',
	CloseBrack		:	']',
	OpenAngle		:	'<',
	CloseAngle		:	'>',
	VerticalBar		:	'|',
	Equal				:	'='
};

var SymType = {
	Nat				: "nat",
	Ident			: "ident",
	String			: "\"",
	RowRep		: "**",
	LessEq			: "leq",
	GreaterEq	: "geq",
    EOF				: "!EOF",
    Unknown		: "?unknown"
};

var KeywordSym = {
	Pattern				:	"pattern",
	CastOn				:	"CO",
	PickUp				:	"PU",
	BindOff	 		:	"BO",
	Join					:	"Join",
	CastOnCirc		:	"circular",
	CastOnProv		:	"provisional",
	Section				:	"section",
	Sample				:	"sample",
	From				:	"from",
	To						:	"to",
	Last					:	"last",
	End					:	"end",
	Row					:	"row",
	Rnd					:	"rnd",
	Repeat				:	"repeat",
	With					:	"with",
	YarnInFront		:	"wyif",
	YarnInBack		:	"wyib",
	ColorMain		:	"MC",
	ColorContrast	:	"CC"
};

var StitchSym = {
	Knit				: /^K$/,
	Purl				: /^P$/,
	KnitTBL		: /^KB$/,
	PurlTBL		: /^PB$/,
	KnitBelow	: /^K[1-9][0-9]*B$/,
	PurlBelow	: /^P[1-9][0-9]*B$/,
	Slip				: /^S$/,
	SlipKW			: /^SK$/,
	SlipPW			: /^SP$/,
	//Increases:
	YarnOver		: /^YO$/,
	KnitFB			: /^KFB$/,
	PurlFB			: /^PFB$/,
	Make			: /^M[1-9][0-9]*$/,
	MakeL			: /^M[1-9][0-9]*L$/,
	MakeR			: /^M[1-9][0-9]*R$/,
	//Decreases:
	KnitTog		: /^K[1-9][0-9]*T$/,
	PurlTog		: /^P[1-9][0-9]*T$/,
	SSK				: /^SSK$/,
	SSP				: /^SSP$/,
	PSSO			: /^PSSO$/
};
\end{lstlisting}\begin{footnotesize} \textsc{Parts of this Block}: \hyperref[Listing142]{1\-4\-2 on page} \pageref{Listing142}, \hyperref[Listing143]{1\-4\-3 on page} \pageref{Listing143}  \end{footnotesize}\vskip 5mm\noindent

\chapter{Built-In Examples and Tests}

A number of pattern examples and tests are provided on the web page. These tests are intended to show the variety of
features available in the language but are not exhaustive.

\begin{lstlisting}[title={<Project Display Board Pattern 144>}, label=Listing144]
sample edging with r:
**
row : *K; to end.
repeat r

sample seedStitchBordered with r:
**
row : K 2, *P, K; to last 3, P, K 2.
repeat r

pattern "Project Display Board":
CO 79.
edging with 4.
seedStitchBordered with 70.
edging with 4.
BO 79.
\end{lstlisting}\begin{footnotesize} \textsc{Used in}: \hyperref[Listing159]{P\-a\-t\-t\-e\-r\-n\- \-T\-e\-s\-t\-s on page} \pageref{Listing159}  \end{footnotesize}\vskip 5mm\noindent

\begin{lstlisting}[title={<Market Bag Pattern 145>}, label=Listing145]
sample circleX with n, max
| n < max:
rnd : [K, YO, K n, YO, K] 4.
rnd : *K; to end.
circleX with n + 2, max.

sample diagonalLace with n:
**
rnd : *K2T, YO; to end.
rnd : *K; to end.
repeat n

sample garterStitchCC with n, type
| type = 0:
**
row CC : *K; to end.
row CC : *P; to end.
repeat n
| type = 1:
**
rnd CC : *K; to end.
rnd CC : *P; to end.
repeat n

pattern "Market Bag":
section "Body":
CO 8 circular.
rnd : *K, YO, K; to end.
rnd : *K; to end.
circleX with 1, 23.
diagonalLace with 30.
garterStitchCC with 4, 1.
BO 100.

section "Handle":
PU 10 from "Body top".
garterStitchCC with 2, 0.
row : K, K2T, K 4, K2T, K.
garterStitchCC with 100, 0.
row : K, M1, K 6, M1, K.
garterStitchCC with 2, 0.
Join 10 to "Body top".
\end{lstlisting}\begin{footnotesize} \textsc{Used in}: \hyperref[Listing159]{P\-a\-t\-t\-e\-r\-n\- \-T\-e\-s\-t\-s on page} \pageref{Listing159}  \end{footnotesize}\vskip 5mm\noindent

\begin{lstlisting}[title={<Shawl Pattern 146>}, label=Listing146]
sample shawlRep with m:
row : K 2, YO, K m, YO, K, YO, K m, YO, K 2.
row : K 2, *P; to last 2, K 2.
row : K 2, YO, P m + 2, YO, K, YO, P m + 2, YO, K 2.
row : K 2, *P; to last 2, K 2. 

sample shawlBody with m
| m > 10:
shawlRep with m.
| m <= 10:
shawlRep with m.
shawlBody with m + 4.

pattern "Shawl":
CO 7.
shawlBody with 1.
BO 39.
\end{lstlisting}\begin{footnotesize} \textsc{Used in}: \hyperref[Listing159]{P\-a\-t\-t\-e\-r\-n\- \-T\-e\-s\-t\-s on page} \pageref{Listing159}  \end{footnotesize}\vskip 5mm\noindent

\begin{lstlisting}[title={<Basic Stitches Test 147>}, label=Listing147]
pattern "Basic Sts":
CO 40.
row : K 40.
row : P 40.
row : KB 40.
row : PB 40.
row : K1B 40.
row : P1B 40.
row : S 40.
row : SK 40.
row : SP 40.
row : K2T 20.
row : P2T 10.
row : SSK 5.
row : SSP 2, P.
row : S, K, PSSO, K.
row : K, YO, K.
row : KFB 3.
row : PFB 6.
row : K, M1, K 11.
row : K, M1L, K 12.
row : K, M1R, K 13.
BO 15.
\end{lstlisting}\begin{footnotesize} \textsc{Used in}: \hyperref[Listing159]{P\-a\-t\-t\-e\-r\-n\- \-T\-e\-s\-t\-s on page} \pageref{Listing159}  \end{footnotesize}\vskip 5mm\noindent

\begin{lstlisting}[title={<Compound Stitch Test 148>}, label=Listing148]
pattern "Compound Stitch Test":
CO 20.
row : &lt K, P &gt, K 19.
BO 20.
\end{lstlisting}\begin{footnotesize} \textsc{Used in}: \hyperref[Listing159]{P\-a\-t\-t\-e\-r\-n\- \-T\-e\-s\-t\-s on page} \pageref{Listing159}  \end{footnotesize}\vskip 5mm\noindent

\begin{lstlisting}[title={<Fixed Stitch Repeat Test 149>}, label=Listing149]
pattern "Fixed Stitch Repeat Test":
CO 18.
row : [K, P, K] 6.
BO 18.
\end{lstlisting}\begin{footnotesize} \textsc{Used in}: \hyperref[Listing159]{P\-a\-t\-t\-e\-r\-n\- \-T\-e\-s\-t\-s on page} \pageref{Listing159}  \end{footnotesize}\vskip 5mm\noindent

\begin{lstlisting}[title={<Undetermined Stitch Repeat Test 150>}, label=Listing150]
pattern "Undetermined Stitch Repeat Test":
CO 100.
row : *K, P; to end.
row : K, *P; to last 1, K.
BO 100.
\end{lstlisting}\begin{footnotesize} \textsc{Used in}: \hyperref[Listing159]{P\-a\-t\-t\-e\-r\-n\- \-T\-e\-s\-t\-s on page} \pageref{Listing159}  \end{footnotesize}\vskip 5mm\noindent

\begin{lstlisting}[title={<Row Repeat Test 151>}, label=Listing151]
pattern "Row Repeat Test":
CO 10.
**
row : *K; to end.
repeat 2
BO 10.
\end{lstlisting}\begin{footnotesize} \textsc{Used in}: \hyperref[Listing159]{P\-a\-t\-t\-e\-r\-n\- \-T\-e\-s\-t\-s on page} \pageref{Listing159}  \end{footnotesize}\vskip 5mm\noindent

\begin{lstlisting}[title={<Section Test 152>}, label=Listing152]
pattern "Section Test":
section "first section":
CO 20.
row : K 20.
BO 20.

section "second section":
CO 5.
row : K, P 3, K.
BO 5.
\end{lstlisting}\begin{footnotesize} \textsc{Used in}: \hyperref[Listing159]{P\-a\-t\-t\-e\-r\-n\- \-T\-e\-s\-t\-s on page} \pageref{Listing159}  \end{footnotesize}\vskip 5mm\noindent

\begin{lstlisting}[title={<Sample Test 153>}, label=Listing153]
sample stockinette with m, n:
**
row : K m.
repeat n

pattern "Sample Test":
CO 20.
stockinette with 20, 3.
BO 20.
\end{lstlisting}\begin{footnotesize} \textsc{Used in}: \hyperref[Listing159]{P\-a\-t\-t\-e\-r\-n\- \-T\-e\-s\-t\-s on page} \pageref{Listing159}  \end{footnotesize}\vskip 5mm\noindent

\begin{lstlisting}[title={<Sample Branch Test 154>}, label=Listing154]
sample sampleBranch with m, n
| m = 0:
row : K n.
| m > 0:
row : P n.

pattern "Branch Test":
CO 4.
sampleBranch with 0, 4.
sampleBranch with 1, 4.
BO 4.
\end{lstlisting}\begin{footnotesize} \textsc{Used in}: \hyperref[Listing159]{P\-a\-t\-t\-e\-r\-n\- \-T\-e\-s\-t\-s on page} \pageref{Listing159}  \end{footnotesize}\vskip 5mm\noindent

\begin{lstlisting}[title={<Recursive Sample Test 155>}, label=Listing155]
sample recursiveSample with m, n
| m <= n:
row : P m, *K; to end. 
recursiveSample with m + 1, n.
| m > n:
row : *K; to end.

pattern "Sample Recursion":
CO 20.
recursiveSample with 1, 10.
BO 20.
\end{lstlisting}\begin{footnotesize} \textsc{Used in}: \hyperref[Listing159]{P\-a\-t\-t\-e\-r\-n\- \-T\-e\-s\-t\-s on page} \pageref{Listing159}  \end{footnotesize}\vskip 5mm\noindent

\begin{lstlisting}[title={<Row Type Test 156>}, label=Listing156]
pattern "Row Type Test":
CO 20.
rnd : K 20.
row : P 20.
BO 20.
\end{lstlisting}\begin{footnotesize} \textsc{Used in}: \hyperref[Listing159]{P\-a\-t\-t\-e\-r\-n\- \-T\-e\-s\-t\-s on page} \pageref{Listing159}  \end{footnotesize}\vskip 5mm\noindent

\begin{lstlisting}[title={<Color Options Test 157>}, label=Listing157]
pattern "Color Test":
CO 20.
row MC : K 20.
row CC : P 20.
BO 20.
\end{lstlisting}\begin{footnotesize} \textsc{Used in}: \hyperref[Listing159]{P\-a\-t\-t\-e\-r\-n\- \-T\-e\-s\-t\-s on page} \pageref{Listing159}  \end{footnotesize}\vskip 5mm\noindent

\begin{lstlisting}[title={<Errors Test 158>}, label=Listing158]
patern "Error Test":
CO 20.
row MC : K 20.
row CC : P 19.
row : knit
BO 20.
\end{lstlisting}\begin{footnotesize} \textsc{Used in}: \hyperref[Listing159]{P\-a\-t\-t\-e\-r\-n\- \-T\-e\-s\-t\-s on page} \pageref{Listing159}  \end{footnotesize}\vskip 5mm\noindent

\begin{lstlisting}[title={<Pattern Tests 159>}, label=Listing159]
<div id="tests">

<input type="button" value="Shawl" onclick="javascript:loadTest(test0);" />
<div id="test0" class="pattern-example">
<<Shawl Pattern 146>>
</div>

<input type="button" value="Market Bag" onclick="javascript:loadTest(test01);" />
<div id="test01" class="pattern-example">
<<Market Bag Pattern 145>>
</div>

<input type="button" value="Project Display Board" onclick="javascript:loadTest(test02);" />
<div id="test02" class="pattern-example">
<<Project Display Board Pattern 144>>
</div>

<input type="button" value="Basic Stitches" onclick="javascript:loadTest(test1);" />
<div id="test1" class="test">
<<Basic Stitches Test 147>>
</div>

<input type="button" value="Compound Stitch" onclick="javascript:loadTest(test2);" />
<div id="test2" class="test">
<<Compound Stitch Test 148>>
</div>

<input type="button" value="Fixed Stitch Rep" onclick="javascript:loadTest(test3);" />
<div id="test3" class="test">
<<Fixed Stitch Repeat Test 149>>
</div>

<input type="button" value="Undetermined Stitch Rep" onclick="javascript:loadTest(test4);" />
<div id="test4" class="test">
<<Undetermined Stitch Repeat Test 150>>
</div>

<input type="button" value="Row Repeat" onclick="javascript:loadTest(test5);" />
<div id="test5" class="test">
<<Row Repeat Test 151>>
</div>

<input type="button" value="Sections" onclick="javascript:loadTest(test6);" />
<div id="test6" class="test">
<<Section Test 152>>
</div>

<input type="button" value="Sample" onclick="javascript:loadTest(test7);" />
<div id="test7" class="test">
<<Sample Test 153>>
</div>

<input type="button" value="Sample Recursion" onclick="javascript:loadTest(test71);" />
<div id="test71" class="test">
<<Recursive Sample Test 155>>
</div>

<input type="button" value="Sample Branch" onclick="javascript:loadTest(test72);" />
<div id="test72" class="test">
<<Sample Branch Test 154>>
</div>

<input type="button" value="Row Type" onclick="javascript:loadTest(test8);" />
<div id="test8" class="test">
<<Row Type Test 156>>
</div>

<input type="button" value="Color Options" onclick="javascript:loadTest(test9);" />
<div id="test9" class="test">
<<Color Options Test 157>>
</div>

<input type="button" value="Errors" onclick="javascript:loadTest(test10);" />
<div id="test10" class="test">
<<Errors Test 158>>
</div>
</div>
\end{lstlisting}\begin{footnotesize} \textsc{Used in}: \hyperref[Listing160]{c\-o\-d\-e\-/\-i\-n\-d\-e\-x\-.\-h\-t\-m\-l on page} \pageref{Listing160}  \textsc{Included Blocks}: \hyperref[Listing146]{1\-4\-6 on page} \pageref{Listing146}, \hyperref[Listing145]{1\-4\-5 on page} \pageref{Listing145}, \hyperref[Listing144]{1\-4\-4 on page} \pageref{Listing144}, \hyperref[Listing147]{1\-4\-7 on page} \pageref{Listing147}, \hyperref[Listing148]{1\-4\-8 on page} \pageref{Listing148}, \hyperref[Listing149]{1\-4\-9 on page} \pageref{Listing149}, \hyperref[Listing150]{1\-5\-0 on page} \pageref{Listing150}, \hyperref[Listing151]{1\-5\-1 on page} \pageref{Listing151}, \hyperref[Listing152]{1\-5\-2 on page} \pageref{Listing152}, \hyperref[Listing153]{1\-5\-3 on page} \pageref{Listing153}, \hyperref[Listing155]{1\-5\-5 on page} \pageref{Listing155}, \hyperref[Listing154]{1\-5\-4 on page} \pageref{Listing154}, \hyperref[Listing156]{1\-5\-6 on page} \pageref{Listing156}, \hyperref[Listing157]{1\-5\-7 on page} \pageref{Listing157}, \hyperref[Listing158]{1\-5\-8 on page} \pageref{Listing158}\end{footnotesize}\vskip 5mm\noindent

\chapter{Test Page DOM}

\lstset{language=HTML}
\begin{lstlisting}[title={<code/index.html 160>}, label=Listing160]
<!DOCTYPE html>

<html>
<head>
	<link rel="stylesheet" type="text/css" href="styles/main.css"/>
	<link rel="stylesheet" type="text/css" href="styles/pattern.css"/>
	<script type="text/javascript" src="http://code.jquery.com/jquery-2.1.0.min.js">
	</script>
	<script type="text/javascript" src="util.js"></script>
	<script type="text/javascript" src="lexer.js"></script>
	<script type="text/javascript" src="parser.js"></script>
	<script type="text/javascript" src="codegen.js"></script>
	<script>
	function loadTest(idStr){
		$("#pattern-entry").val($(idStr).text().trim());
	}
	
	function showMessages(msgArr) {
	
		for (var i = 0; i < msgArr.length; i++) {
			var msgObj = msgArr[i];
			var msg = [];
			
			if (msgObj.sectionName != null && msgObj.sectionName != "") {
				msg.push("Section: \'" + msgObj.sectionName + "\'");
			}
			
			if (msgObj.rowIndex > 0) {
				msg.push("Row: " + msgObj.rowIndex);
			}
			
			var cursorPos = 0;
			if (msgObj.line != null) {
				msg.push("Line: " + msgObj.line.num + ":" + msgObj.line.pos);
				cursorPos = msgObj.line.charPos;
			}
			
			msg.push(msgObj.message);
			
			var classStr = " class=\"" + msgObj.messageType + "\"";
			var valueStr = " value=\"" + msg.join(", ") + "\"";
			var onclick = " onclick=\"javascript:moveCursor(" + cursorPos + ");\"";
			var tagStr = "<input type=\"button\"" + classStr  + valueStr + onclick + "/>";
			
			$("#message-display ul").append("<li>" + tagStr + "</li>");
		}
	}
	
	function moveCursor(pos) {
		document.getElementById("pattern-entry").selectionStart = pos;
		document.getElementById("pattern-entry").selectionEnd = pos;
		$("#pattern-entry").focus();
	}
	
	$(document).ready(function() {
		$("#compile input").click(function() {
			$("#pattern-display").empty();
			$("#message-display ul").empty();
			var root = Parser.Parse($("#pattern-entry").val())
			
			if (root.messages.length > 0) {
				showMessages(root.messages);
			}
			var output = PatternTextWriterHTML.Generate(root);
			$("#pattern-display").prepend(output);
		});
	});
	</script>
</head>

<body>

	<div id="header">
		<h1>Purl</h1>
	</div>
	
		<div id="left-content">
			<div id="left-content-wrapper">
				<div id="entry-wrapper">
					<textarea id="pattern-entry" spellcheck="false">pattern "Example Pattern":
section "This is a section of a pattern":
CO 4.
row : P, K, P, K.
row : [P, K] 2.
**
row : *P, K; to end.
repeat 2
BO 4.
					</textarea>
					
					<div id="compile">
						<input type="button" value="Compile" />
				</div>
				</div>
				
				<div id="message-display">
					<ul id="message-list"></ul> 
				</div>
			</div>
		</div>
		
		<div id="center-content">
			<div id="center-content-wrapper">
				<div class="weave-shadow-top"></div>
				<div class="weave-shadow-bottom"></div>
				<div id="tests-wrapper">
					<<Pattern Tests 159>>
				</div>
			</div>
		</div>
		
		<div id="right-content">
			<div id="right-content-wrapper">
				<div id="pattern-display">
					
				</div>
			</div>
		</div>
	
	<div id="ast1-display" class="ast-display"></div>
	<div id="ast2-display" class="ast-display"></div>
	<div id="ast3-display" class="ast-display"></div>
	
	<div id="footer">
	</div>
	
</body>
</html>

\end{lstlisting}\begin{footnotesize} \textsc{Included Blocks}: \hyperref[Listing159]{1\-5\-9 on page} \pageref{Listing159}\end{footnotesize}\vskip 5mm\noindent

\chapter{Style Sheets}

\section{Test Page Styles}

\begin{lstlisting}[title={<code/styles/main.css 161>}, label=Listing161]
*
{
  margin : 0;
  padding : 0;
  border : 0;
  box-sizing : border-box;
}

html, body { height : 100%; }

html
{
  background-image : url("../img/knitting.png");
  padding-top : 50px;
}

#header
{
  position : absolute;
  top : 0;
  left : 0;
  right : 0;
  text-align : center;
  font : 25px overlockblackit;
  height : 50px;
}


#left-content, #right-content
{
  width : 44%;
  height : 100%;
  min-width : 150px;
  min-height : 300px;
  padding : 10px;
}

#center-content { float : left; width : 12%; height : 100%; padding : 0px 0; }
#center-content-wrapper { height : 100%; position : relative; }

#left-content { float : left; }
#right-content { float : right; }

#left-content-wrapper, #right-content-wrapper
{
  width : 100%;
  height : 100%;
  box-shadow : 0px 1px 10px 1px #888888;
}

/*Left content*/

#pattern-entry
{
  display : block;
  width : 100%;
  height : 100%;
  padding : 10px;
  background-color : #333333;
  color : #E6E6E6;
  box-shadow : inset 0px 1px 10px 1px black;
  font : 16px convergence;
  line-height : 150%;
  letter-spacing : 1px;
}

@font-face { font-family : "shadows"; src: url("../res/ShadowsIntoLightTwo-Regular.ttf"); }
@font-face { font-family : "sofia"; src: url("../res/Sofia-Regular.ttf"); }

#entry-wrapper
{
  position : relative;
  height : 85%;
  padding-bottom : 80px;
}

#compile
{
  position : absolute;
  bottom : 0;
  width : 100%;
}

#compile input, #tests input
{
  color : white;
  background-color : #6194FF;
  border : 2px solid #286CFC;
}

#compile input:hover, #tests input:hover { background-color : #709eff; }

#compile input:active, #tests input:active 
{
  box-shadow : inset 0px 0px 1px 1px #286CFC;
}

#compile input
{
  display : table;
  margin : 0 auto;
  padding : 5px 0;
  width : 100%;
  height : 80px;
  font : 30px  overlockblackit;
}

#message-display
{
  width : 100%;
  height : 15%;
  min-height : 60px;
  background-color : #DBDBDB;
  border : 1px solid #CCCCCC;
  overflow-y : auto;
}

ul#message-list li input
{
  white-space : normal;
  width : 100%;
  text-align : left;
}

#message-display .error-message { border : 1px solid red; }
#message-display .warning-message { border : 1px solid yellow; }
#message-display .verification-message { border : 1px solid purple; }

/*Center content*/

#tests-wrapper { height : 100%; padding : 10px 0; }
#tests { width : 100%; height : 100%; margin : 0 auto; padding: 10px; overflow-y : scroll; box-shadow : inset 0px 0px 10px 1px #525252; }
#tests::-webkit-scrollbar { display : none; }

#tests input
{
  width : 100%;
  min-height : 75px;
  padding : 10px;
  margin-bottom : 10px;
  font : 20px overlockblackit;
  white-space : normal;
}

.test, .pattern-example { visibility : hidden; display : none; }


/*Right content*/

#pattern-display
{
  width : 100%;
  height : 100%;
  overflow-y : auto;
  padding : 10px 10px 0 10px;
  background-color : white;
  border : 1px solid #CCCCCC;
}

/*Other Styles*/

.weave-shadow-top
{
	position : absolute;
	width : 100%;
	height: 10px;
	z-index: 5;
	border-bottom : 1px solid #b9b9b9;
	-webkit-box-shadow : 0px 6px 10px -3px #525252;
	box-shadow : 0px 6px 10px -3px #525252;
}

.weave-shadow-bottom
{
	position : absolute;
	width : 100%;
	height: 10px;
	bottom: 0%;
	z-index: 5;
	border-bottom : 1px solid #b9b9b9;
	-webkit-box-shadow: 0px -6px 10px -3px #525252;
	box-shadow: 0px -6px 10px -3px #525252;
}

/*Fonts*/

@font-face { font-family : "artifica"; 	src: url("../res/Artifika-Regular.ttf"); }
@font-face { font-family : "delius"; src: url("../res/Delius-Regular.ttf"); }
@font-face { font-family : "novacut"; src: url("../res/NovaCut.ttf"); }
@font-face { font-family : "novaslim"; src: url("../res/NovaSlim.ttf"); }
@font-face { font-family : "radley"; src: url("../res/Radley-Italic.ttf"); }
@font-face { font-family : "convergence"; src: url("../res/Convergence-Regular.ttf"); }
@font-face { font-family : "overlockreg"; src: url("../res/Overlock-Regular.ttf"); }
@font-face { font-family : "overlockblackit"; src: url("../res/Overlock-BlackItalic.ttf"); }
@font-face { font-family : "overlockit"; src: url("../res/Overlock-Italic.ttf"); }
@font-face { font-family : "overlockblack"; src: url("../res/Overlock-Black.ttf"); }
@font-face { font-family : "overlockbold"; src: url("../res/Overlock-Bold.ttf"); }
@font-face { font-family : "novaround"; src: url("../res/NovaRound.ttf"); }
\end{lstlisting}

\section{Target Language Styles}

\begin{lstlisting}[title={<code/styles/pattern.css 162>}, label=Listing162]
.pattern * { font : 20px overlockreg; }

.patternname { font : 30px overlockblack; }
.sectionname { font : 24px overlockbold; margin-top : 30px; }

.caston, .bindoff, .join, .pickup { margin : 20px 0; }

.body { margin : 10px 0 15px 0; }

.row { margin : 5px 0; }

.rowrepeat { margin : 20px 0; }
.rowrepeat > * { padding-left : 10px; }

.stitch { font-family : overlockit; }

.stitchcount {  font : 16px overlockit; }

.verification, .error, .warning { font : bold 16px arial; float : left; clear : left; margin-right : 5px; }
.verification { color : purple; }
.error { color : red; }
.warning { color : yellow; }

.ast-display { background-color : white; clear : both; display : none; visibility : hidden; }
#ast1-display, #ast2-display { margin-bottom : 110px; }
#ast3-display { margin-bottom : 110px; }

\end{lstlisting}

\end{appendices}

\bibliography{Purl}{}
\bibliographystyle{plain}

\end{document}